\documentclass[aps,prd,groupedaddress,preprintnumbers,%
              showpacs,showkeys,floatfix]{revtex4-1}  
\usepackage{textcomp}
\usepackage[latin1]{inputenc}                    
\usepackage{graphicx}                            
\usepackage{epsfig}
\usepackage{epsf}
\usepackage{latexsym}                            
\usepackage{amsfonts}                            
\usepackage{amssymb}                             
\usepackage{amsmath}                             
\usepackage[mathscr]{eucal}                      
\usepackage{dcolumn}                             
\usepackage{multirow}                               
\usepackage{bm}                                  
\usepackage{hyperref}                            
\usepackage[usenames]{color}
\usepackage{array}
\usepackage{appendix}
\usepackage{bbold}

\graphicspath{{.}{Graphics/}}                    

\newcommand{\be}{\begin{equation}}
\newcommand{\ee}{\end{equation}}
\newcommand{\bea}{\begin{eqnarray}}
\newcommand{\eea}{\end{eqnarray}}

\def\eq#1{Eq.~(\ref{#1})}

\def \3{\ss }

\newcommand{\tr}{\operatorname{Tr}}

\newcommand{\beqn}{\begin{eqnarray}}
\newcommand{\eeqn}{\end{eqnarray}}

\newcommand{\idnty}{\hbox{1$\!\!$1}}

\renewcommand{\arraystretch}{1.8}

\def\cyp{a}
\def\cyi{b}
\def\teml{c}
\def\desy{d}
\def\infn{e}
\begin{document}

\begin{titlepage}
  \begin{flushright}
    {\footnotesize DESY 17-064}
  \end{flushright}
  {\vspace{-0.5cm} \normalsize
  \hfill \parbox{60mm}{
}}\\[10mm]
  \begin{center}
    \begin{LARGE}
      \textbf{Nucleon axial form factors using $N_f=2$ twisted mass fermions with a physical value of the pion mass}
    \end{LARGE}
  \end{center}

 \vspace{.5cm}

 \vspace{-0.8cm}
  \baselineskip 20pt plus 2pt minus 2pt
  \begin{center}
    \textbf{
      C.~Alexandrou$^{(\cyp, \cyi)}$,
	  M.~Constantinou$^{(\teml)}$,
      K.~Hadjiyiannakou$^{(\cyi)}$,
      K.~Jansen$^{(\desy)}$,
      C.~Kallidonis$^{(\cyi)}$,
      G.~Koutsou$^{(\cyi)}$,
      A.~Vaquero Aviles-Casco$^{(\infn)}$ 
      }
\end{center}

  \begin{center}
    \begin{footnotesize}
      \noindent 	
 	$^{(\cyp)}$ Department of Physics, University of Cyprus, P.O. Box 20537, 1678 Nicosia, Cyprus\\	
 	$^{(\cyi)}$ Computation-based Science and Technology Research Center, The Cyprus Institute, 20 Kavafi Str., Nicosia 2121, Cyprus \\
 	$^{(\teml)}$ Temple University,1925 N. 12th Street, Philadelphia, PA 19122-1801, USA \\
 	$^{(\desy)}$ NIC, DESY, Platanenallee 6, D-15738 Zeuthen, Germany \\
 	$^{(\infn)}$  Department of Physics and Astronomy, University of Utah, Salt Lake City, UT 84112, USA\\
     \vspace{0.2cm}
    \end{footnotesize}
  \end{center}

  \begin{abstract}
    We present results on the nucleon axial and induced pseudo-scalar
    form factors using an ensemble of two degenerate twisted mass
    clover-improved fermions with mass yielding a pion mass of
    $m_\pi=130$~MeV. We evaluate the isovector and the isoscalar, as
    well as, the strange and the charm axial form factors. The
    disconnected contributions are evaluated using recently developed
    methods that include deflation of the lower eigenstates, allowing
    us to extract the isoscalar, strange and charm axial form factors.
    We find that the disconnected quark loop contributions are
    non-zero and particularly large for the induced pseudo-scalar form
    factor.
    \begin{center}
      \today
    \end{center}
  \end{abstract}
  \keywords{Nucleon axial form factors, Nucleon Structure, Lattice QCD}
  \maketitle 
\end{titlepage}

\section{Introduction}

Understanding the structure of the nucleon from first principles
constitutes one of the key endeavors of both nuclear and particle
physics. Despite the long history of experimental activity its
structure is not yet fully understood. This includes the portion of
its spin carried by quarks as well as the charge radius of the
proton. While electromagnetic form factors have been well studied
experimentally, the axial form factors are known to less accuracy. An
exception is the nucleon axial charge, which has been measured from
$\beta$-decays to very high precision. Two methods have been
extensively used to determine the momentum dependence of the nucleon
axial form factor. The most direct method is using elastic scattering
of neutrinos and protons, typically $\nu_\mu \; + p \; \longrightarrow
\; \mu^+ \;+ \; n$ ~\cite{Ahrens:1988rr}. The second method is based
on the analysis of charged pion electro-production
data~\cite{Bernard:1992ys} off the proton, which is slightly above the
pion production threshold. The induced pseudo-scalar form factor
$G_p(q^2)$ is even harder to measure experimentally. For the case of
the induced pseudo-scalar coupling $g_p$, a range of muon capture
experiments, as proposed in Ref.~\cite{Bardin:1980mi}, have been
carried out for its determination (see Ref.~\cite{Gorringe:2002xx} for
a review). The form factor $G_p(q^2)$ is less well known, and has only
been determined at three values of the momentum transfer from the
longitudinal cross section in pion
electro-production~\cite{Choi:1993vt}.


Lattice QCD presents a rigorous framework for computing the axial form factors from first principles, in particular in light of the tremendous progress made in simulating the theory at near physical values of the quark masses, large enough volumes and small enough lattice spacings. Having simulations using the physical values of the light quarks eliminates chiral extrapolations, which for the baryon sector introduced a large systematic uncertainty.  In addition, improved algorithms and novel computer architectures have enabled the computation of contributions due to  disconnected quark loops, which previously were mostly neglected.  

In this work we present results for the nucleon axial and induced pseudo-scalar form factors from an ensemble generated with two degenerate quarks with masses fixed approximately to their physical value~\cite{Abdel-Rehim:2015pwa}. We study  both the isovector and isoscalar combinations  as well as the strange and charm form factors, which receive only disconnected contributions. 

The paper is organized as follows: in Section~\ref{sec:ffs} we introduce the axial form factors and the nucleon axial matrix element and in Section~\ref{sec:action} we give details of the lattice action used. In Section~\ref{sec:correlators} we explain our set-up, the correlation functions used and the methods employed to extract the nucleon matrix elements from the lattice data. The renormalization process is described in Section~\ref{sec:renormalization} and in Section~\ref{sec:results} we present our results. Finally, in Section~\ref{sec:conclusions} we conclude.

\section{Axial form factors}
\label{sec:ffs}
\begin{figure}[h!]
  \begin{minipage}[t]{0.4\linewidth} 
    \includegraphics[width=\linewidth]{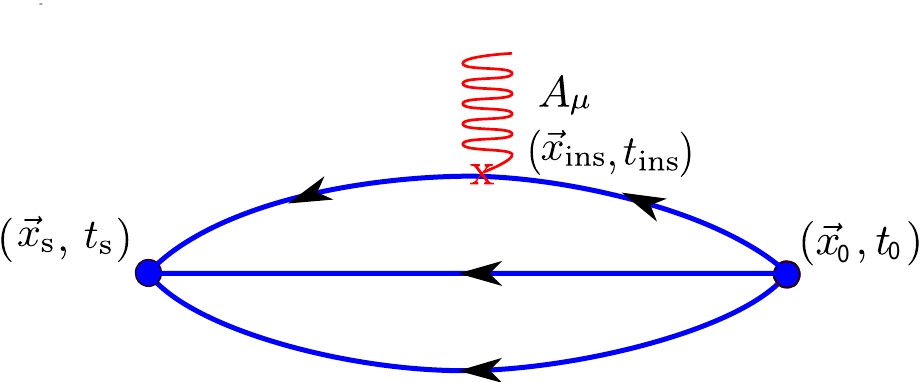}
  \end{minipage} \hfill
  \begin{minipage}[t]{0.4\linewidth}
    \includegraphics[width=\linewidth]{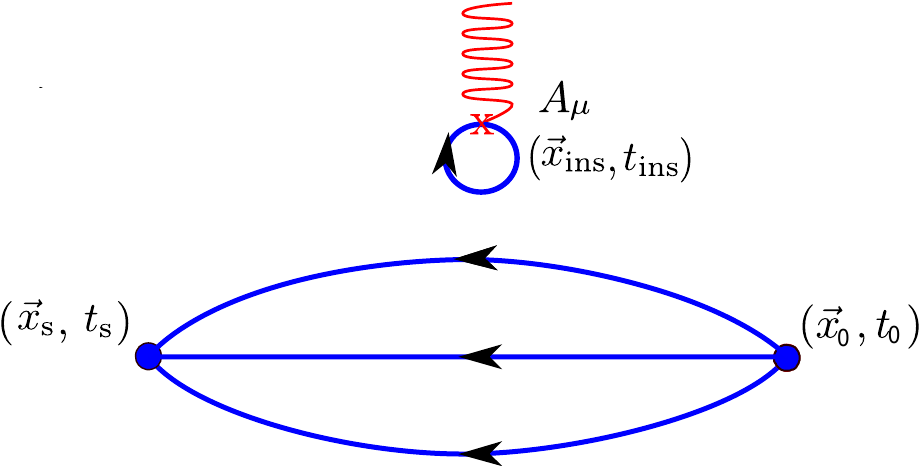}
  \end{minipage}
    \caption{\small Diagrams for the connected (left) and disconnected (right) three-point function. The solid lines represent quark propagators.} 
        \label{fig:3pt_diag}
\end{figure}

To extract the axial and pseudo-scalar form factors one needs to evaluate the nucleon matrix element
\begin{equation}
\langle N(p',s') \vert A_\mu \vert N(p,s) \rangle
\end{equation}
where $A_\mu(x)=\bar{\psi}(x)\gamma_\mu\gamma_5\tau^a\psi(x)$ is the axial-vector current with $\psi(x)=(u(x),d(x))$ the doublet of up- and down-quarks, $\tau^a$ a Pauli matrix acting on flavor components and $p,s$ ($p',s'$) are the momentum and spin of the initial (final) nucleon state, $N$.
The nucleon matrix element of the axial-vector current decomposes into two form factors, $G_A(q^2)$ and $G_p(q^2)$, which are functions of the momentum transfer squared $q^2=(p'_\mu-p_\mu)^2=-Q^2$.
 In a lattice QCD computation one performs a  Wick rotation to imaginary time. Working in Euclidean space the nucleon matrix element of the axial-vector operator can be written in the continuum as 
\begin{equation}
\langle N(p',s') \vert A_\mu \vert N(p,s) \rangle = i \sqrt{ \frac{m_N^2}{E_N(\vec{p}') E_N(\vec{p})} } \bar{u}_N(p',s') \left( \gamma_\mu G_A(Q^2) - i \frac{Q_\mu}{2 m_N} G_p(Q^2) \right) \gamma_5 u_N(p,s),
\label{Eq:matrix_element}
\end{equation}
where $u_N$ are nucleon spinors and $m_N$ and $E_N(\vec{p})$ are the nucleon mass and energy at momentum $\vec{p}$. In this work, we consider the isovector,  isoscalar as well as strange and charm combinations
\begin{equation}
A^{\rm isov}_\mu = \bar{\psi}(x) \gamma_\mu \gamma_5 \tau^3 \psi(x), \,\, A^{\rm isos}_\mu=\bar{\psi}(x) \gamma_\mu \gamma_5 {\mathbb{1}} \psi(x), \,\, A^s_\mu = \bar{s}(x) \gamma_\mu \gamma_5 s(x)\,\,\,\textrm{and}\,\,\,A^c_\mu = \bar{c}(x) \gamma_\mu \gamma_5 c(x).
\end{equation}
In the  isovector case disconnected contributions cancel in the isospin limit. For the isoscalar combination both connected and disconnected contributions enter, while  for the strange and charm form factors we have  only  disconnected contributions. In this work the disconnected contributions  are computed for the first time using simulations with a physical value of the pion mass. The connected and disconnected three-point functions are represented schematically in Fig.~\ref{fig:3pt_diag}.

\section{Lattice action}
\label{sec:action}
In this work we use a single gauge ensemble of two degenerate
($N_f=2$) up and down twisted mass quarks with mass tuned to reproduce
approximately the physical pion mass~\cite{Abdel-Rehim:2015pwa}. The
parameters of our calculation are shown in Table~\ref{table:sim}. The
``Iwasaki" improved gauge action is
used~\cite{Iwasaki:1983ck,Abdel-Rehim:2013yaa} for the gluonic
part. In the fermion sector, the twisted mass fermion action for a
doublet of degenerate quark
flavors~\cite{Frezzotti:2000nk,Frezzotti:2003ni} is employed including
in addition a clover-term~\cite{Sheikholeslami:1985ij}. 

\begin{table}
  \caption{Simulation parameters of the ensemble used here.  The
    nucleon and pion mass and the lattice spacing have been determined
    in Ref.~\cite{Alexandrou:2017xwd}.}
  \label{table:sim}
  \begin{tabular}{c|r@{=}l}
    \hline\hline
    \multicolumn{3}{c}{$\beta$=2.1, $c_{\rm SW}$=1.57751, $a$=0.0938(3)~fm, $r_0/a$=5.32(5)} \\
    \hline
    \multirow{4}{*}{48$^3\times$96, $L$=4.5~fm} & $a\mu_l$  & 0.0009       \\
      & $m_\pi$      & 0.1304(4)~GeV\\
      & $m_\pi L$    & 2.98(1)      \\
    & $m_N$        & 0.932(4)~GeV \\
    & $m_N/m_\pi$  &  7.15(4)\\
      \hline\hline
  \end{tabular}
\end{table}

The fermionic action is given by \be\label{eq:S_tml}
S_F\left[\chi,\overline{\chi},U \right]= a^4\sum_x
\overline{\chi}(x)\left(D_W[U] + i \mu_l \gamma_5\tau^3 - \frac{1}{4}
c_{\rm SW}\sigma^{\mu\nu}\mathcal{F}^{\mu\nu}[U] \right) \chi(x)\;,
\ee where $D_W$ is the Wilson-Dirac operator, $\mu_l$ is the bare
twisted light quark mass and $ - \frac{1}{4} c_{\rm
  SW}\sigma^{\mu\nu}\mathcal{F}^{\mu\nu}[U]$ is the clover-term, with
$c_{\rm SW}$ the so-called Sheikoleslami-Wohlert improvement
coefficient. The field strength tensor $\mathcal{F}^{\mu\nu}[U]$ is
given by~\cite{Sheikholeslami:1985ij}

\be{\label{eq:F_clover}}
\mathcal{F}^{\mu\nu}[U] = \frac{1}{8}\left[P_{\mu,\nu}(x)+P_{\nu,-\mu}(x)+P_{-\mu,-\nu}(x)+P_{-\nu,\mu}(x) - ({\rm h.c.}) \right]\quad,
\ee
where $P_{\mu,\nu}(x)$ is a fundamental $1\times 1$ Wilson plaquette and $\sigma^{\mu\nu} = \frac{1}{2} [\gamma_\mu,\gamma_\nu]$. We take  $c_{\rm SW}=1.57551$ from Ref.~\cite{Aoki:2005et}.
The quark fields denoted by $\chi$ in \eq{eq:S_tml} are in the so-called ``twisted basis". The fields in the ``physical basis" denoted by  $\psi$, are obtained at maximal twist by the  transformation
\be
\psi(x)=\frac{1}{\sqrt{2}}\left(\idnty+ i \tau^3\gamma_5\right) \chi(x),\qquad
\overline{\psi}(x)=\overline{\chi}(x) \frac{1}{\sqrt{2}}\left(\idnty + i \tau^3\gamma_5\right).
\ee
In this paper, unless otherwise stated, the quark fields will be understood as ``physical fields", $\psi$, in particular when we define the interpolating fields.

Twisted mass fermions (TMF) provide an attractive formulation for lattice QCD allowing for automatic ${\cal O}(a)$ improvement, infrared regularization of small eigenvalues and fast dynamical simulations~\cite{Frezzotti:2003ni}. However, the $\mathcal{O}(a^2)$ lattice artifacts that the twisted mass action exhibits lead to instabilities in the numerical simulations, particularly at lower values of the quark masses and influence the phase structure of the lattice theory~\cite{Aoki:1983qi,Sharpe:1998xm,Farchioni:2004us}. The clover-term was added in the TMF action to allow for smaller $\mathcal{O}(a^2)$ breaking effects between the neutral and charged pions that lead to the stabilization of simulations with light quark masses close to the  physical pion mass retaining at the same time the particularly significant ${\cal O}(a)$ improvement that the TMF action features. The reader interested in more details regarding the twisted mass formulation is referred to Refs.~\cite{Frezzotti:2000nk,Frezzotti:2003ni,Frezzotti:2003xj,Frezzotti:2005gi,Boucaud:2008xu} and for the simulation strategy to Refs.~\cite{Abdel-Rehim:2015pwa,Abdel-Rehim:2015owa}.

\section{Lattice evaluation of the nucleon matrix elements}
\label{sec:correlators}
In order to extract the nucleon matrix elements, we need an appropriately defined three-point function and the nucleon  two-point function. To construct these correlation functions one  creates states with the quantum numbers of the nucleon  from the vacuum at some initial time (source) and annihilates them at a later time (sink). The commonly used nucleon interpolating field is given by
\begin{equation}
J(x) = \epsilon^{abc} \left(u^{a\intercal} (x) \mathcal{C} \gamma_5 d^b(x) \right) u^c(x),
\end{equation}
where $\mathcal{C}$ is the charge conjugation matrix. To improve the overlap of this operator with the ground state we employ Gaussian smearing~\cite{Gusken:1989qx,Alexandrou:1992ti} to the quark fields at the source and the sink. In addition, we apply APE smearing~\cite{Albanese:1987ds} to the gauge links entering the hopping matrix in order to reduce unphysical ultra-violet fluctuations.

The three-point function in momentum space can be written as 
\begin{equation}
G_\mu(\Gamma_\nu,\vec{q},\vec{p}\,';t_s,t_{\rm ins},t_0) = \sum_{\vec{x}_{\rm ins},\vec{x}_s} e^{i (\vec{x}_{\rm ins} - \vec{x}_0)  \cdot \vec{q}} \tr \left[ \Gamma_\nu \langle J(t_s,\vec{x}_s) A_\mu(t_{\rm ins},\vec{x}_{\rm ins}) \bar{J}(t_0,\vec{x}_0) \rangle \right] e^{-i(\vec{x}_s - \vec{x}_0)\cdot \vec{p}\,'}\;,
\end{equation}
and the two-point function is given by
\begin{equation}
C(\Gamma_0,\vec{p};t_s,t_0) = \sum_{\vec{x}_s} e^{-i (\vec{x}_s-\vec{x}_0) \cdot \vec{p}} \tr \left[ \Gamma_0  \langle J(t_s,\vec{x}_s) \bar{J}(t_0,\vec{x}_0) \rangle \right]\;,
\end{equation}
where $\vec{q}=\vec{p}\,' -\vec{p}$ is the momentum transfer. For
the two-point function we use the projector $\Gamma_0 =
\frac{1}{4}(\idnty+\gamma_0)$, whereas for the three-point function
the projector is $\Gamma_j=i\Gamma_0 \gamma_5 \gamma_j$, $j=1,2,3$,
which permits the extraction of the axial $G_A(Q^2)$ and the induced
pseudo-scalar $G_p(Q^2)$ form factors. The matrix element can be
extracted by taking appropriate combinations of three- and two-point
functions. An optimal ratio \cite{Hagler:2003jd}, which cancels
unknown overlap terms and time dependent exponentials is,
\begin{equation}
R_\mu(\Gamma_\nu,\vec{p}\,',\vec{p};t_s,t_{\rm ins}) = \frac{G_\mu(\Gamma_\nu,\vec{p}\,',\vec{p};t_s,t_{\rm ins})}{C(\Gamma_0,\vec{p}\,';t_s)} \sqrt{\frac{C(\Gamma_0,\vec{p};t_s-t_{\rm ins}) C(\Gamma_0,\vec{p}\,';t_{\rm ins}) C(\Gamma_0,\vec{p}\,';t_s)}{C(\Gamma_0,\vec{p}\,';t_s-t_{\rm ins}) C(\Gamma_0,\vec{p};t_{\rm ins}) C(\Gamma_0,\vec{p};t_s)}} \;\;
\label{Eq:ratio}
\end{equation}
where we measure all times relative to the time of the source,
i.e. $t_{\rm ins}$ and $t_s$ measure the time separation of the
current insertion and the sink, respectively, from the source. The
ratio becomes time independent in the large time limit yielding a
plateau $\Pi_\mu$ from where the matrix element of the ground state is
extracted, defined via:
\begin{equation}
R_\mu(\Gamma_\nu,\vec{p}\,',\vec{p};t_s,t_{\rm ins})  \xrightarrow[t_s-t_{\rm ins} \gg 1]{t_{\rm ins} \gg 1 \;\;}
 \;\; \Pi_\mu(\Gamma_\nu, \vec{p}\,',\vec{p}).
\label{Eq:plateau}
\end{equation}
In practice, the source-insertion and insertion-sink time
separations cannot be chosen arbitrarily large because the gauge noise
becomes dominant, thus several time separations must be tested to
ensure convergence to the ground state. It is expected that different
matrix elements have different sensitivity to excited states. In the
case of the scalar operator, it has been shown that source-sink
separations larger than $t_s = 1.5$~fm are required in order to damp
out sufficiently excited state effects~\cite{Abdel-Rehim:2016won}. For
the axial-vector current excited state contamination is found to be
less severe at least for pion masses larger than physical used in
previous calculations~\cite{Alexandrou:2010hf,Alexandrou:2007xj}. In
this work we use three values of $t_s$ in the case of the connected
contributions to assess the influence of excited states. As will be
explained, for the case of the disconnected contributions all $t_s$
and $t_{\rm ins}$ values are available. We also employ different
methods to analyze the ratio of Eq.~(\ref{Eq:ratio}) as explained
below. Identifying a time-independent window in this ratio and
extracting the desired matrix element by fitting to a constant is
referred to as the \textit{plateau method}. We seek convergence of the
extracted value as we increase $t_s$.

Instead of using the aforementioned \textit{plateau method} to extract the matrix element of the ground state, another option is to take into account the contribution of the first excited state. The three-point function can  then be expressed as
\begin{eqnarray}
G_\mu(\vec{p}\,',\vec{p},t_s,t_{\rm ins}) &=& A_{00}(\vec{p}\,',\vec{p}) e^{-E_0(\vec{p}\,')(t_s-t_{\rm ins})-E_0(\vec{p})t_{\rm ins}} + A_{01}(\vec{p}\,',\vec{p}) e^{-E_0(\vec{p}\,')(t_s-t_{\rm ins})-E_1(\vec{p})t_{\rm ins}} \nonumber \\
&+& A_{10}(\vec{p}\,',\vec{p}) e^{-E_1(\vec{p}\,')(t_s-t_{\rm ins})-E_0(\vec{p})t_{\rm ins}} + A_{11}(\vec{p}\,',\vec{p}) e^{-E_1(\vec{p}\,')(t_s-t_{\rm ins})-E_1(\vec{p})t_{\rm ins}},
\label{Eq:Thrp_tsf}
\end{eqnarray}
while the two-point function is
\begin{equation}
C(\vec{p},t_s) = c_0(\vec{p}) e^{-E_0(\vec{p}) t_s} + c_1(\vec{p}) e^{-E_1(\vec{p}) t_s}.
\label{Eq:Twp_tsf}
\end{equation}
$E_0(\vec{p})$ and $E_1(\vec{p})$ are the energies of the ground state and first excited state at momentum $\vec{p}$, respectively. For non-zero momentum transfer, fitting to the two- and three-point functions taking into account the contribution of the first excited state  involves twelve fit parameters, namely
$A_{00},\,A_{01},\,A_{10},\,A_{11},\,E_0(\vec{p}\,'),E_0(\vec{p}),\,E_1(\vec{p}\,'),\,E_1(\vec{p}),\,c_0(\vec{p}\,'),\,c_0(\vec{p}),\,c_1(\vec{p}\,'),$ and $c_1(\vec{p})$. 
We note that $A_{01} \neq A_{10}$ for non-zero momentum transfer. The desired nucleon matrix element $\mathcal{M}$ is obtained via
\begin{equation}
{\cal M}=\frac{A_{00}(\vec{p}\,',\vec{p})}{\sqrt{c_0(\vec{p}\,') c_0(\vec{p})}}.
\label{Eq:Tsf}
\end{equation}
In what we refer to as the \textit{two-state fit method} a simultaneous fit is performed to the three- and two-point functions for several values of $t_s$ to obtain $\mathcal{M}$. For the connected three-point function we have three values of $t_s$, namely $t_s/a=$10, 12 and 14, while for the disconnected we have all values since in our approach the loops are computed for all time slices. We find practical to use a maximal time separation $t_s/a$=18, since beyond this separation the correlation functions have large errors and do not contribute to the fit. 
An alternative technique to study excited state effects is the \textit{summation method} \cite{Maiani:1987by,Capitani:2012gj}. Summing over the insertion time $t_{\rm ins}$ of the ratio in Eq.~(\ref{Eq:ratio}) we obtain
  \begin{equation}
R^{\rm{sum}}_\mu(\Gamma_\nu,\vec{p}\,',\vec{p},t_s) \equiv \sum_{t_{\rm ins}=a}^{t_s-a} R_\mu(\Gamma_\nu,\vec{p}\,',\vec{p},t_s,t_{\rm ins})=C + t_s {\cal{M}} + \mathcal{O}(e^{-\Delta t_s}) + \cdots,
\label{Eq:Summ}
  \end{equation}
where we omit the source and sink time slices and sum over the
geometric series of exponentials. The constant $C$ is independent of
$t_s$ and $\Delta$ is the energy gap between the first excited state and the ground state, while the matrix
element of interest ${\cal M}$ is extracted from a linear fit to
Eq.~(\ref{Eq:Summ}) with fit parameters $C$ and ${\cal
  M}$. Alternatively, as described in Ref.~\cite{Savage:2016kon},
one can fit to the finite difference,
\begin{equation}
dR^{\rm sum}_\mu (\Gamma_\nu, \vec{p}\,', \vec{p}, t_s) =   \frac{R^{\rm sum}_\mu (\Gamma_\nu, \vec{p}\,',\vec{p}, t_s+dt_s)-R^{\rm sum}_\mu (\Gamma_\nu, \vec{p}\,', \vec{p}, t_s)}{dt_s}
\end{equation}
of the summation method, which cancels $C$. We have checked that the
two analyses yield consistent results and errors for ${\cal M}$. In
the results we quote for the summation method here, we use a linear
fit to Eq.~(\ref{Eq:Summ}).

We now briefly describe the so-called fixed-sink method employed to compute the connected
contributions to the three-point functions depicted in 
 Fig.~\ref{fig:3pt_diag}. Within this method, the sink time,
momentum, projector and final and initial hadron states are fixed, but
any insertion time-slice and operator with any momentum transfer is
allowed, making it the most appropriate method for the study of form
factors.  An alternative approach is to use stochastic methods to
compute the all-to-all quark propagator from the current insertion to
the sink, which would allow for both varying the current as well as
the sink parameters. This more versatile approach, however, introduces
stochastic noise that has to be controlled~\cite{Alexandrou:2013xon}.
Since in this work we are interested in nucleon form factors, we use
the fixed-sink approach for the connected three-point function, which
allows for obtaining all insertion momenta with practically no
additional computational cost. 
New sets of inversions are
needed for each of the three sink times and each of the three
projectors, while the sink momentum is fixed to zero $\vec{p}\,'=0$. 
To increase further our statistics, we average over sixteen randomly
  selected source positions per gauge configuration.

The disconnected contribution to the three-point function requires the computation of the  disconnected quark loop given by
\begin{equation}
L^{(f)}(t_{\rm ins},\vec{q}) = \sum_{\vec{x}_{\rm ins}} \mathrm{Tr} \left[ M_f^{-1}(x_{\rm ins};x_{\rm ins})\mathcal{G} \right] e^{+i \vec{q} \cdot \vec{x}_{\rm ins}}
\label{Eq:Loop}
\end{equation}
correlated with the nucleon two-point function. With $M_f^{-1}$ we denote the inverse of the twisted mass clover-improved Dirac matrix for the quark flavor $q_f$ and with $\mathcal{G}$ a general $\gamma$-structure. For the axial-vector current $\mathcal{G} = \gamma_\mu\gamma_5$. Eq.(\ref{Eq:Loop}) requires information from the all-to-all propagator, which is prohibitively expensive to calculate by the standard inversion of the Dirac matrix. For a typical lattice size one needs $\mathcal{O}(10^8)$  inversions to compute the disconnected quark loop exactly.
The standard approach to overcome this difficulty  is to use  stochastic techniques to obtain an unbiased estimate for the quark loop at the expense of introducing  stochastic noise~\cite{Bitar:1989dn}. Stochastic techniques have been employed successfully in recent studies including our previous studies as for example in Refs.~\cite{Alexandrou:2012zz,Alexandrou:2013wca}. 
For certain flavor and operator combinations, such as for example the isoscalar of a flavor doublet of the scalar operator,  the twisted mass formulation has a powerful advantage. Such an operator transforms into an isovector
of the pseudo-scalar operator in the twisted mass formulation
at maximal twist. For the u- and d-flavor doublet
we have $\bar{u}u+\bar{d}d = i\bar{\chi}_u\gamma_5\chi_u-i\bar{\chi}_d\gamma_5\chi_d$ where ${\chi}_u$ and ${\chi}_d$
are the two degenerate light quark fields in the twisted
mass basis. 
The disconnected quark loop contribution to $\sigma_{\pi
  N}$ therefore becomes~\cite{Michael:2007vn}
\begin{eqnarray}
& & \sum_{\vec{x}_{\rm ins}}{\rm Tr} \left[i\gamma_5M^{-1}_{\chi_u}(x_{\rm ins};x_{\rm ins})-i\gamma_5M^{-1}_{\chi_d}(x_{\rm ins};x_{\rm ins})\right]\nonumber \\
& =&2\mu_l\sum_{y,\vec{x}_{\rm ins}}{\rm Tr}\left[\gamma_5M^{-1}_{\chi_u}(x_{\rm ins};y)\gamma_5M^{-1}_{\chi_d}(y;x_{\rm ins})\right].
\label{mu-reduction}
\end{eqnarray}
In other words a subtraction of propagators is replaced by a multiplication resulting in increasing the signal-to-noise ratio from $1/\sqrt{V}$ to $V/\sqrt{V^2}$ due to the appearance of an effective double sum over the volume. 
In this form, stochastic
techniques can be employed to obtain the trace via the so-called
\textit{one-end trick}~\cite{McNeile:2006bz} enabling the accurate
computation of the quark loops at all time insertions $t_{\rm
  ins}$~\cite{Alexandrou:2013wca,Abdel-Rehim:2013wlz}.
This method was  applied to compute the light, strange and charm $\sigma$-terms with good accuracy~\cite{Abdel-Rehim:2016won}. In the case of the axial-vector operator the isoscalar combination does not result into a subtraction in the twisted basis.  However, we can generalize the one-end trick to convert the addition of propagators appearing inside a trace into a multiplication. Namely, one can write
\begin{eqnarray}\label{Eq:prop_sum}
  L^{u+d}(t_{\rm ins};\vec{q}) &=& \sum_{\vec{x}_{\rm ins}} \tr \left[ (M_{\chi_u}^{-1}(x_{\rm ins};x_{\rm ins}) + M_{\chi_d}^{-1}(x_{\rm ins};x_{\rm ins}))\mathcal{G} \right]e^{+i \vec{q} \cdot \vec{x}_{\rm ins}} \nonumber\\
  &=& 2 \sum_{\vec{x}_{\rm ins}}\sum_{y,y'} \tr \left[ M_{\chi_d}^{-1^\dagger}(y',x_{\rm ins}) \gamma_5 \mathcal{G} \gamma_5 D_{WC}(x_{\rm ins};y) M_{\chi_d}^{-1}(y;y') \right] e^{+i \vec{q} \cdot \vec{x}_{\rm ins}},
\end{eqnarray}
where $D_{WC}$ is the Wilson-Clover operator with bare quark mass set to its critical value.
Introducing the stochastic noise vectors $\xi_r$ with the properties
\begin{equation}
\frac{1}{N_r} \sum_r \vert \xi_r \rangle \langle \xi_r \vert = \mathbb{1} + \mathcal{O}\left(\frac{1}{\sqrt{N_r}}\right),
\end{equation}
where $N_r$ is the number of stochastic vectors, the solution vectors $\phi_r = M^{-1}_u \xi_r$, in Eq.~(\ref{Eq:prop_sum}) can be written as
\begin{eqnarray}\label{Eq:one_end_trick}
L^{u+d}(t_{\rm ins};\vec{q}) &=& \frac{2}{N_r} \sum_r \sum_{\vec{x}_{\rm ins}}\sum_y \left[ \phi^\dag_r(x_{\rm ins}) \gamma_5 \mathcal{G} \gamma_5 D_{WC}(x_{\rm ins};y) \phi_r(y) \right] e^{+i \vec{q} \cdot \vec{x}_{\rm ins}} + \mathcal{O}\left(\frac{1}{\sqrt{N_r}}\right).
\end{eqnarray}
Computing the loop in this way still results in increasing the signal-to-noise ratio from $1/\sqrt{V}$ to $V/\sqrt{V^2}$.  We refer to the specific application of the trick as in Eq.~(\ref{Eq:one_end_trick}) as the \textit{generalized one-end trick}, applicable in the case of the axial-vector current where the relative sign between u- and d-quarks does not change in the twisted mass basis. As already pointed out,  the one-end trick allows for the evaluation of  the quark loops for all insertion time-slices, enabling us to couple them with two-point functions for any value of $t_s$ and therefore study thoroughly the excited states behavior.

For computing the strange and charm axial form factors, we use Osterwalder-Seiler \cite{Osterwalder:1977pc} valence strange and charm quarks with masses tuned to reproduce the $\Omega^-$ and $\Lambda_c$ mass respectively. The values we obtain are $a\mu_s=0.0259(3)$ and $a\mu_c = 0.3319(15)$ following the procedure described in Ref.~\cite{Alexandrou:2017xwd}. Since we use Osterwalder-Seiler quarks we have the choice to consider doublets  with $\pm \mu$-value. We thus construct the axial-vector current as $\frac{1}{2}(\bar{f}^+ \gamma_\mu \gamma_5 f^+ + \bar{f}^- \gamma_\mu \gamma_5 f^-)$, where $f=s,c$ and $f^{\pm}$ refers to $\pm \mu_f$, yielding the same expressions as for the light quark doublets $(u,d)$ and thus allowing us to apply the generalized one-end trick.

\begin{figure}[!ht]
 \includegraphics[scale=0.7]{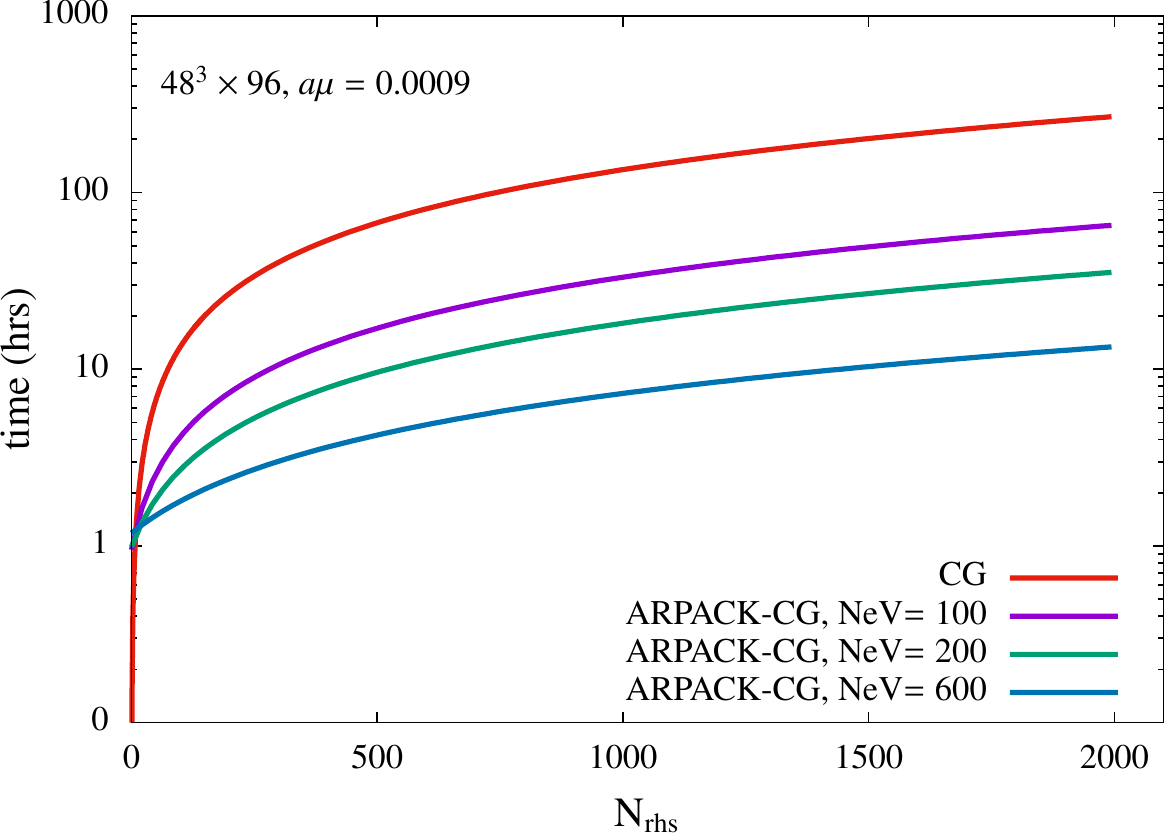}
 \caption{Time in hours needed for the calculation of the disconnected quark loops as a function of the number of right hand sides (rhs). The time includes the buildup of eigenvectors. With the red line we show the standard CG without deflation while with the  purple, green and blue lines we show  the time using deflated CG with 100, 200 and 600 eigenvectors, respectively.}
 \label{Fig:Arpack_CG}
\end{figure}

As the pion mass approaches its physical value, the condition number of the Dirac operator increases, hence the conjugate gradient (CG) algorithm requires a larger number of iterations to converge. One can speedup the solver by calculating the lowest eigenvectors of the Dirac operator and then use them to precondition the conjugate gradient (CG) algorithm, by deflating the Dirac operator. In our calculations, we use the implicitly restarted Lanczos algorithm to calculate the eigenvectors. We found that deflating 600 eigenvectors results in a factor of about $20\times$ speedup for the light quark masses as shown in Fig.~\ref{Fig:Arpack_CG}. For the light fermion loops we calculate 2250 stochastic noise vectors per configuration to high precision (HP), i.e. to a solver precision of 10$^{-9}$. Note that no dilution has been employed, therefore one inversion per noise source is performed.

\begin{figure}[!ht]
\begin{center}
  \begin{minipage}[t]{0.47\linewidth}
    \includegraphics[width=\textwidth]{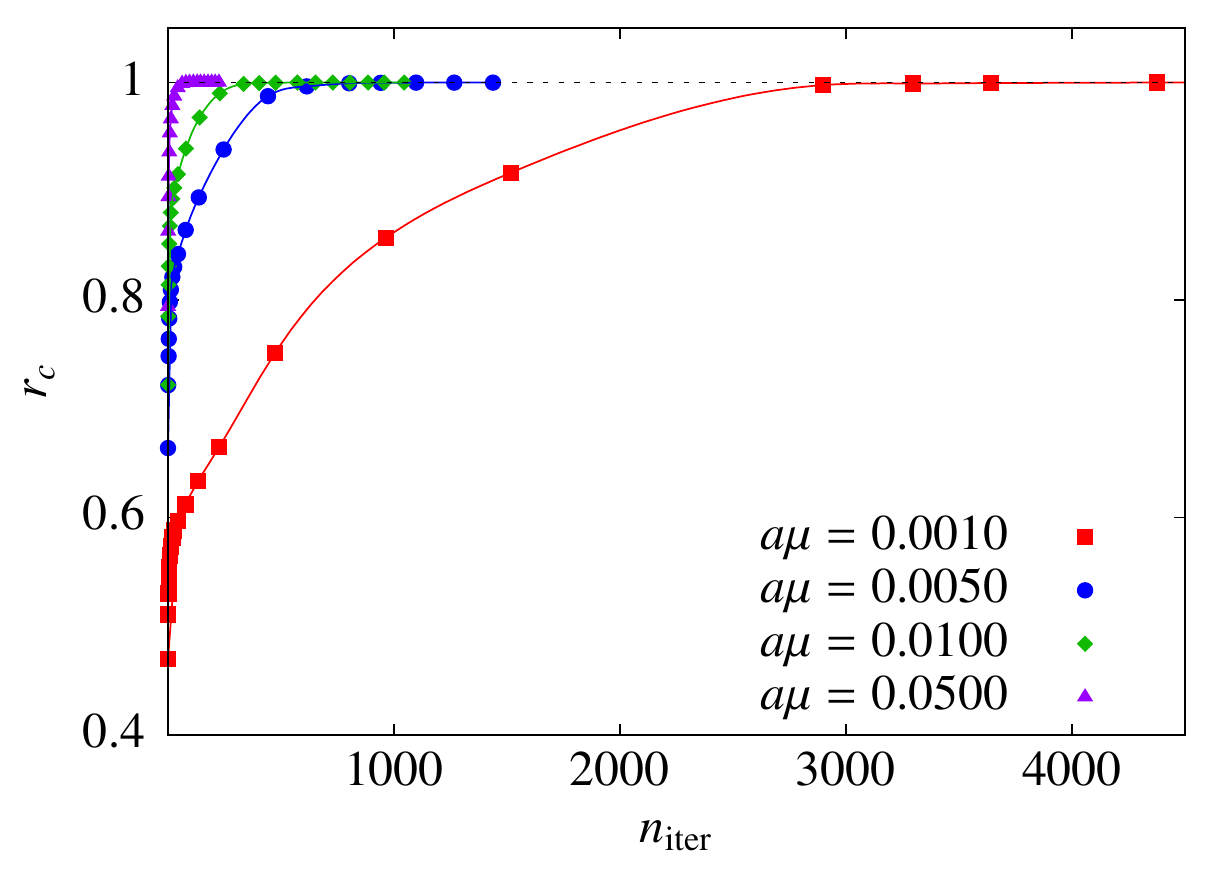}
  \end{minipage}
  \hfill
  \begin{minipage}[t]{0.47\linewidth}
    \includegraphics[width=\textwidth]{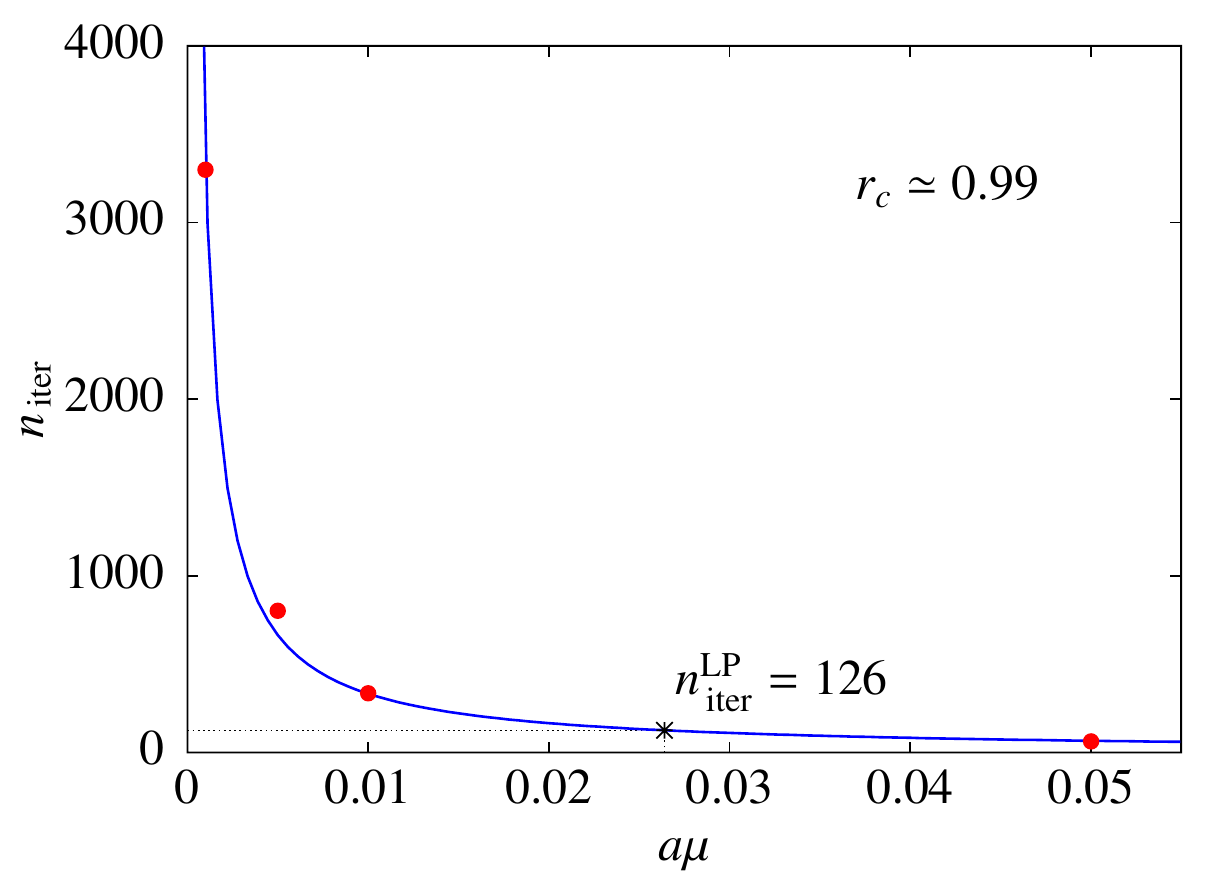}
  \end{minipage}
\hfill
\end{center}  
\caption{Left: The correlation between low and high precision quark loops for a range of twisted mass values. The dashed line shows $r_c=1$. Right: The number of iterations of the low precision inversions as a function of $a\mu$ to yield $r_c=0.99$.}
\label{Fig:TSM}
\end{figure}

For the strange and the charm quarks the condition number of the Dirac operator is significantly smaller, thus there is no need for deflation. Instead, we employ the truncated solver method (TSM)~\cite{Bali:2009hu} where a large number of low precision (LP) noise vectors is used to reduce the stochastic variance and the bias is corrected by a small number of HP inversions. The number of iterations for a LP solve ($n_{\rm LP}$), as well as, the number of low ($N^{\rm LP}_r$) and high ($N^{\rm HP}_r$) precision inversions needs to be tuned in order to produce an unbiased estimate of the disconnected quark loop at optimal computational cost. Namely, the variance $\sigma^2$ can be approximated by~\cite{Blum:2012uh}:
\begin{equation}
  \sigma^2 \propto 2(1-r_c) + \frac{N^{\rm HP}_r}{N^{\rm LP}_r},
\end{equation}
where $r_c$ is the correlation between the targeted observable computed to high and low precision. A compromise is necessary that keeps  the ratio $N^{\rm HP}_r/N^{\rm LP}_r$ small, while having $r_c \simeq 1$. For the strange and charm loops used in this work, we take $n_{\rm LP}$ such that we obtain $r_c\simeq 0.99$. We investigate the dependence of $r_c$ on $n_{\rm LP}$ in the left panel of Fig.~\ref{Fig:TSM}, for various values of the twisted mass parameter. One can see that as $\mu$ decreases a larger number of iterations is needed to obtain the same value for $r_c$. For $a\mu=0.001$, which is very close to our value of $a\mu_l=0.0009$, the number of iterations needed to reach a good correlation is very large, indicating that the TSM is not efficient for light quark masses. In the right panel of Fig.~\ref{Fig:TSM} we show the number of iterations needed to have $r_c\simeq 0.99$ for a given bare quark mass. This figure shows that about $\approx 100$ iterations are sufficient for the case of the strange quark mass $a\mu_s\simeq 0.03$, resulting in a solver precision of $\simeq 10^{-3}$. With the values of $r_c$ and $n_\textrm{LP}$ at hand, we use:
\begin{equation}
  \frac{N^{\rm LP}_r}{N^{\rm HP}_r} = \sqrt{\frac{n_{\rm HP}}{2(1-r_c)n_{\rm LP}}},\label{eq:TSMeq}
\end{equation}
from Ref.~\cite{Bali:2009hu} to determine the ratio $N^{\rm LP}_r/N^{\rm HP}_r$. Eq.~(\ref{eq:TSMeq}) is obtained by requiring minimization of the stochastic variance for equal cost. For the strange quark we test that there is no bias by increasing the resulting $N^{\rm HP}_r$ and observing whether the central value of our observable changes. For the charm quark, the inverter reaches very quickly our target $r_c$ after $\mathcal{O}(10)$ iterations. We therefore increase $n_{\rm LP}$ to yield $r_c\simeq 0.999$ since this increases minimally the total computational cost. This allows us to use a larger value for the $N^{\rm LP}_r/N^{\rm HP}_r$ ratio for the charm loops.
  
The statistics used and the parameters used for the TSM for the strange and the charm quark loops are listed in Table~\ref{Tab:statistics}, with the disconnected fermion loops  calculated for all time-slices. For the connected three-point functions three source-sink time separations have been analyzed for 16 source-positions per gauge configuration, while for the two-point functions 100 source-positions per gauge configuration have been produced  in order to accumulate enough statistics for the disconnected three-point function.

\begin{table}[ht!]
\begin{center}
\renewcommand{\arraystretch}{1.2}
\renewcommand{\tabcolsep}{3.5pt}
\begin{tabular}{ccc||ccccc}
\multicolumn{3}{c||}{Connected} & \multicolumn{5}{c}{Disconnected} \\
\hline\hline
$t_s/a$ & $N_{\rm conf}$ & $N_{\rm src}$ & Flavor & $N_{\rm conf}$ & $N_r^{\rm HP}$ & $N_r^{\rm LP}$ & $N_{\rm src}$ \\
\hline
10 & 579 & 16 & light   & 2120 & 2250 &   -  & 100 \\
12 & 579 & 16 & strange & 2057 & 63   & 1024 & 100 \\
14 & 579 & 16 & charm   & 2034 & 5    & 1250 & 100 \\
\hline
\end{tabular}
\caption{The statistics of our calculation. $N_{\rm conf}$ is the number of gauge configurations analyzed and $N_{\rm src}$ is the number of source positions per configuration. For the disconnected contributions, $N_r^{\rm HP}$ is the number of high-precision stochastic vectors produced, and $N_r^{\rm LP}$ is the number of low-precision vectors used when employing the TSM.} 
\label{Tab:statistics}
\end{center}
\end{table}

\section{Renormalization}
\label{sec:renormalization}
In order to make a comparison of form factors calculated from lattice QCD with experimental and phenomenological results, one must renormalize the lattice results. 
The renormalization functions can be calculated perturbatively as well as non-perturbatively. In this work, we use the  non-perturbatively calculated renormalization functions~\cite{Alexandrou:2012mt} where lattice artifacts are computed perturbatively~\cite{Alexandrou:2015sea} and subtracted from the non-perturbative  results before taking the continuum limit. The Rome-Southampton method~\cite{Martinelli:1994ty}, also known as the ${\rm RI^\prime_{MOM}}$ scheme, is used for the calculation of the renormalization functions. Note that the renormalization function $Z_A$ for the axial current is scheme and scale independent in the chiral limit.
\begin{figure}[!ht]
\begin{minipage}{0.49\linewidth}
\includegraphics[width=\linewidth]{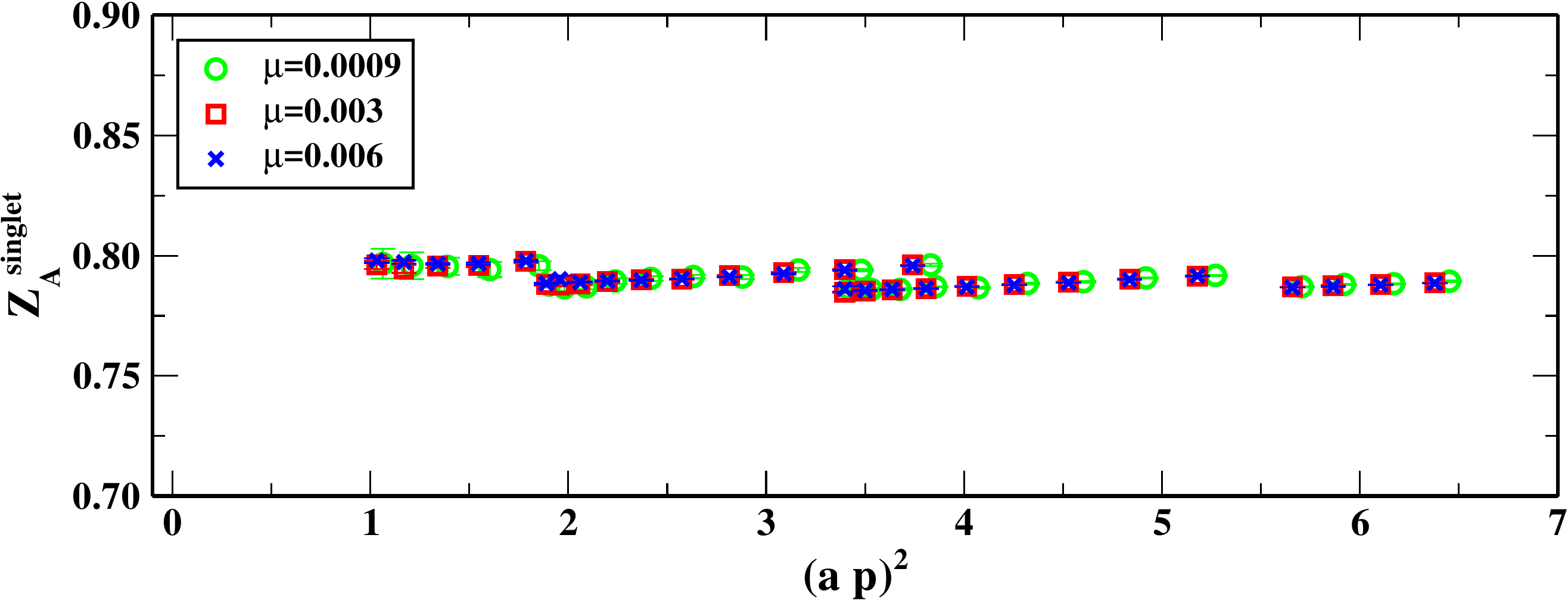}
\end{minipage}\hfill
\begin{minipage}{0.49\linewidth}
\includegraphics[width=\linewidth]{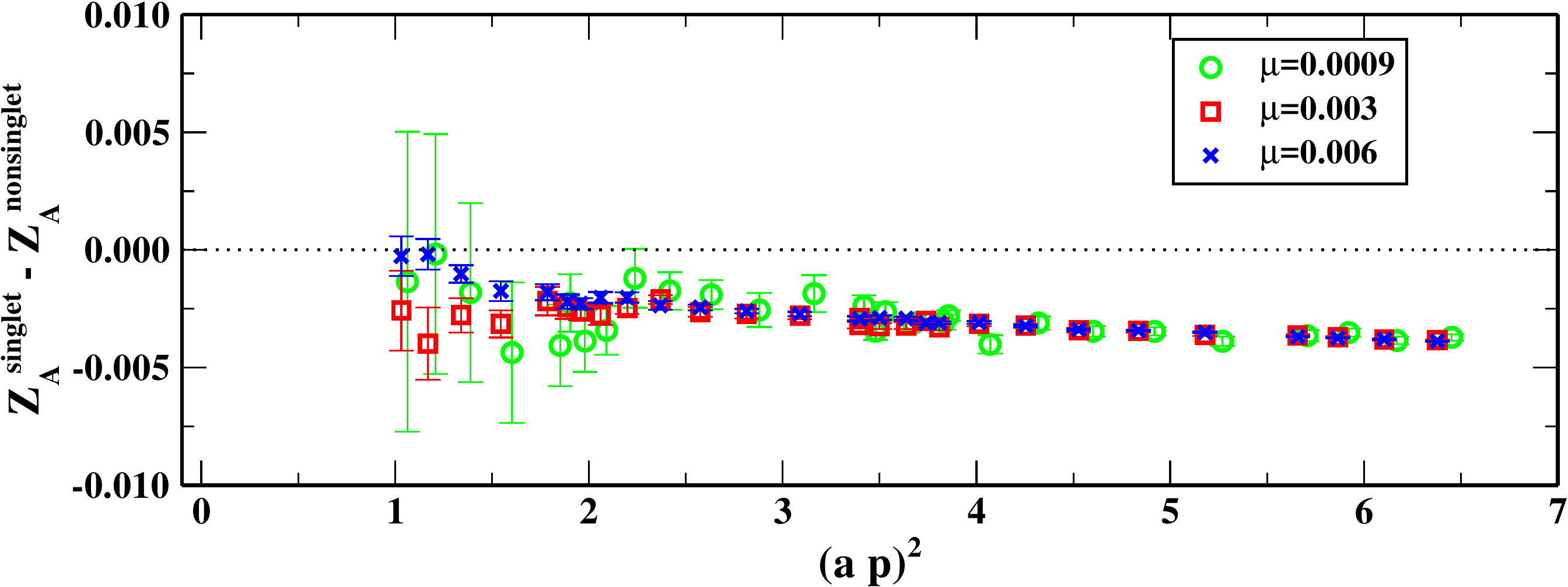}
\end{minipage}
\caption{ The axial singlet renormalization function (left) and the difference between singlet and non-singlet axial renormalization (right)  as a function of $(ap)^2$ for three values of the twisted mass parameter, namely $\mu$ = 0.0009 (open green circles), 0.003 (open red squares) and 0.006 (blue crosses) corresponding to $m_\pi$ = 130, 241, and 331~MeV respectively. }
\label{Fig:ZA_singlet_all_mu}
\end{figure}

\begin{figure}[!ht]
\begin{minipage}{0.42\linewidth}
\includegraphics[width=\linewidth]{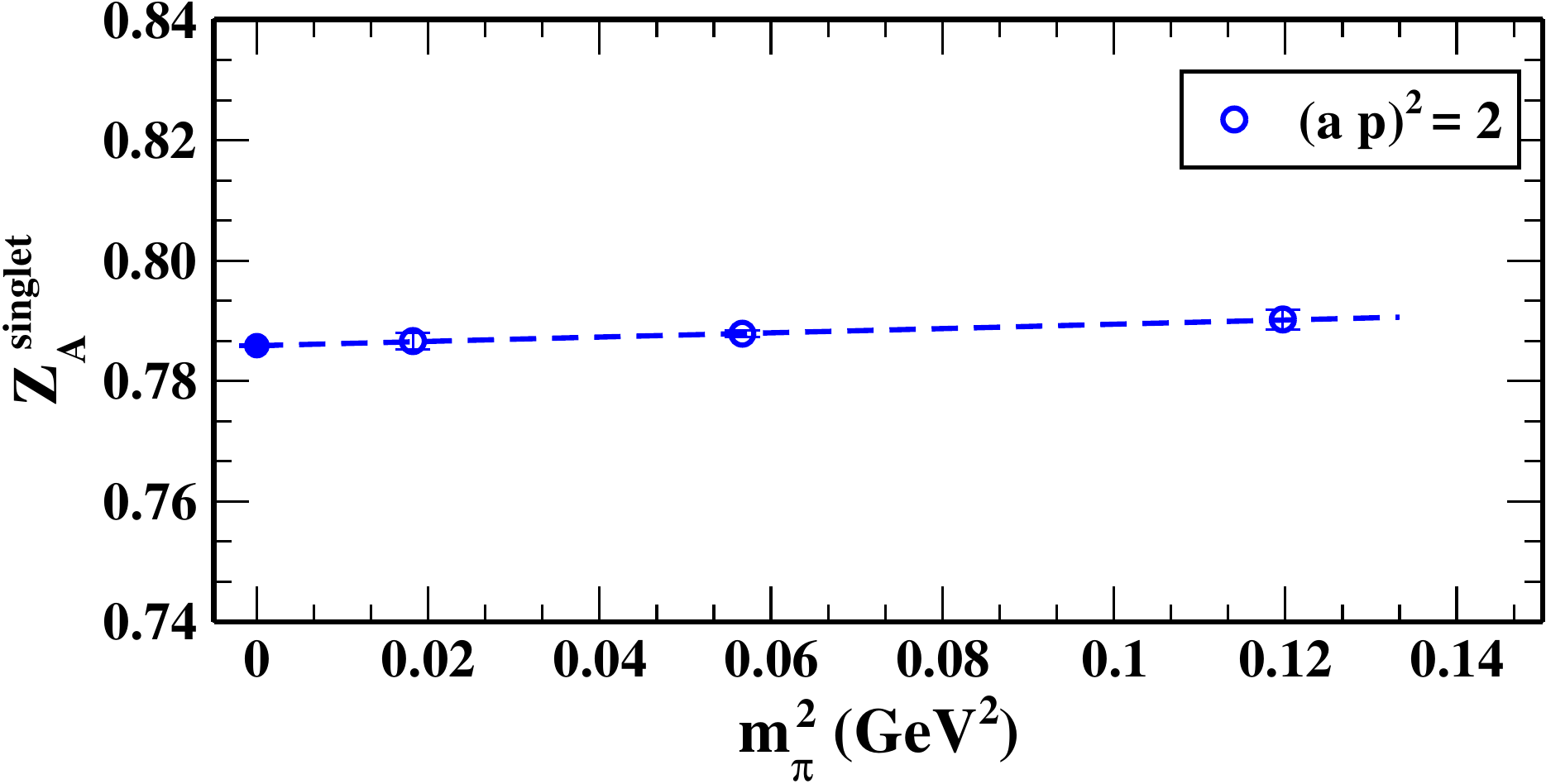}
\end{minipage}
\begin{minipage}{0.56\linewidth}
\includegraphics[width=\linewidth]{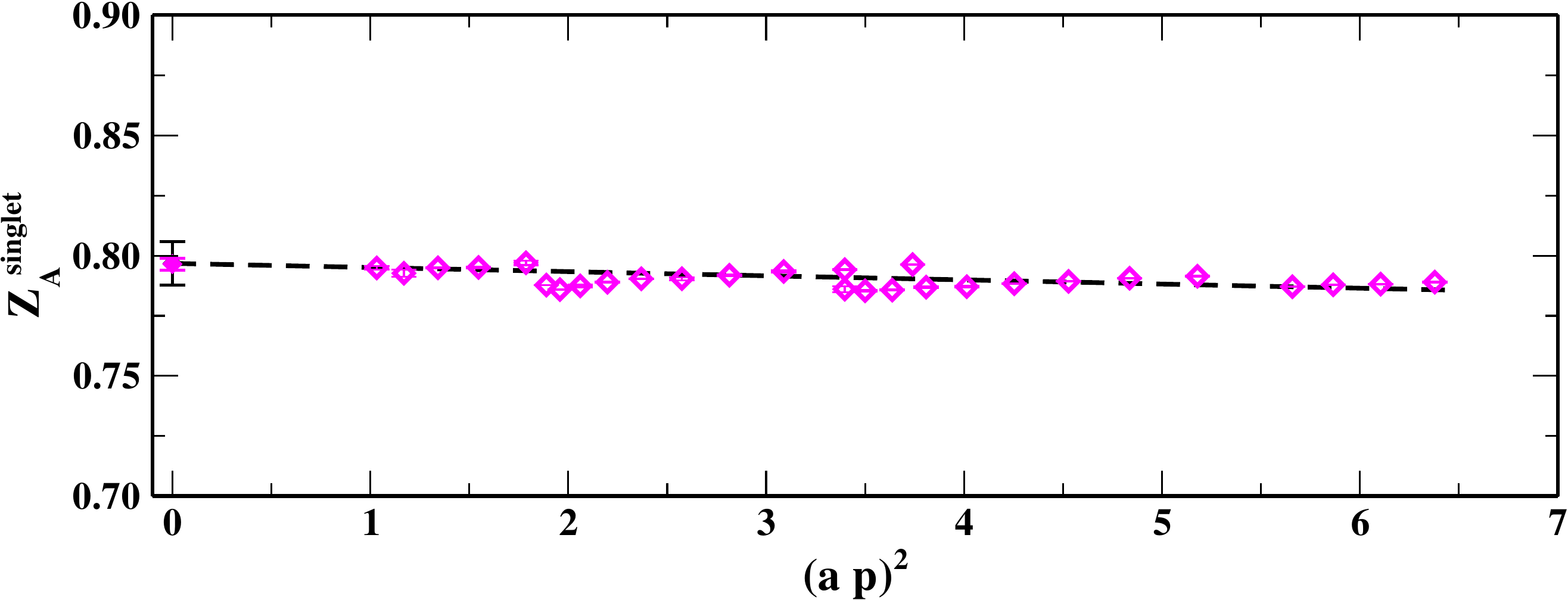}
\end{minipage}
\caption{Left: The axial singlet renormalization function for $(ap)^2=2$ as a function of $m_\pi^2$ (open circles). The dashed line shows a linear fit, and the filled blue circle shows the value at the chiral limit. Right: Continuum extrapolation of the axial renormalization function
  using a linear fit. The extrapolated value is presented by a filled
  diamond and its statistical error is shown with the magenta error
  bar, while the systematic due to the fit range is shown with black.}
\label{Fig:ZA_singlet_continuum_limit}
\end{figure}

In the case of flavor non-singlet operators such as the isovector axial operator, the renormalization functions can be calculated accurately with a relatively low cost, whereas the isoscalar combination receives contributions from a disconnected diagram, leading to significant increase in the computational effort. In order to calculate the renormalization functions non-perturbatively, we consider the bare vertex functions~\cite{Gockeler:1998ye}
\begin{equation}
G^{ns}_{\mathcal{G}}(p) = \frac{a^{12}}{V} \sum_{x,y,z} \langle u(x) \bar{u}(z) \mathcal{G} d(z) \bar{d}(y) \rangle e^{-ip(x-y)}, \;\; G^{s}_{\mathcal{G}}(p) = \frac{a^{12}}{V} \sum_{x,y,z} \langle u(x) \bar{u}(z) \mathcal{G} u(z) \bar{u}(y) \rangle e^{-ip(x-y)}  
\end{equation}  
where $G^{ns}_{\mathcal{G}}$ and $G^{s}_{\mathcal{G}}$ are the
non-singlet and singlet cases, respectively, $V$ is the lattice volume
and, in our case, $\mathcal{G}=\gamma_\mu\gamma_5$. We employ the
momentum source method which offers a high statistical accuracy. In
particular, statistical errors are of the order of \textperthousand~with ${\cal O}(10)$ measurements. The amputated vertex function can be
derived from the vertex function as
\begin{equation}
\Lambda_{\mathcal{G}}(p) = (S(p))^{-1} G_{\mathcal{G}}(p) (S(p))^{-1}
\end{equation}
where $S(p)$ is the propagator in momentum space. For the singlet vertex function the disconnected contribution is amputated using one inverse propagator as the closed quark loop does not have an open leg.

In the ${\rm RI^\prime_{MOM}}$ scheme the renormalization functions are computed by imposing that the amputated vertex function $\Lambda_{\mathcal{G}}(p)$ at large Euclidean scale $p^2=\mu^2$, is equal to its tree-level value in the chiral limit. The renormalization condition is given by
\begin{equation}
Z_q^{-1} Z_{\mathcal{G}} \tr \left[ \Lambda_{\mathcal{G}}(p) \Lambda_{\mathcal{G}}^{\text{tree}} \right] = \tr \left[ \Lambda_{\mathcal{G}}^{\text{tree}} \; \Lambda_{\mathcal{G}}^{\text{tree}}\right]\;\;\; \text{with} \;\;\; Z_q = - \frac{i}{4} \tr \left[ \frac{\frac{1}{a} \gamma_\rho \sin(ap_\rho)}{\frac{1}{a^2} \sum_\rho \sin^2(ap_\rho)} S^{-1}(p)\right] \Bigg \vert_{p_\rho=\mu_\rho}.
\end{equation}
The non-singlet renormalization functions for the ensemble used in
this work can be found in Ref.~\cite{Alexandrou:2015sea}. In
Fig.~\ref{Fig:ZA_singlet_all_mu} we show our results for the axial
singlet renormalization function for three pion masses and for several
initial momenta. As can been seen, the dependence on the light quark
mass is very mild. In Fig.~\ref{Fig:ZA_singlet_continuum_limit} we show
an example of a chiral extrapolation we perform at $(ap)^2=2$, which
corroborates that the pion mass dependence is very weak. In
Fig.~\ref{Fig:ZA_singlet_all_mu} we also show the difference between the
singlet and the non-singlet case for different pion masses and
$(ap)^2$. We observe a small but non-zero difference. The chirally
extrapolated values are shown in
Fig.~\ref{Fig:ZA_singlet_continuum_limit} and they are used to perform
the continuum limit. In general, the momentum source method leads to
small statistical errors and thus a careful investigation of
systematic uncertainties is required. We eliminate the systematic
effect that comes from the asymmetry of our lattices, such as due to
the larger time extent and the antiperiodic boundary conditions in
time, by averaging over the different components corresponding to the
same renormalization function. Furthermore, remaining lattice
artifacts are partially removed by the subtraction of the ${\cal
  O}(g^2\,a^\infty)$ terms as was done in
Refs.~\cite{Constantinou:2009tr,Alexandrou:2015sea}. However, the
largest systematic error comes from the choice of the momentum range
to use for the extrapolation to $(ap)^2 \to 0$. To address this effect
we use different intervals for the $(ap)^2 \to 0$ fit and obtain the
systematic error, shown by the black error bar in
Fig.~\ref{Fig:ZA_singlet_continuum_limit}, by taking the largest
difference in the values of the renormalization function extracted
from different fit ranges. We find for the non-singlet operator that
$Z_A^{ns}=0.7910(4)(5)$, as was originally reported in
Ref.~\cite{Alexandrou:2015sea}, while for the singlet $Z_A^s =
0.7968(25)(91)$. Due to the large systematic error $Z_A^{ns}$ and
$Z_A^s$ are compatible.

\section{Results}
\label{sec:results}
\subsection{Axial charge}
\label{sec:charge}

\begin{figure}[!ht]
\begin{center}
  \begin{minipage}[t]{0.47\linewidth}
    \includegraphics[width=\textwidth]{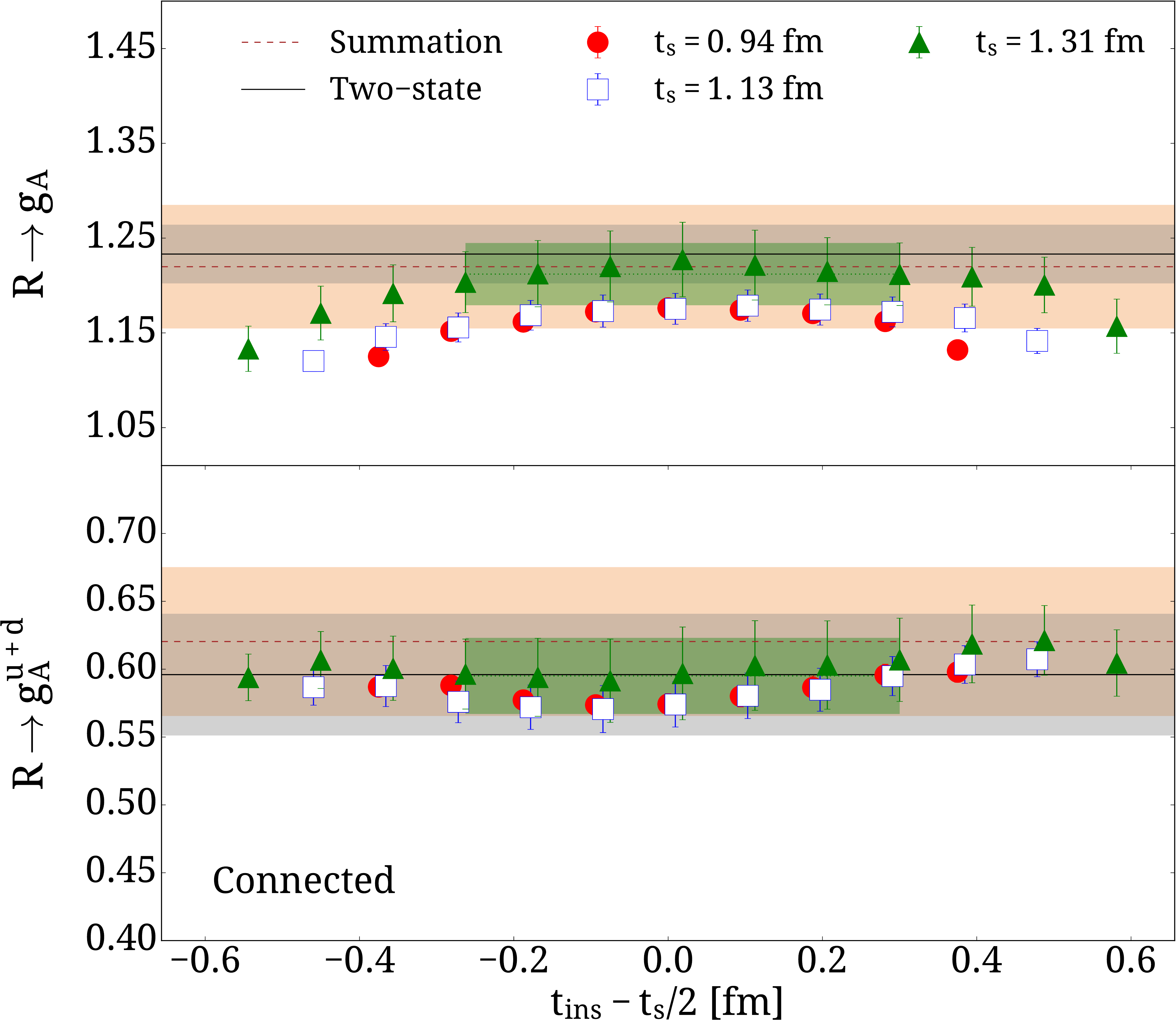}
  \end{minipage}
  \hfill
  \begin{minipage}[t]{0.47\linewidth}
    \includegraphics[width=\textwidth]{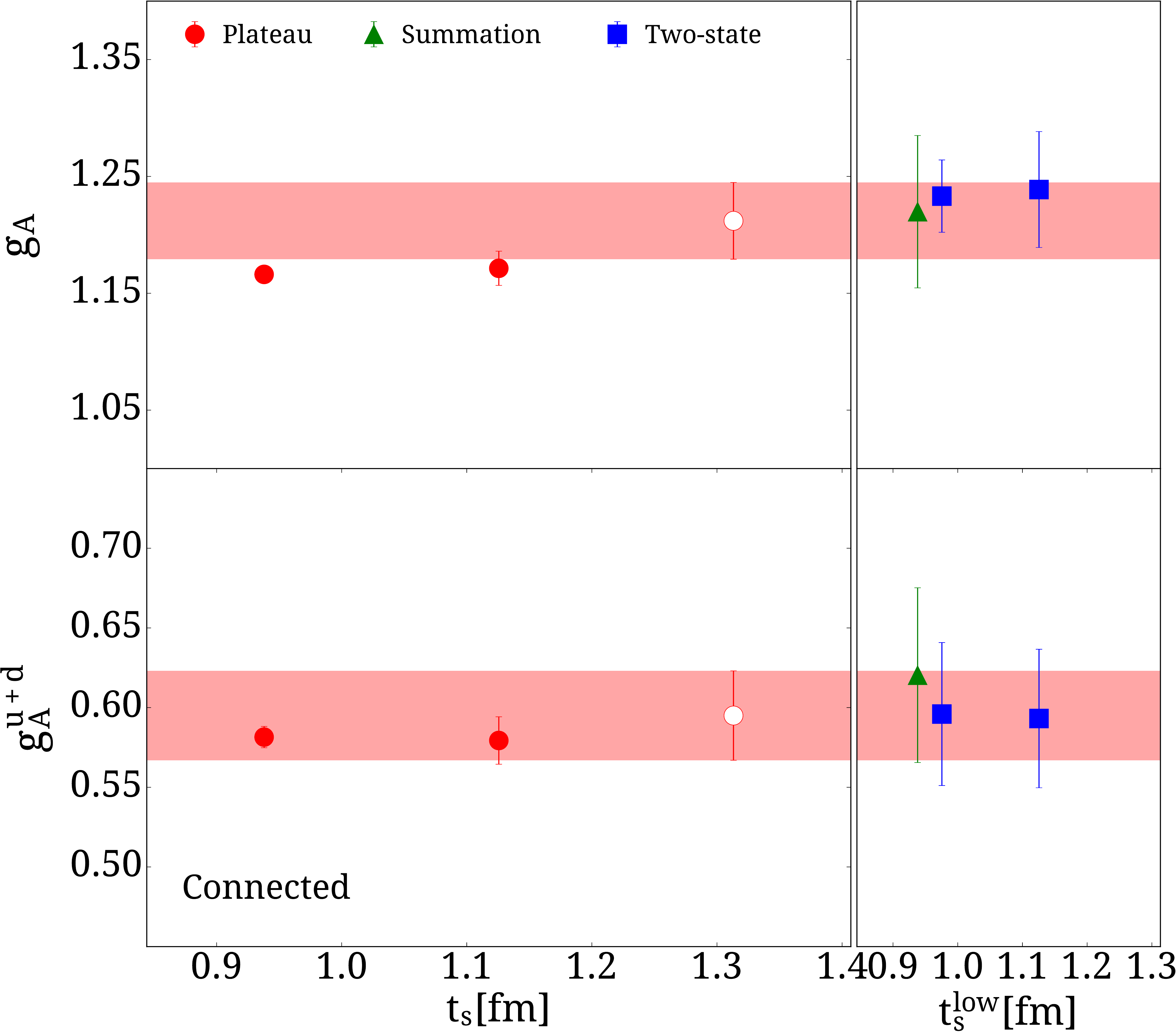}
  \end{minipage}
\caption{Left: The ratio from where we extract the values for $g_A$ and the connected part of $g_A^{u+d}$. Results for the ratio are presented for three source-sink time separations, namely $t_s=0.94,\; 1.13$ and 1.31~fm shown with the filled red circles, open blue squares and filled green triangles, respectively. A fit to the plateau is shown with the dotted line spanning from the initial to final fit $t_{\rm ins}$ and its corresponding error band. Results extracted from the summation method are shown with the brown dashed line and corresponding error band, while results using two-state fits are shown with the solid black line spanning the entire horizontal axis and its corresponding error band. Right: The left column shows the extracted values using the plateau method for $t_s=$ 0.94, 1.13 and 1.31~fm. The open red circle and band shows the plateau value that we take as our final result and its error. The right column  shows the values extracted from the summation method (filled green triangles) and the two-state fits (filled blue squares) as one varies the lowest value of $t_s$, $t_s^{\rm low}$, entering in the fits. Results are slightly shifted to the right for clarity.}
  \label{Fig:ratios_gA_conn}
\hfill
\end{center}  
\end{figure}

\begin{figure}[!ht]
\begin{center}
  \begin{minipage}[t]{0.47\linewidth}
    \includegraphics[width=\textwidth]{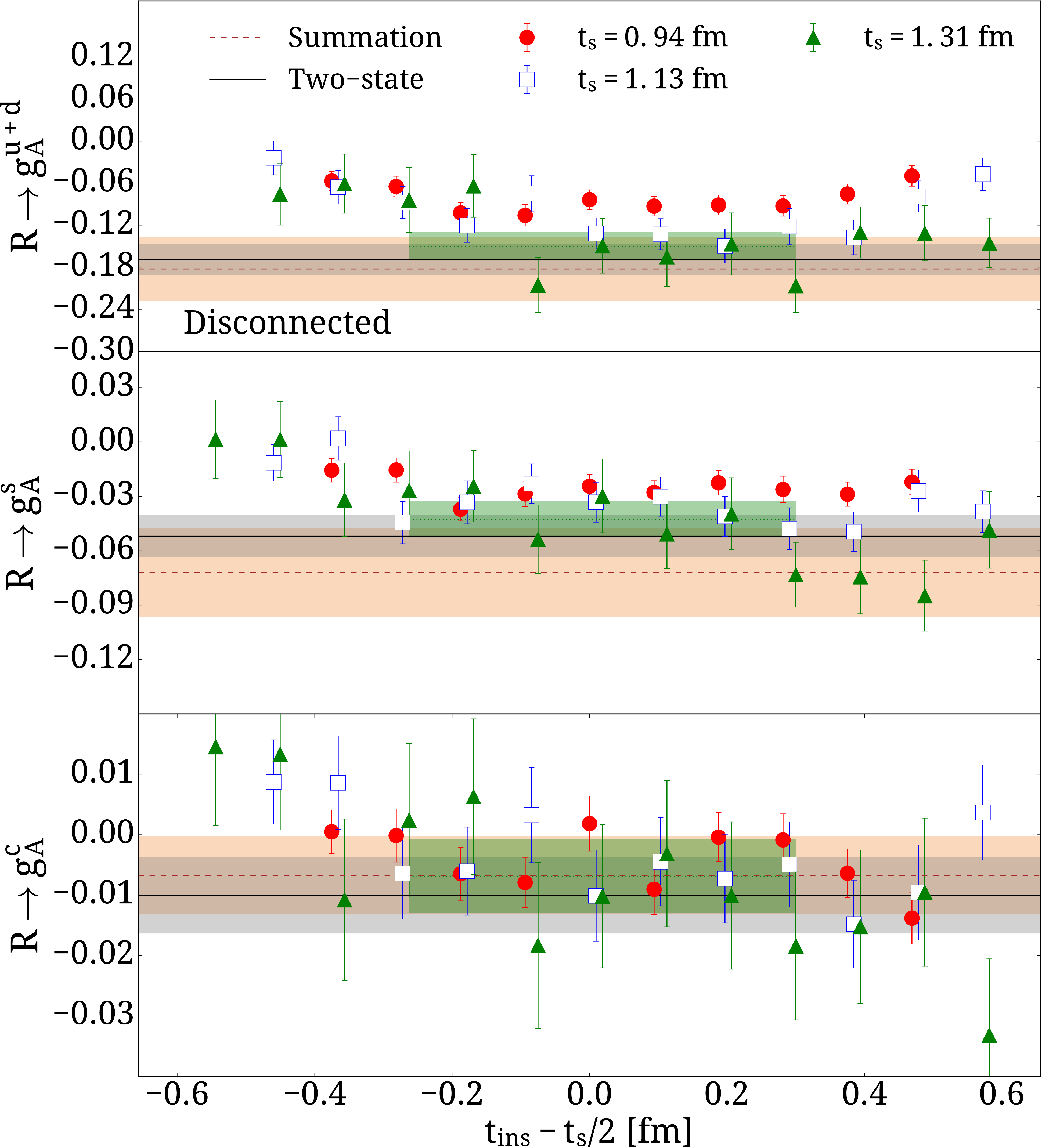}
  \end{minipage}
  \hfill
  \begin{minipage}[t]{0.47\linewidth}
    \includegraphics[width=\textwidth]{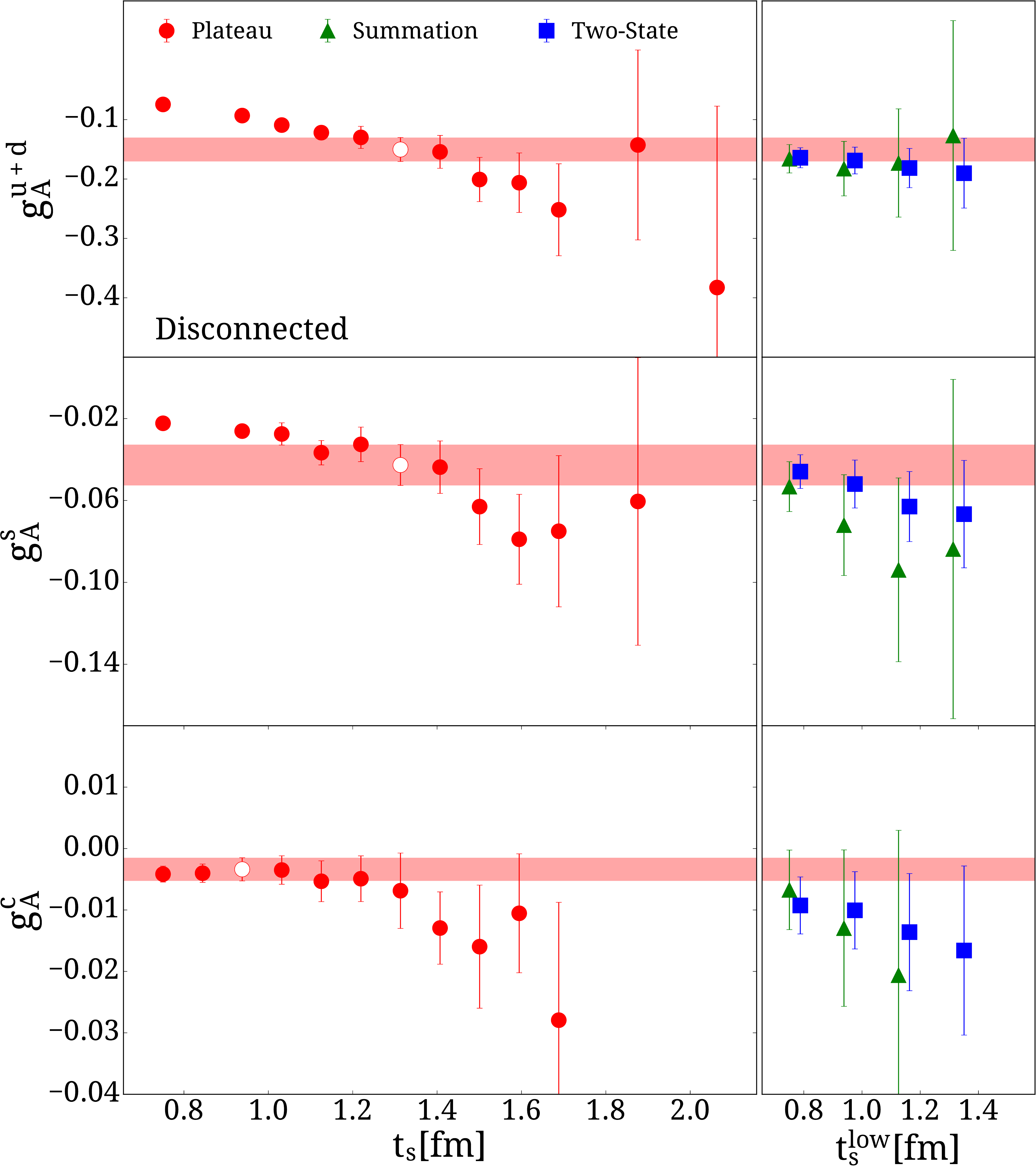}
  \end{minipage}
\caption{Left: The ratio from where we extract the values for $g_A^{u+d}$ disconnected, $g_A^s$ and $g_A^c$. Right: The first column shows the extracted values using the plateau method from $t_s=0.8$~fm to $t_s=2.0$~fm in increments of the lattice spacing $a=0.0938$~fm. The notation is the same as in Fig.~\ref{Fig:ratios_gA_conn}.
  }
  \label{Fig:ratios_gA_disc}
\hfill
\end{center}  
\end{figure}

We first examine the extraction of the axial charge of the nucleon, which is given by $g_A \equiv G_A^{u-d}(0)$. In order to assess the effect of the excited states  we study the ratio of Eq.~(\ref{Eq:ratio}) for various source-sink time separations. In Figs.~\ref{Fig:ratios_gA_conn} and \ref{Fig:ratios_gA_disc} we show the ratio from which we extract the nucleon isovector axial charges $g_A$ and the isoscalar $g_A^{u+d}$ including the disconnected contribution. We also show the corresponding ratios from where $g_A^s$ and $g_A^c$ are determined.

In the case of zero momentum transfer the square root of
Eq.~(\ref{Eq:ratio}) reduces to unity, and the matrix element of
Eq.~(\ref{Eq:matrix_element}) directly yields the axial charge. In
Fig.~\ref{Fig:ratios_gA_conn} we show the ratio of
Eq.~(\ref{Eq:ratio}) for various values of $t_s$ as a function of the
insertion time. The values extracted from the plateau, summation and
two-state fits are collected in the right panel of the figure. As can
be seen, as $t_s$ increases the plateau value converges to a constant
indicating that excited states become very small. When the plateau
value is in agreement with the value extracted from the two-state fit
we consider that contributions from excited states are sufficiently
suppressed. We take the plateau value for the smallest $t_s$ where
agreement with the two-state fit is observed as our final value for
the matrix element. This value is always consistent with the result
from the summation method since the statistical error of the latter is
usually larger as compared to the two-state fit. As a systematic error
due to excited states we take the difference between the plateau value
that demonstrates convergence with $t_s$ and that extracted from the
two-state fit.
 
\begin{figure}[!ht]
\begin{center}
  \includegraphics[scale=0.6]{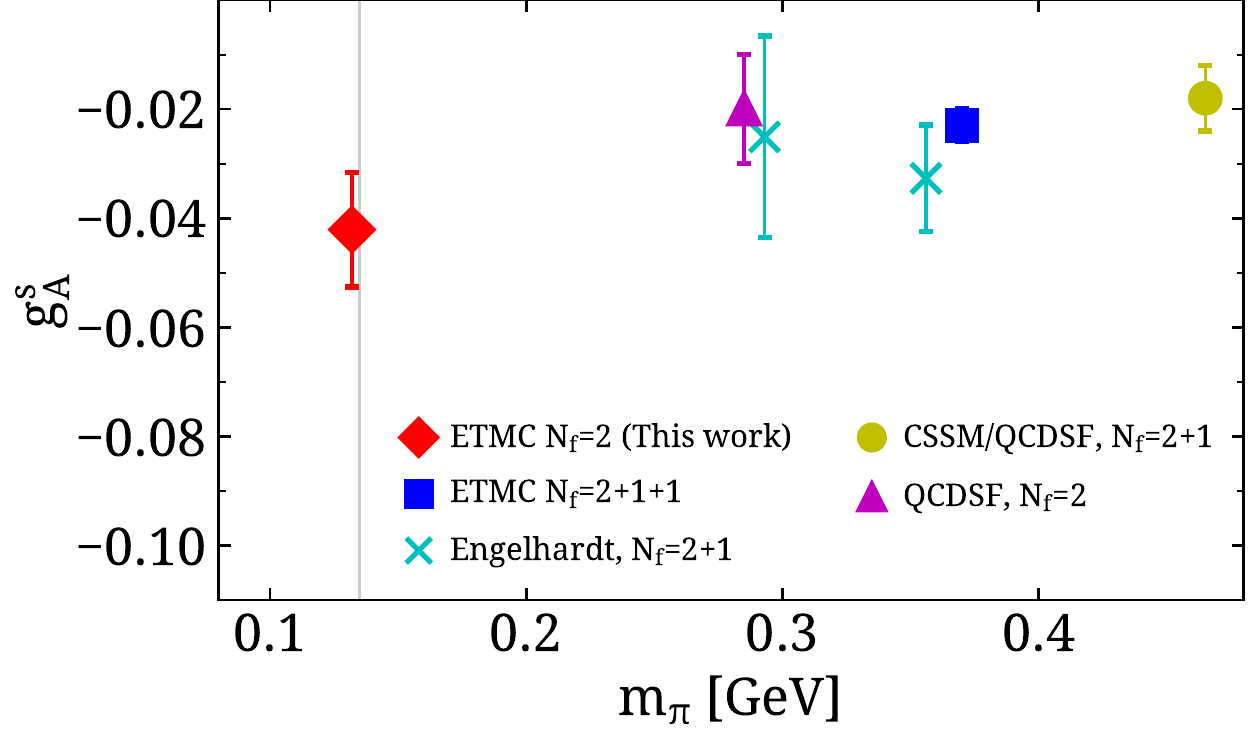}
  \caption{ Comparison of lattice results for $g_A^s$. The current
    study is shown by the red diamond and the result using $N_f=2+1+1$
    twisted mass fermions at heavier than physical pion mass is shown with the blue
    square from Ref.~\cite{Abdel-Rehim:2013wlz}. In addition, we
    compare to results using $N_f=2+1$ clover fermions (yellow circle)
    from Ref.~\cite{Chambers:2015bka}, using $N_f=2$ clover fermions
    (magenta triangle) from Ref.~\cite{QCDSF:2011aa}, and using domain
    wall valence fermions on asqtad configurations (cyan crosses) from
    Ref.~\cite{Engelhardt:2012gd}.  }
  \label{Fig:gAs_world}
\end{center}
\end{figure}

As can be clearly seen from Fig.~\ref{Fig:ratios_gA_disc}, the
disconnected contributions are non-zero and negative. The value of
$g_A^s$ is smaller as compared to the disconnected contribution to
$g_A^{u+d}$. $g_A^c$, although still negative, has a large error and a
small value, namely $|g_A^c| < 0.005$. We note here that the value of
the disconnected contribution to $g_A^{u+d}$ extracted from our
previous study \cite{Abdel-Rehim:2013wlz} using a TMF ensemble
simulated at a pion mass of $m_\pi=370$ MeV is about twice smaller,
namely -0.07(1), compared to the physical point value obtained here.
Lattice artifacts for nucleon observables such as the ones calculated
here are expected to be small. A comparison of results for the axial
charge from various lattice actions including $N_f=2$, $N_f=2+1$ and
$N_f=2+1+1$ flavors of quarks, as well as various lattice spacings and
volumes shows that volume, cut-off and strange quark quenching effects
are smaller than current statistical errors~\cite{Alexandrou:2017oeh}. 
In Fig.~\ref{Fig:gAs_world}, we show a comparison of
lattice results for $g^A_s$. In particlar, we compare results using  $N_f=2$
  clover fermions at a pion mass of about $300$~MeV clover fermions
     from Ref.~\cite{QCDSF:2011aa} with results using domain
    wall valence fermions on $N_f=2+1$ asqtad gauge configurations (hybrid action)   from
    Ref.~\cite{Engelhardt:2012gd}.  Both $N_f=2$ and $N_f=2+1$ results are
    compatible indicating that strange sea quark effects and lattice artifacts are small compared with the statistical errors. The $N_f=2+1$ hybrid action  result at about 370~MeV is also in agreement with the $N_f=2+1+1$ twisted mass fermion result, which was a high accuracy computation. Since we do expect charm quark effects to be negligible, this agreement corroborates between calculations with different actions corroborates  that lattice artifacts are indeed smaller that
    the current statistical errors.

Our values for the nucleon axial charges are tabulated in
Table~\ref{Tab:Axial_charges}. In the case of $g_A$ our result is
compatible with recent results from the
lattice~\cite{Berkowitz:2017gql,Bhattacharya:2016zcn,Green:2012ud,Capitani:2017qpc,Bali:2014nma,Horsley:2013ayv}
and slightly underestimates the experimental value of
1.2723(23)~\cite{Olive:2016xmw}. In the case of $g_A^{u+d}$ there is a
good agreement with the experimental value of
0.416(18)~\cite{Olive:2016xmw} within the current statistics.

\begin{table}[!ht]
	\begin{center}
		\begin{tabular}{ c  c  c }
			\hline\hline
			& This work & Experiment \\
			\hline
			$g_A$ & 1.212(33)(22) & 1.2723(23) \\
			$g_A^{u+d}$ (Conn.) & 0.595(28)(1) & - \\
			$g_A^{u+d}$ (Disc.)& -0.150(20)(19) & - \\
			$g_A^{u+d}$ & 0.445(34)(19) & 0.416(18) \\
			$g_A^u$ & 0.827(30)(5) & 0.843(12) \\
			$g_A^d$ &  -0.380(15)(23) & -0.427(12) \\
			$g_A^s$ & -0.0427(100)(93) & - \\
			$g_A^c$ & -0.00338(188)(667) & - \\
			\hline\hline
		\end{tabular}
	\end{center}
	\caption{We give the values extracted from the plateau method for the isovector and isoscalar axial charges and for the axial charge of  the individual quarks.  The first error is the statistical error determined using jackknife  and the second is the  systematic error due to the excited states  computed as the difference in the mean value between the plateau fit and the two-state fit. The experimental values have been taken from Ref.~\cite{Olive:2016xmw}.}
		\label{Tab:Axial_charges}	
\end{table}

\subsection{Axial and induced pseudo-scalar form factor}
\label{sec:fromfactors}

For non-zero momentum transfer, both $G_A$ and $G_p$ enter in Eq.~(\ref{Eq:matrix_element}), namely the large time limit of Eq.~(\ref{Eq:ratio}) is related to the form factors via
\begin{eqnarray}
\Pi_k(\Gamma_k,\vec{p^\prime},\vec{p}) &=& i G_p(Q^2) C \left[\frac{ \Big( p^\prime_k - p_k \Big) \Big[ \Big(E(\vec{p}) + m_N \Big) p^\prime_k  - \Big(E(\vec{p^\prime}) + m_N \Big)p_k \Big] }{8m_N^3}\right] \nonumber \\
&-& i G_A(Q^2) C \left[\frac{ \Big( E(\vec{p^\prime}) + E(\vec{p}) \Big) m_N + m_N^2 + 2 p^\prime_k p_k - p^\prime_\rho p_\rho }{4m_N^2} \right]
\label{Eq:Axial_i_eq_k}
\end{eqnarray}
in the case where $i=k$, and
\begin{eqnarray}
\Pi_i(\Gamma_k,\vec{p^\prime},\vec{p}) &=& i G_p(Q^2) C \left[\frac{ \Big( p^\prime_i - p_i \Big) \Big[ \Big(E(\vec{p}) + m_N \Big) p^\prime_k  - \Big(E(\vec{p^\prime}) + m_N \Big) p_k \Big] }{8m_N^3}\right] \nonumber \\
&-& i G_A(Q^2) C \left[\frac{ p^\prime_i p_k - p^\prime_k p_i }{4m_N^2} \right]
\label{Eq:Axial_i_neq_k}
\end{eqnarray}
for $i \neq k$ with
\begin{equation}
  C = \frac{2 m_N^2}{E(\vec{p})(E(\vec{p^\prime}) +m_N)} \times \sqrt{\frac{E(\vec{p}) (E(\vec{p^\prime}) +m_N) }{E(\vec{p^\prime}) (E(\vec{p}) +m_N)}}.
  \label{Eq:C_genFrame}
\end{equation}
Since the form factors depend only on the momentum transfer squared ($Q^2$),
while the plateau of Eq. (25) depends on $\vec{p}\,'$ and $\vec{p}$, the extraction of
the form factors is over-constrained. In practice, we form the system:
\begin{equation}
  \Pi_i(k, \vec{p}\,',\vec{p}) = D_i(k, \vec{p}\,',\vec{p})F(Q^2),
\label{Eq:OC}
\end{equation}
where $D$ is an array of kinematic coefficients according to Eqs.~(\ref{Eq:Axial_i_eq_k},~\ref{Eq:Axial_i_neq_k}) and $F$ is the vector: $F^\intercal=(G_A, G_p)$. The system is solved for $F$ by taking the Singular Value Decomposition (SVD) of $D$ in order to minimize:
\begin{equation}
  \chi^2 = \sum_{i,k,\vec{p}\,',\vec{p} \; \in Q^2}\left[\frac{D_i(k,\vec{p}\,',\vec{p})F(Q^2) - \Pi_i(k,\vec{p}\,',\vec{p})}{w_i(k,\vec{p}\,',\vec{p})}\right]^2
\label{Eq:SVD_chisq}
\end{equation}
for each $Q^2$, where $w$ is the statistical error of $\Pi$.

\begin{figure}[!ht]
\begin{center}

  \begin{minipage}[t]{0.47\linewidth}
    \includegraphics[width=\textwidth]{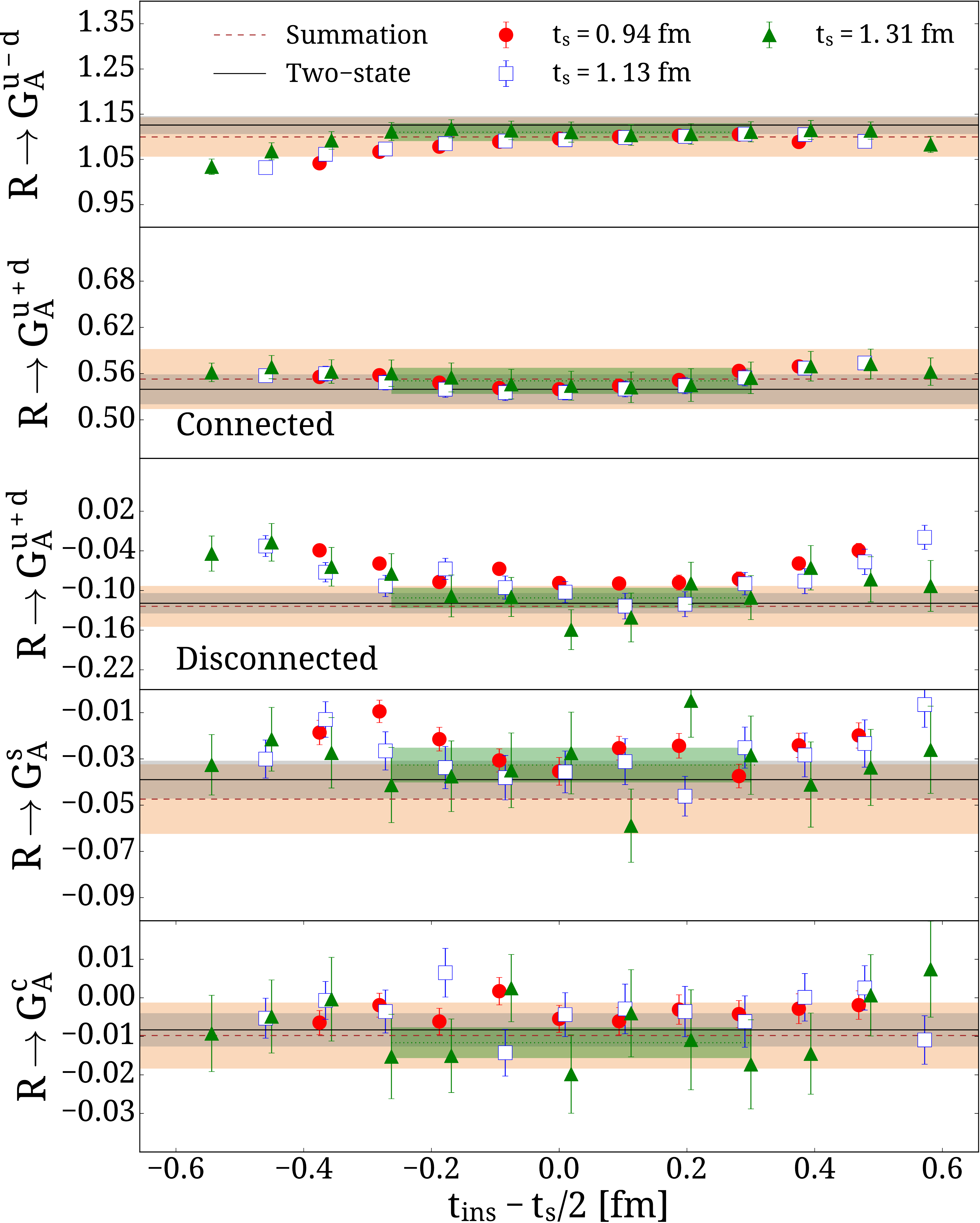}
  \end{minipage}
  \hfill
  \begin{minipage}[t]{0.47\linewidth}
    \includegraphics[width=\textwidth]{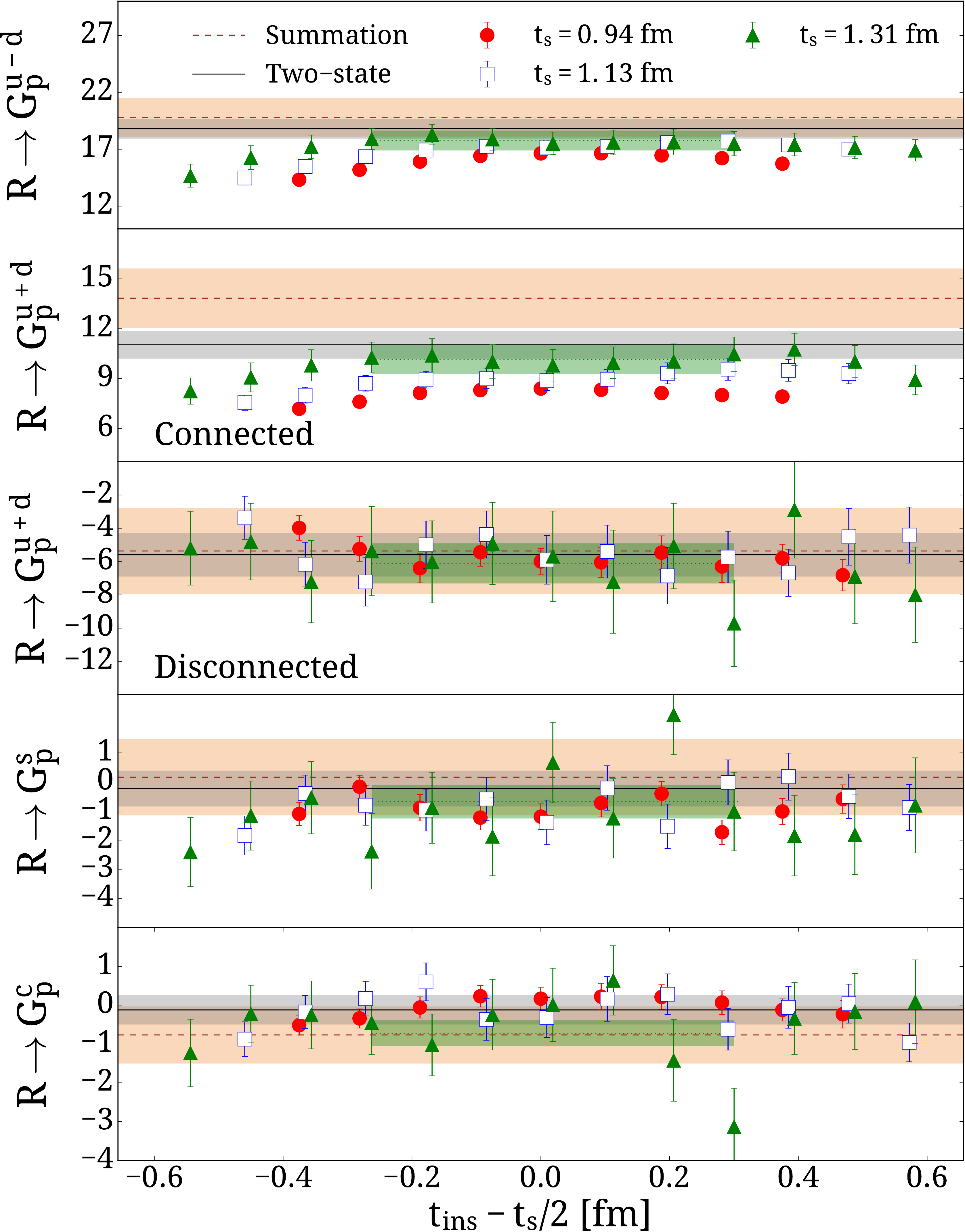}
  \end{minipage}
\caption{The ratio for $G_A$ (left) and $G_p$ (right) obtained as
  explained in the text, for $Q^2=0.0753\; \mathrm{GeV^2}$. From top
  to bottom we present the isovector, connected isoscalar,
  disconnected isoscalar, strange and charm contributions. The
  notation is as in the left panel of Fig.~\ref{Fig:ratios_gA_conn}.}
  \label{Fig:plat_mom1}
\hfill
\end{center}  
\end{figure}

\begin{figure}[!ht]
\begin{center}
  \begin{minipage}[t]{0.47\linewidth}
    \includegraphics[width=\textwidth]{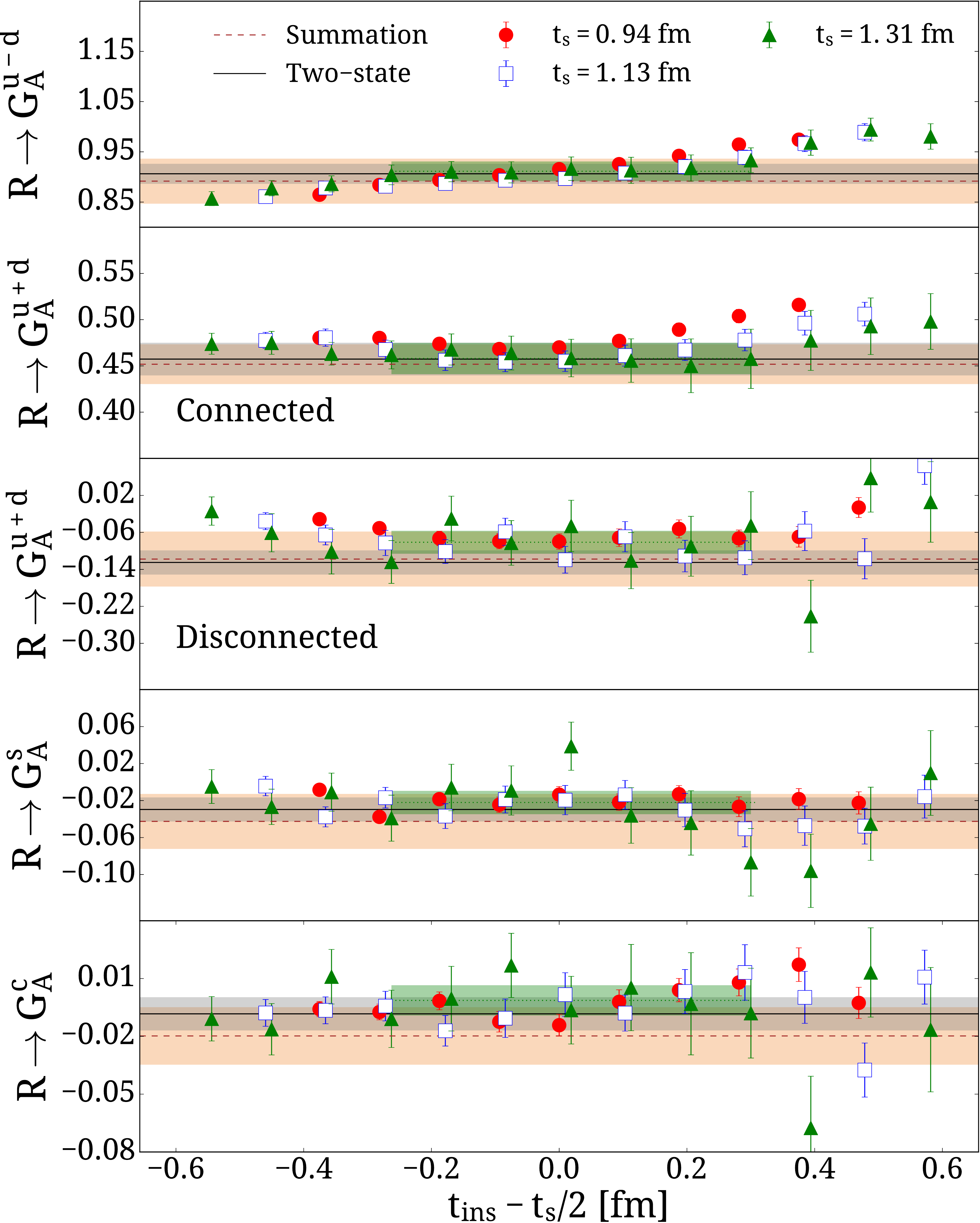}
  \end{minipage}
  \hfill
  \begin{minipage}[t]{0.47\linewidth}
    \includegraphics[width=\textwidth]{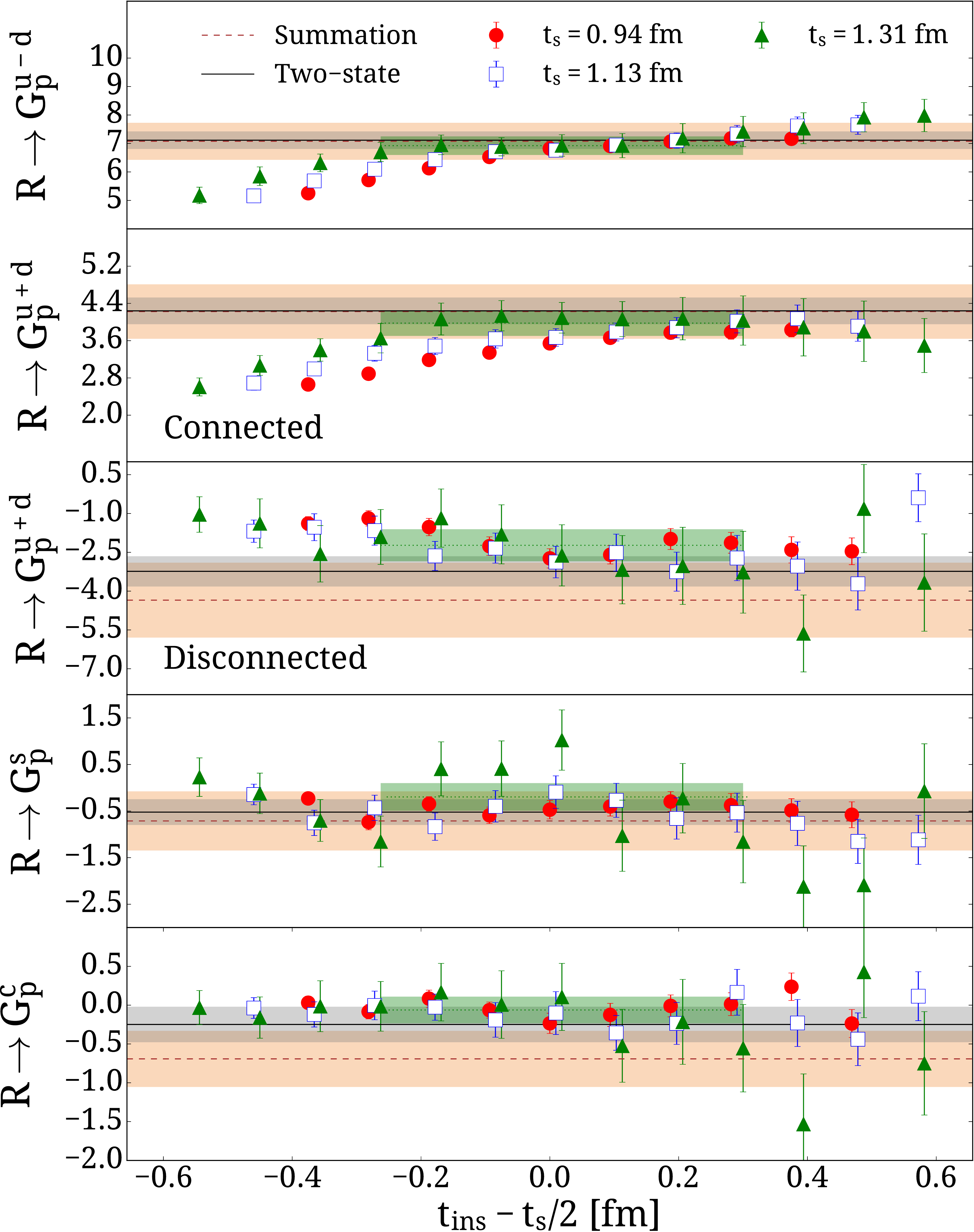}
  \end{minipage}
\caption{The ratio for $G_A$ (left) and $G_p$ (right) obtained as
  explained in the text, for $Q^2=0.2848\; \mathrm{GeV^2}$. The
  notation is as in Fig.~\ref{Fig:plat_mom1}.}
  \label{Fig:plat_mom4}
\hfill
\end{center}  
\end{figure}

\begin{figure}[!ht]
\begin{center}

  \begin{minipage}[t]{0.47\linewidth}
    \includegraphics[width=\textwidth]{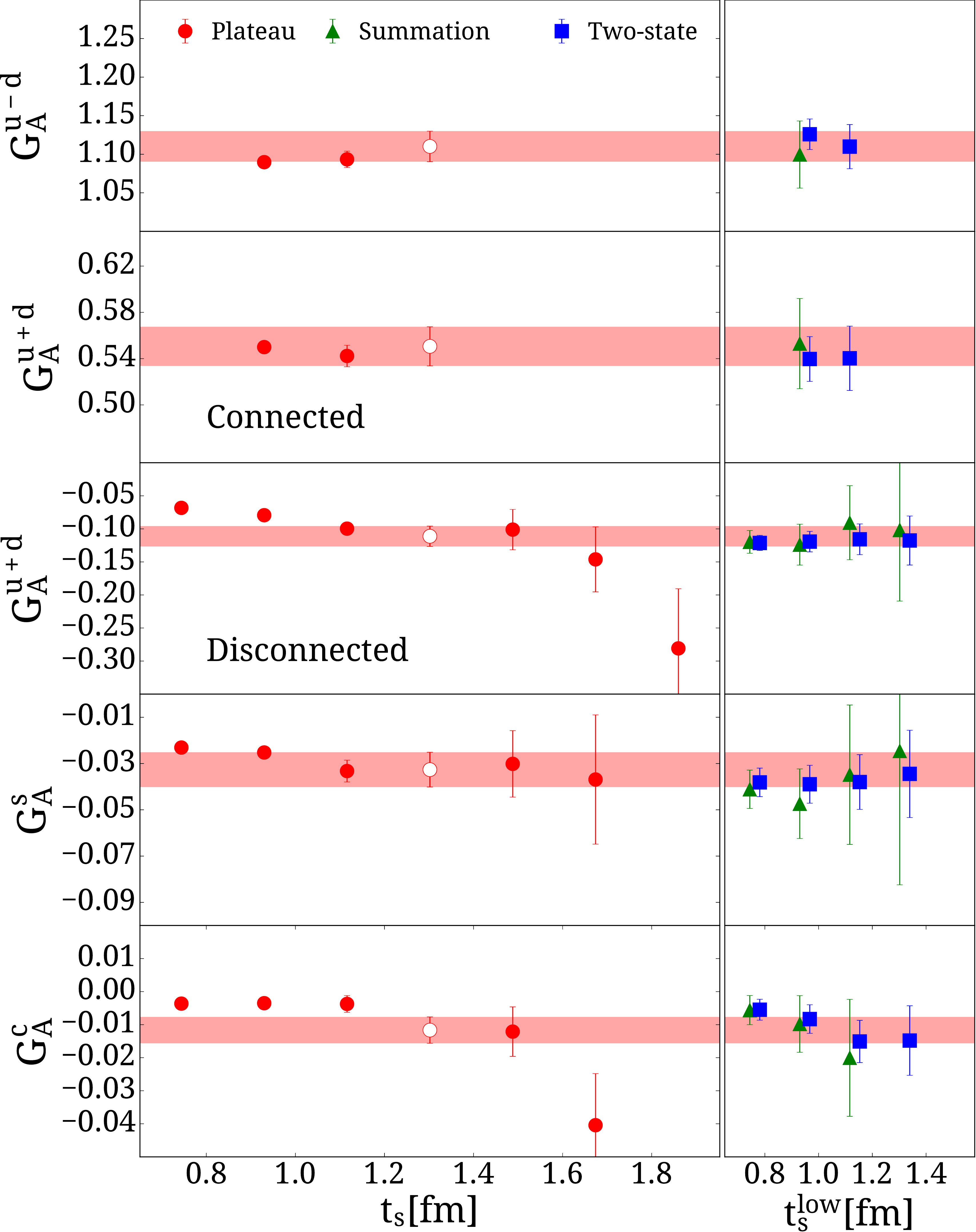}
  \end{minipage}
  \hfill
  \begin{minipage}[t]{0.47\linewidth}
    \includegraphics[width=\textwidth]{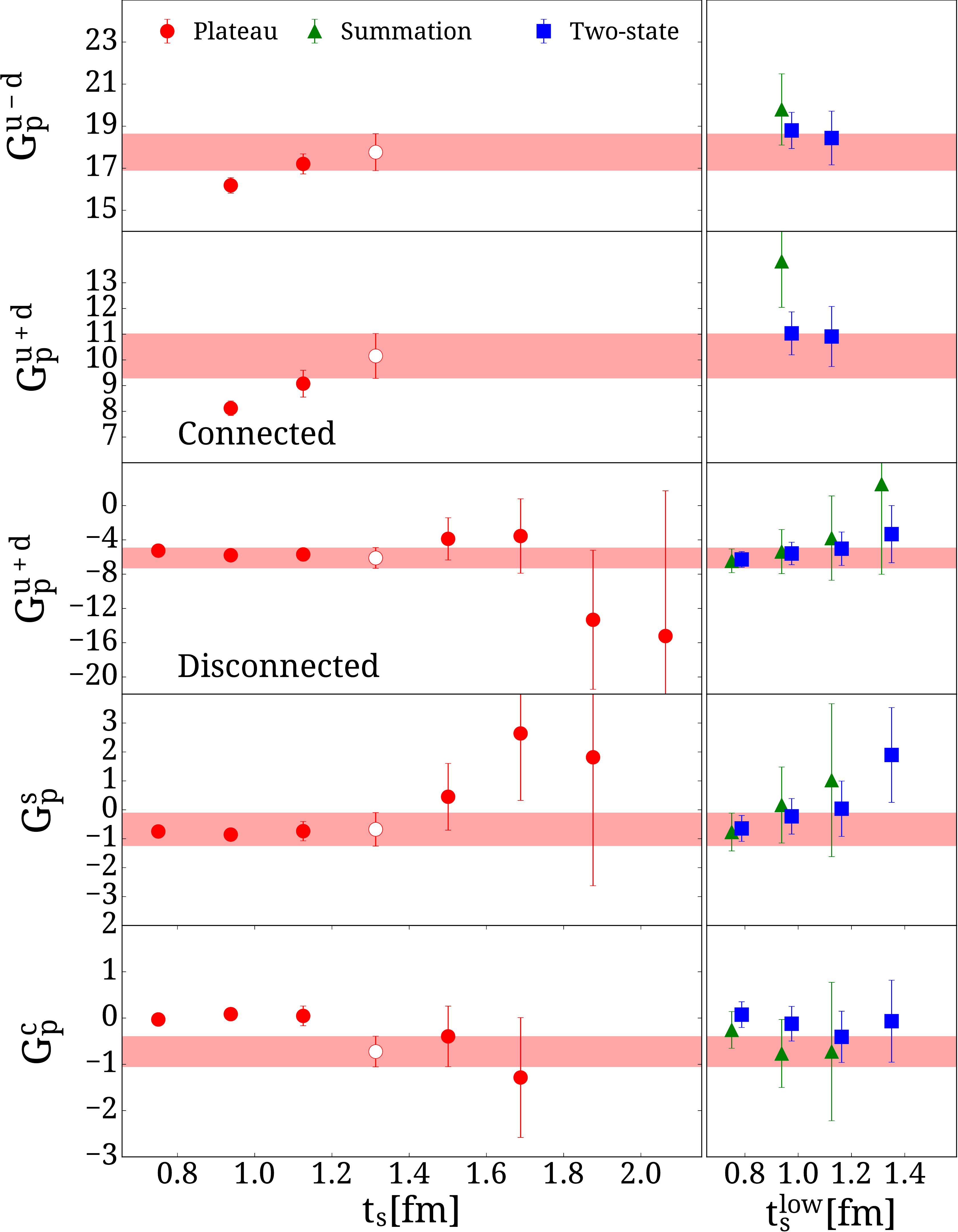}
  \end{minipage}
  \caption{Results for $G_A(Q^2)$ (left) and $G_p(Q^2)$ (right)for momentum
    transfer $Q^2=0.0753\; \mathrm{GeV^2}$. From top to bottom we
    present the isovector, connected isoscalar, disconnected
    isoscalar, strange and charm contributions. The remaining notation
    is as for the right panel of Fig.~\ref{Fig:ratios_gA_conn}.}
  \label{Fig:fits_mom1}
\hfill
\end{center}  
\end{figure}

\begin{figure}[!ht]
\begin{center}
  \begin{minipage}[t]{0.47\linewidth}
    \includegraphics[width=\textwidth]{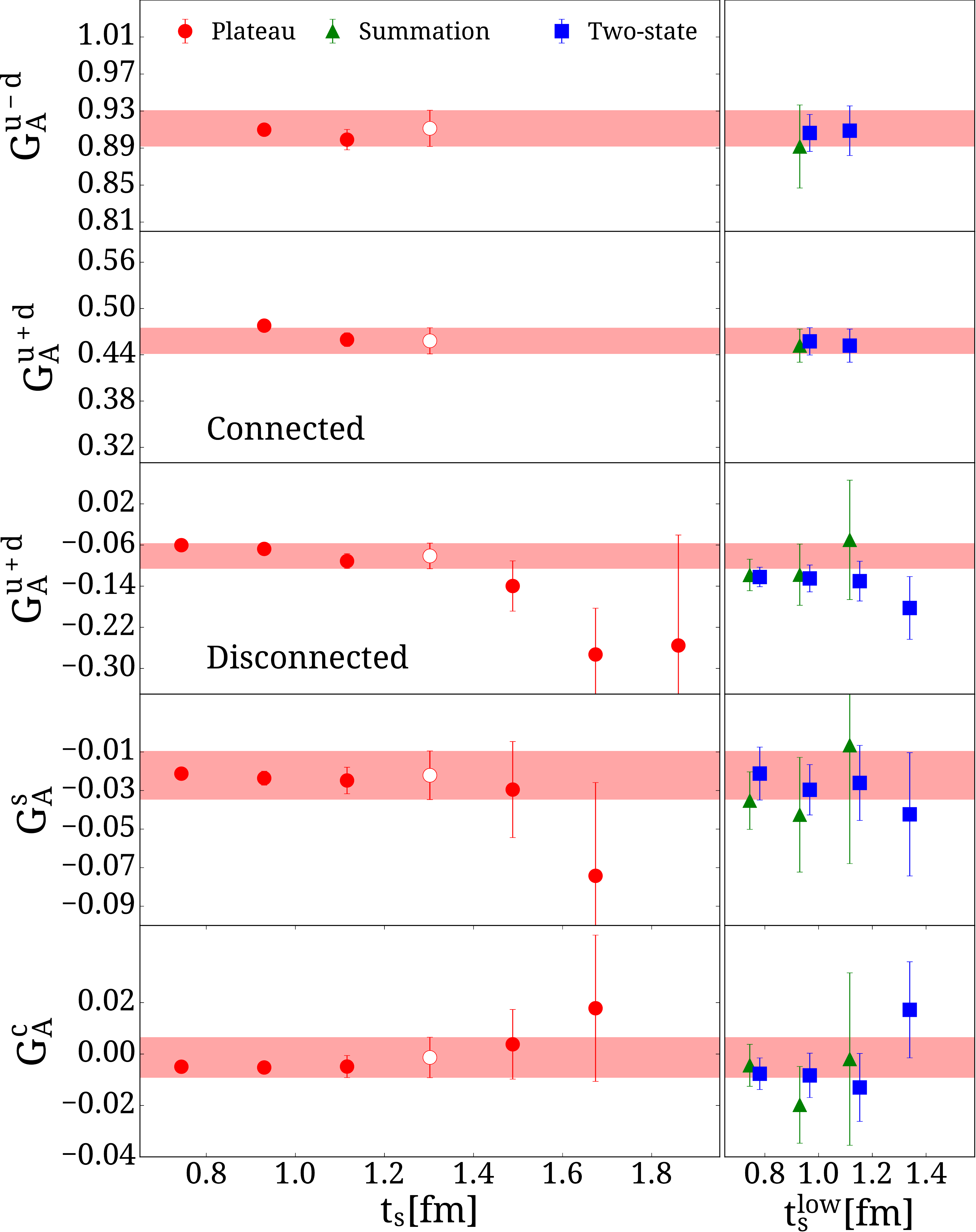}
  \end{minipage}
  \hfill
  \begin{minipage}[t]{0.47\linewidth}
    \includegraphics[width=\textwidth]{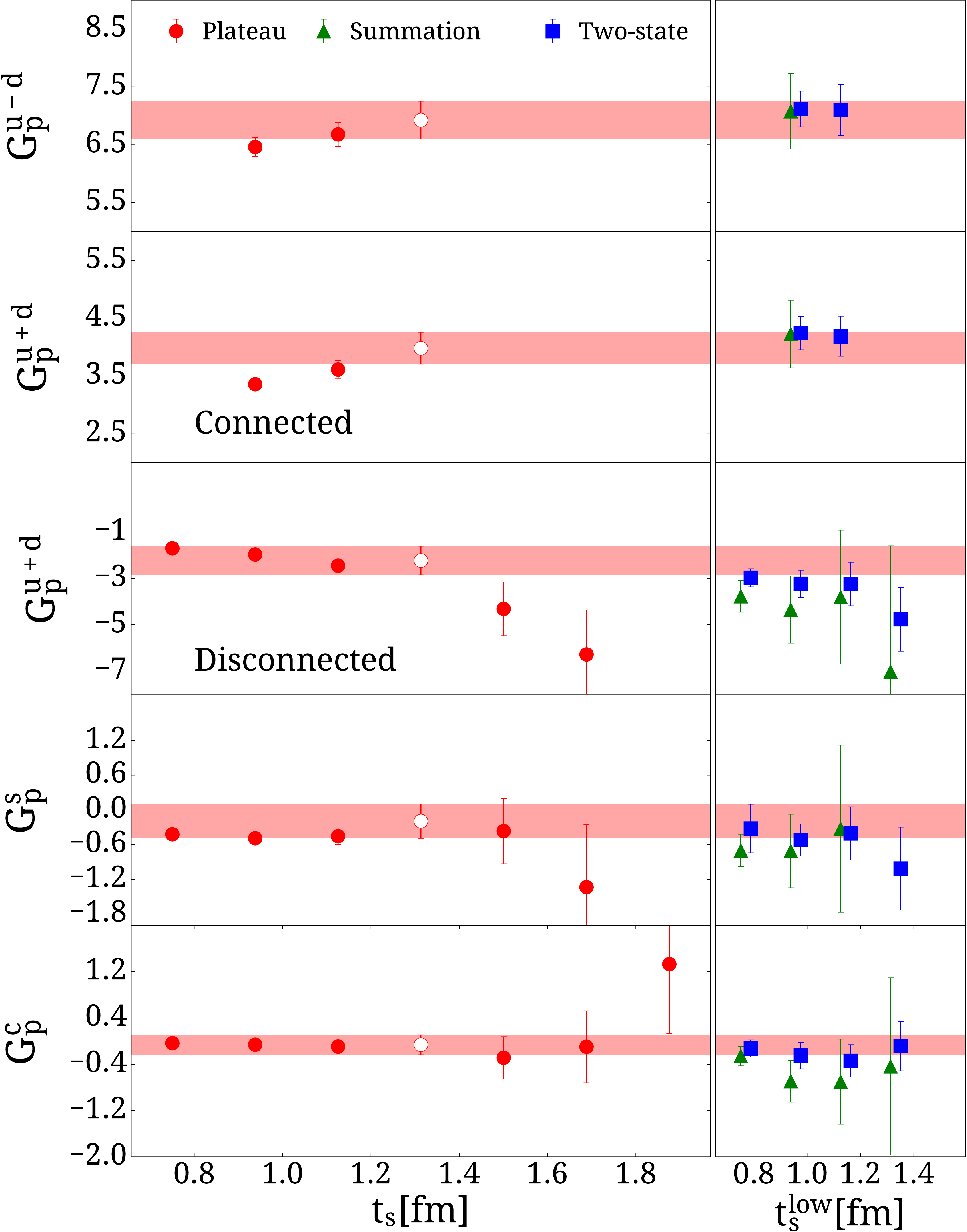}
  \end{minipage}
  \caption{Results for $G_A(Q^2)$ (left) and $G_p(Q^2)$ (right) for
    momentum $Q^2=0.2848 \;\mathrm{GeV^2}$. The notation is as in Fig.~\ref{Fig:fits_mom1}.}
  \label{Fig:fits_mom4}
  \hfill
\end{center}  
\end{figure}

All results quoted in this paper are computed by first fitting the ratio $R_i(\Gamma_k,\vec{p}\,',\vec{p};t_s,t_{\rm ins})$ with either the plateau, two-state or summation method to obtain $\Pi_i(k,\vec{p}\,',\vec{p})$ and subsequently minimize Eq.~(\ref{Eq:SVD_chisq}) to obtain $F(Q^2)$ without a time dependence.  In order to  demonstrate these plateaus we carry out an analysis in a different order. Namely we apply the SVD and minimization of Eq.~(\ref{Eq:SVD_chisq}) by inserting the ratio $R_i(\Gamma_k,\vec{p}\,',\vec{p};t_s,t_{\rm ins})$ instead of  $\Pi_i(k,\vec{p}\,',\vec{p})$.
In Figs.~\ref{Fig:plat_mom1} and~\ref{Fig:plat_mom4} we show representative examples of our obtained plateaus  for a small momentum transfer, namely for $Q^2=0.0753\; \mathrm{GeV^2}$ and for a higher momentum transfer, namely $Q^2$=0.2848~GeV$^2$. The corresponding results for the form factors are shown in Figs.~\ref{Fig:fits_mom1} and~\ref{Fig:fits_mom4} for the same momentum transfers as for Figs.~\ref{Fig:plat_mom1} and~\ref{Fig:plat_mom4}. We observe a similar behavior with respect to the excited states as that for $Q^2=0$ shown in Figs.~\ref{Fig:ratios_gA_conn} and \ref{Fig:ratios_gA_disc}. We thus take the plateau value at $t_s=1.31$~fm for both $G_A^{u-d}$ and $G_A^{u+d}$ as our final values since they are in good agreement with the two-state and summation fits. For $G_A^s$, $t_s=1.31$~fm is still a reasonable choice, but for $G_A^c$ due to the large statistical uncertainty, $t_s=0.94$~fm is enough.

For $G_p^{u-d}$ and the connected part of $G_p^{u+d}$ we observe excited state contributions for the two smaller values of $t_s$. For $t_s=1.31$~fm the plateau value is in  agreement with the two-state fit, however, indicating partial convergence. For the disconnected contribution we find better convergence and we take the value also at $t_s=1.31$~fm. 
What is particularly notable are the large disconnected contributions to the isoscalar induced pseudo-scalar form factor that are comparable in magnitude to the connected part, but  with opposite sign. This has already been observed in Ref.~\cite{Green:2017keo}, which used an ensemble simulated with a pion mass $m_\pi=317$~MeV. The explanation of such large disconnected contributions is that they are needed to cancel the pion pole of the connected isoscalar form factor in order to yield the expected $\eta$-meson pole mass dependence. Since the connected isoscalar shows a sharp rise consistent with a pion pole the disconnected contributions must also be large at small $Q^2$ to cancel it. This would be analogous to the case of the $\eta$-meson mass extraction on the lattice, where disconnected contributions are important since the connected contribution alone of the two-point correlation function has the mass of the pion as ground state~\cite{Alexandrou:2012zz}. From Fig.~\ref{Fig:fits_mom4} where results are shown for a relatively high $Q^2$ value, the overall observation is that excited state contributions tend to be less severe but non-negligible. This trend continues as we increase $Q^2$ at least for the connected contributions where statistical uncertainties are small enough for such an investigation.

\begin{figure}[!ht]
\begin{center}
  \begin{minipage}[t]{0.47\linewidth}
    \includegraphics[width=\textwidth]{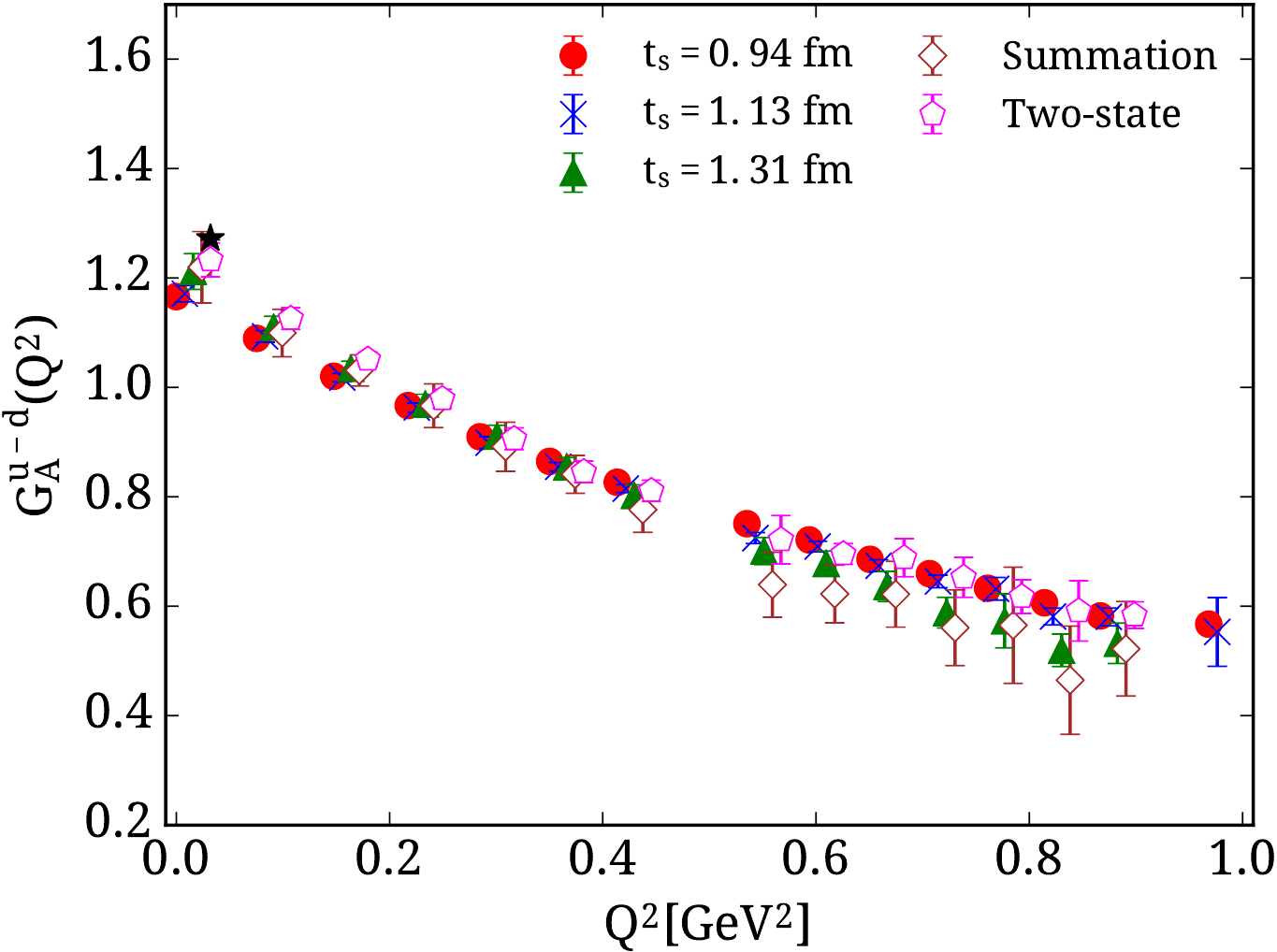}
  \end{minipage}
  \hfill
  \begin{minipage}[t]{0.47\linewidth}
    \includegraphics[width=\textwidth]{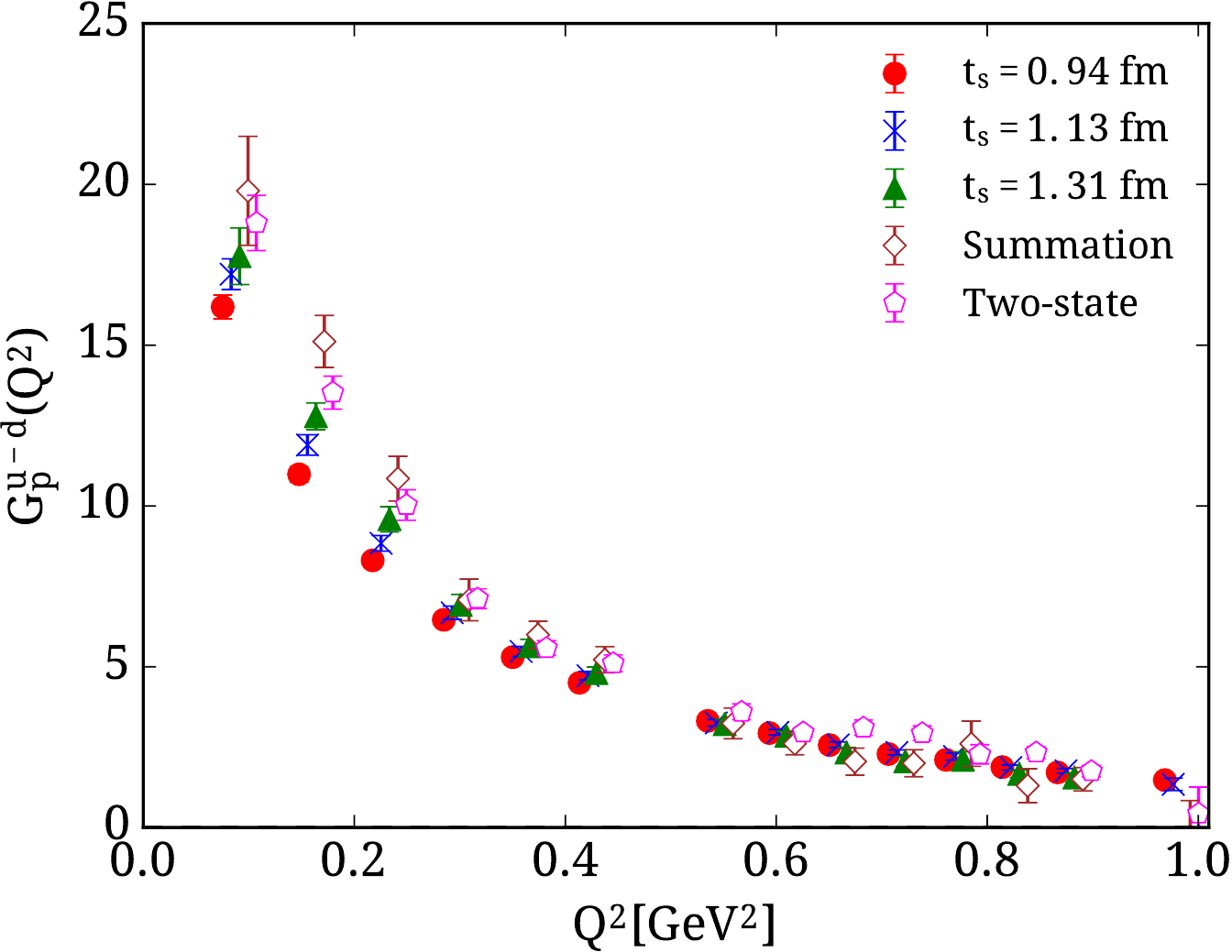}
  \end{minipage}
\hfill
\caption{Results for $G_A^{u-d}(Q^2)$ (left) and  $G_p^{u-d}(Q^2)$  (right) as a function of $Q^2$ for three source-sink time separations, namely  $t_s=0.94$~fm (red filled circles), $t_s=1.13$~fm (blue crosses) and $t_s= 1.31$~fm (green filled triangles). We also show results extracted from the summation method (open brown diamonds) and two-state fit (open magenta pentagons). The experimental value of $g_A$ is shown with the black asterisk.  Results  are slightly shifted to the right for clarity.}
\label{Fig:GA_Gp}
\end{center}  
\end{figure}

In Fig.~\ref{Fig:GA_Gp} we show the isovector form factors  up to $Q^2=1\; \mathrm{GeV^2}$ extracted from the plateau at the three values of   $t_s$ considered, from the two-state and summation methods~\cite{Alexandrou:2017msl}. As already noted, for $G_p^{u-d}(Q^2)$, excited states contributions are notably more severe for small values of $Q^2$, which tend to decrease its value. Nevertheless, the  values extracted from the plateau at $t_s=1.31$~fm are in agreement with the value extracted from the two-state fit for all $Q^2$-values. We thus take the plateau value at $t_s=1.31$~fm as our final value for the form factors with a systematic error the difference between the mean value from fitting the plateau at $t_s=1.31$~fm and that extracted from the two-state fit. This systematic error may be underestimated for $G_p(Q^2)$ at low $Q^2 \leqslant 0.2$~GeV$^2$ where a larger time separation may be needed to ensure convergence.  

Having a determination of the axial form factors we proceed to 
examine their $Q^2$-dependence. As customarily done in experiment we fit the axial form factor $G_A(Q^2)$ to a dipole form given by
\begin{equation}
G_A(Q^2) = \frac{g_A}{(1+\frac{Q^2}{m_A^2})^2},
\label{Eq:GA_dipole}
\end{equation}
where $m_A$ is the so-called axial mass and the axial radius, $\langle r_A^2\rangle$, is related to $m_A$ by
\begin{equation}
  \langle r_A^2\rangle = -\frac{6}{G_A(0)}\frac{\partial}{\partial Q^2} G_A(Q^2)|_{Q^2=0} = \frac{12}{m_A^2}.
  \label{Eq:AxialRadius}
\end{equation}
We note that experimentally, one of the determinations of $m_A$ is obtained by fitting the axial form factor $G_A(Q^2)$  extracted from pion electroproduction data, yielding a value of 
 $m_A=1.077(39)$ GeV~\cite{Liesenfeld:1999mv}. Recent results from charged-current muon-neutrino scattering events produced from the MiniBooNE experiment report a value of  $m_A=1.350(170)$ GeV using a similar fit~\cite{AguilarArevalo:2010zc}, which is significantly higher than the historical world average. Recent results from neutrino-nucleus cross sections using deuterium target data report a smaller value of $m_A=1.010(240)$~GeV~\cite{Meyer:2016oeg}.  

\begin{figure}[!ht]
\begin{center}
  \begin{minipage}[t]{0.47\linewidth}
    \includegraphics[width=\textwidth]{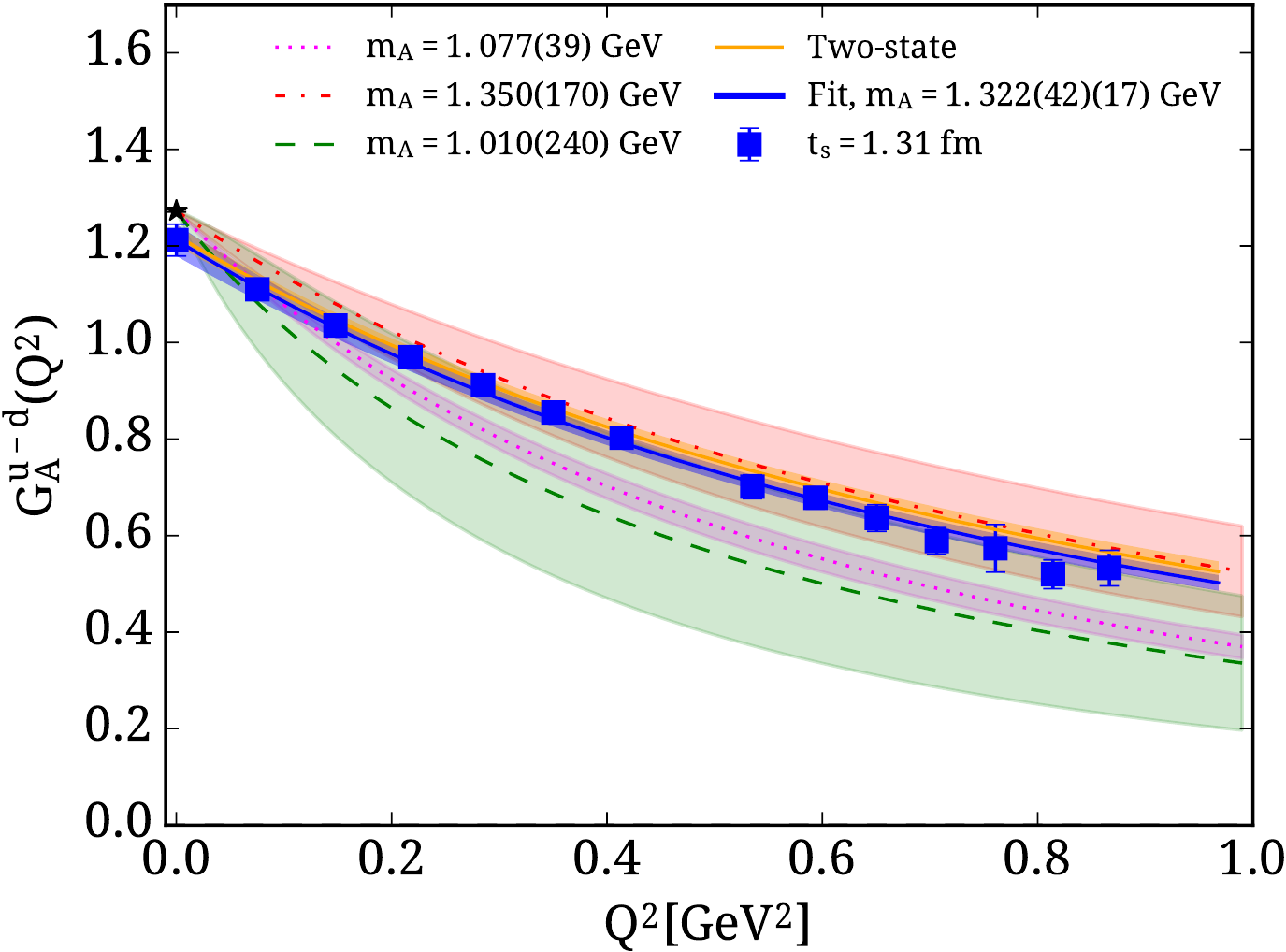}
  \end{minipage}
  \hfill
  \begin{minipage}[t]{0.47\linewidth}
    \includegraphics[width=\textwidth]{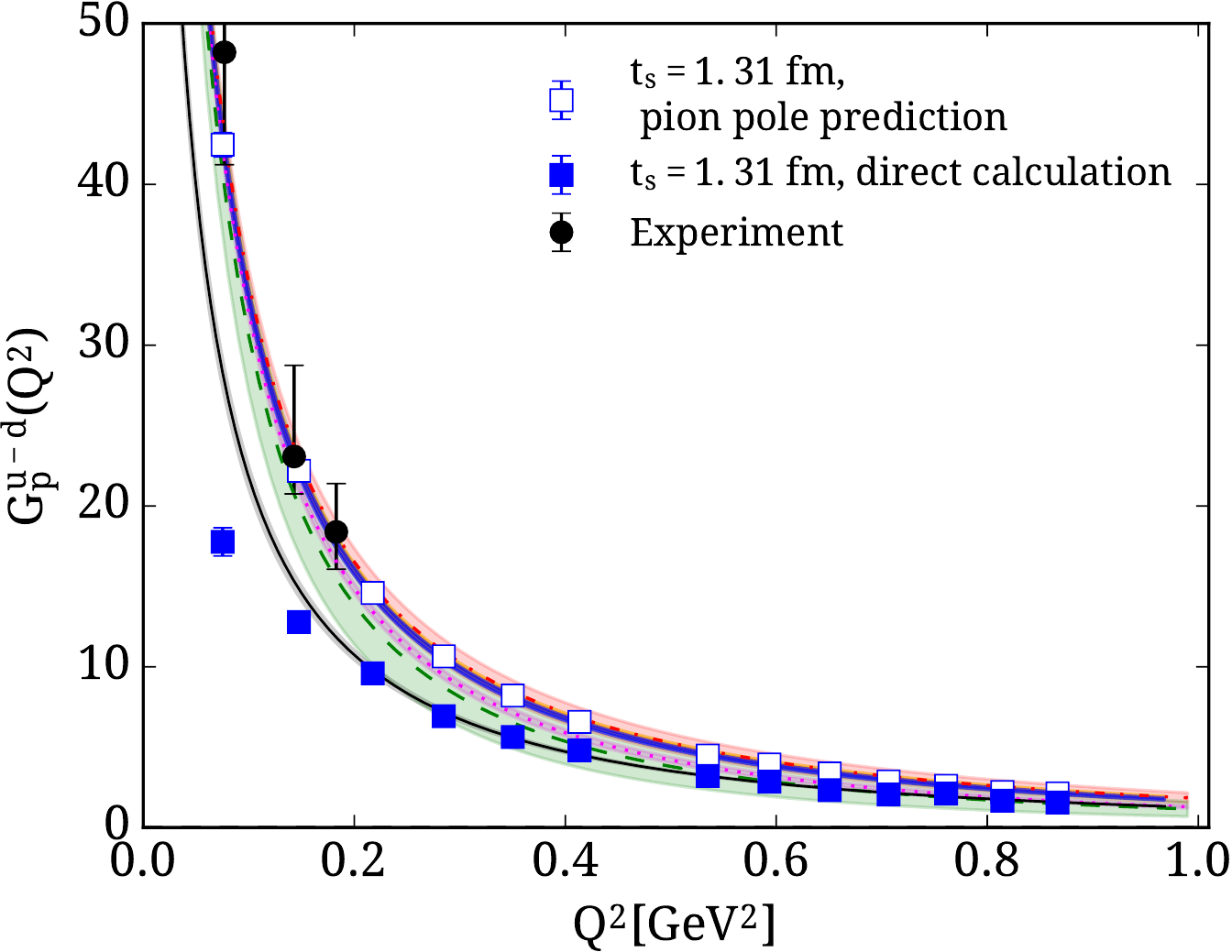}
  \end{minipage}
\hfill
\caption{
  Our results for $G_A^{u-d}(Q^2)$ (left) and $G_p^{u-d}(Q^2)$ (right) using the plateau method for $t_s=1.31$~fm (filled blue squares). In the left panel,  the solid blue (orange) line shows the fit to our lattice QCD results extracted from the plateau at $t_s=1.31$~fm (from the two-state fit) using Eq.~(\ref{Eq:GA_dipole}). The experimental value of $g_A$ is shown with the black asterisk. The purple, red and green bands are experimental results for $G_A^{u-d}(Q^2)$ taken from Refs.~\cite{Liesenfeld:1999mv}, \cite{AguilarArevalo:2010zc} and~\cite{Meyer:2016oeg} respectively.
  In the right panel, the open blue squares show the prediction for $G_p^{u-d}(Q^2)$ assuming pion-pole dominance and using Eq.~(\ref{Eq:Gp_dipole}) to extract  $G_p^{u-d}(Q^2)$ from our lattice results for $G_A^{u-d}(Q^2)$ shown in the left panel, together with the corresponding fits, blue (orange) band is a fit to the predicted  $G_p^{u-d}(Q^2)$ using $G_A^{u-d}(Q^2)$ extracted from  the plateau (two-state). The two fits are overlapping. The filled blue squares show  $G_p^{u-d}(Q^2)$ extracted directly from the nucleon matrix element with a fit to Eq.~(\ref{Eq:Gp_expanded}) (solid black line)  after omitting the two lowest $Q^2$ values. The filled black circles are direct measurements of $G_p^{u-d}(Q^2)$ from Ref.~\cite{Choi:1993vt}. The purple, red and green bands use the  experimental results for $G_A^{u-d}(Q^2)$ and pion pole to infer $G_p^{u-d}(Q^2)$. }
\label{Fig:GA_Gp_bands}
\end{center}  
\end{figure}

Fitting the momentum dependence of our results for $G_A^{u-d}(Q^2)$ using Eq.~(\ref{Eq:GA_dipole})  we obtain a value of $m_A=1.322(42)(17)$~GeV, which is consistent with the larger value extracted from $\nu_\mu$-interactions~\cite{AguilarArevalo:2010zc}.   The fit is performed by fixing the value for $g_A$ directly from our lattice result for $G_A^{u-d}(0)$. We have checked that allowing $g_A$ to vary as a fit parameter yields consistent results.  We also  extract  a consistent
value for $m_A$ using the results from the two-state fit. We quote the difference in the mean value of $m_A$ extracted from fitting the $t_s=1.31$~fm plateau results for the form factors  and that extracted from the results of the two-state fits as the systematic error due to excited states.
In the left panel of Fig.~\ref{Fig:GA_Gp_bands} we show a comparison of the fits to our lattice QCD results and the experimental ones. The spread in the mean values is an indication of remaining excited state contributions, which are small and which we quote as our systematic error. The bands produced using the values from Refs.~\cite{Liesenfeld:1999mv,Meyer:2016oeg} are lower and have  a steeper slope than our results, albeit with  large errors.

In Fig.~\ref{Fig:GA_Gp_bands} we  show our lattice QCD results for $G_p^{u-d}(Q^2)$. As expected from pion pole dominance, this form factor has a much stronger $Q^2$ dependence as compared to $G_A^{u-d}(Q^2)$.   Using the partially conserved axial current relation (PCAC) and pion pole dominance one can relate the induced pseudo-scalar form factor $G_p^{u-d}(Q^2)$ to $G_A^{u-d}(Q^2)$ by
\begin{equation}
G_p(Q^2) = G_A(Q^2) \frac{C}{Q^2+m_p^2},
\label{Eq:Gp_dipole}
\end{equation}
where $C=4 m_N^2$ and $m_p=m_\pi$. This relation is used to extract the induced pseudo-scalar form factor using  the experimental determination of $G_A^{u-d}(Q^2)$. We perform the same analysis for our lattice QCD results. Namely, in Fig.~\ref{Fig:GA_Gp_bands} we include results for $G_p^{u-d}(Q^2)$ obtained by applying the pion-pole dominance hypothesis to the lattice results on $G_A^{u-d}$ using the lattice pion mass of $m_\pi=130$~MeV in  Eq.~(\ref{Eq:Gp_dipole}). At low $Q^2$-value we observe a much steeper rise as compared to the direct lattice computation of  $G_p(Q^2)$, and agreement both with the experimentally determined bands taken by applying the pion-pole assumption as well as with the directly determined values of $G_p^{u-d}(Q^2)$ from Ref.~\cite{Choi:1993vt}. As noted above, for $Q^2<0.2$~GeV$^2$ where the discrepancy is largest, excited states tend to produce smaller values. In addition, similar discrepancy at low $Q^2$ has been observed in previous lattice studies at heavier pion masses between multiple volumes~\cite{Alexandrou:2010hf,Alexandrou:2007xj}, indicating that volume effects may also need to be investigated to resolve this tension. We plan to investigate both these systematics using a larger volume of $64^3\times 128$ in a future work. For the current analysis we will discard  $G_p^{u-d}(Q^2)$ at the two lowest values of $Q^2$.

\begin{figure}[!ht]
  \begin{center}
    \includegraphics[scale=0.6]{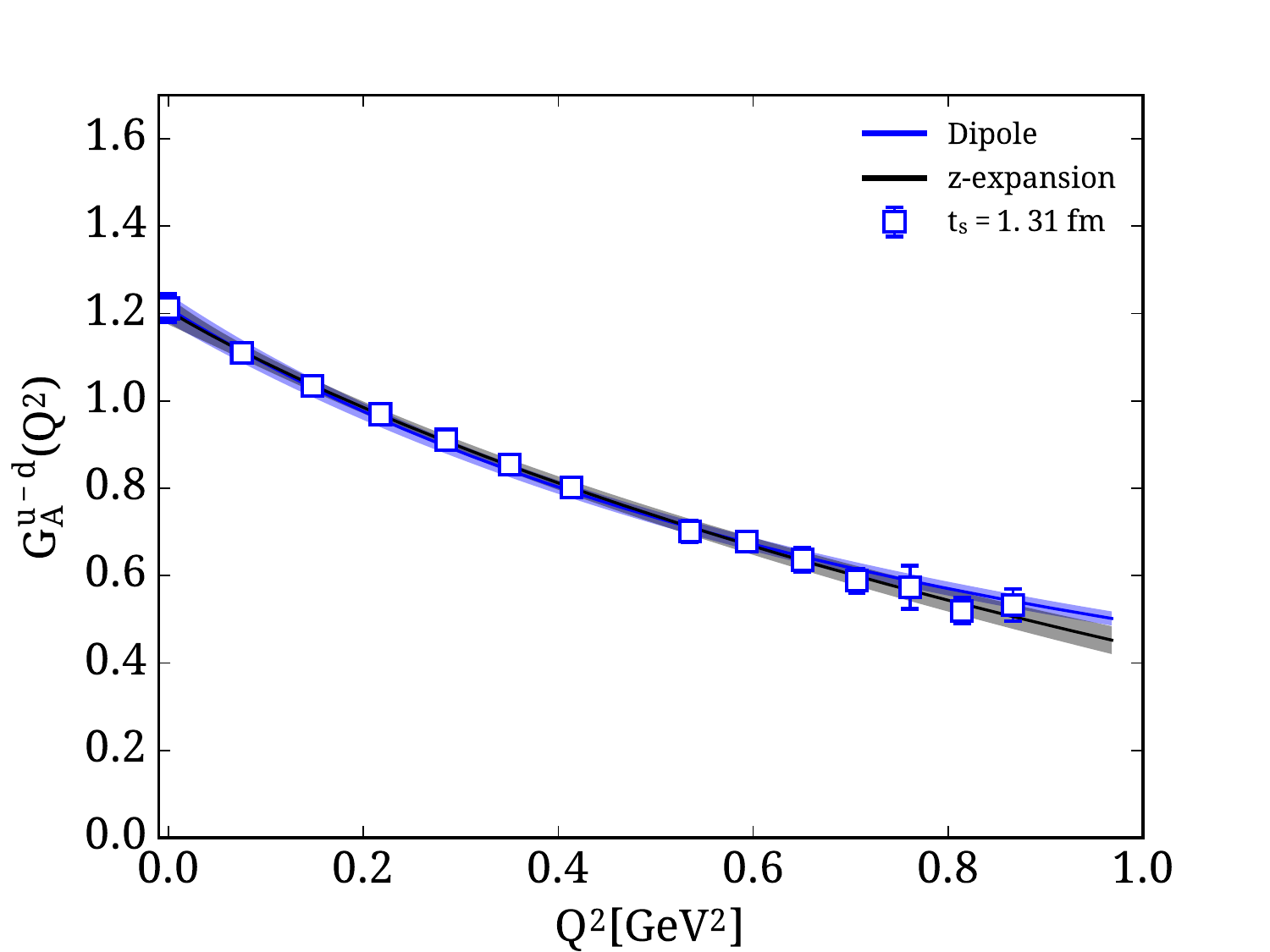}
    \caption{$G_A^{u-d}(Q^2)$ extracted from the plateau method at
      $t_s=1.31$~fm, fitted to the dipole form (grey band) and to the
      z-expansion (blue band).}
    \label{Fig:DipoleVSzExpansion}
  \end{center}
\end{figure}

 \begin{figure}[!ht]
\begin{center}
  \begin{minipage}[t]{0.47\linewidth}
    \includegraphics[width=\textwidth]{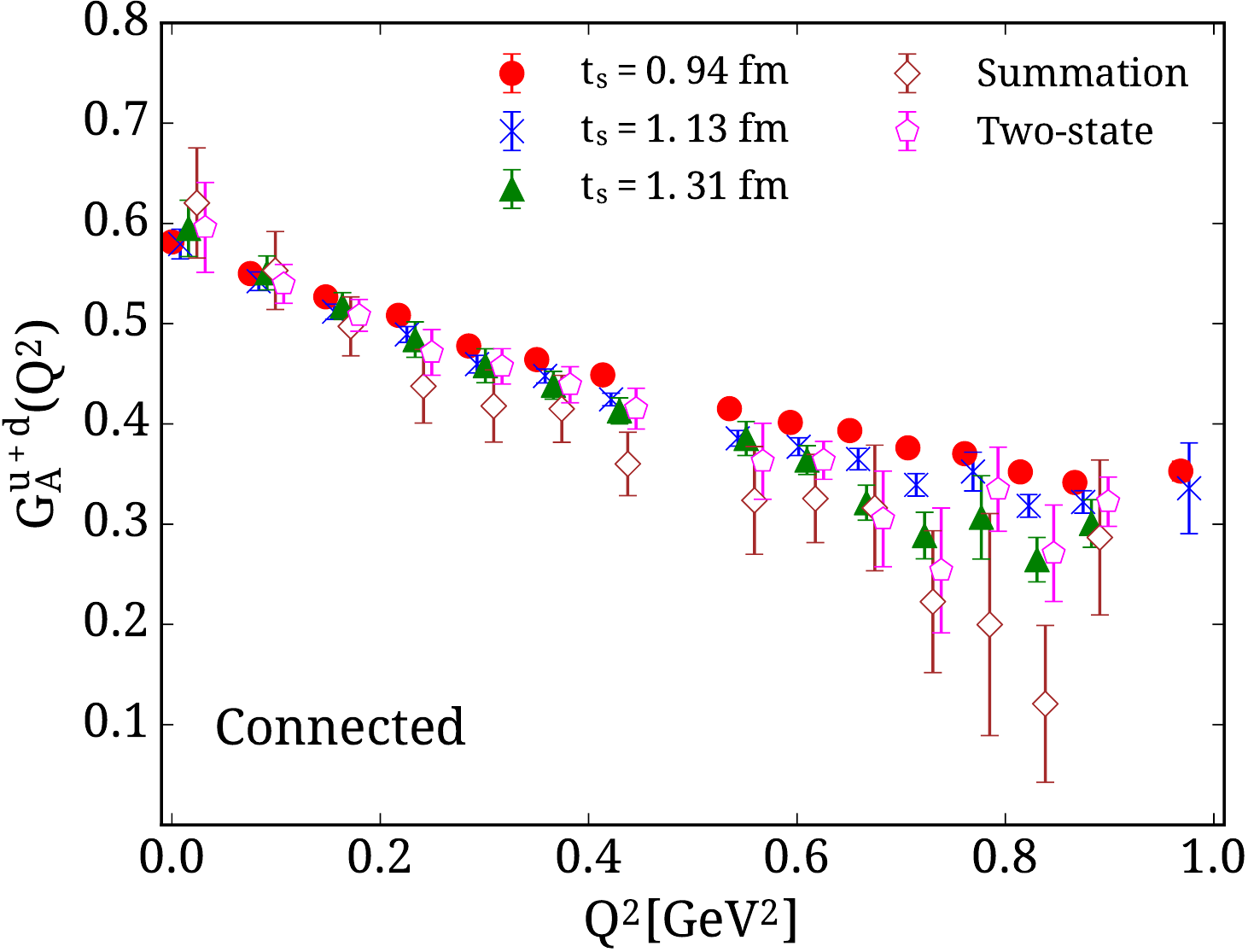}
  \end{minipage}
  \hfill
  \begin{minipage}[t]{0.47\linewidth}
    \includegraphics[width=\textwidth]{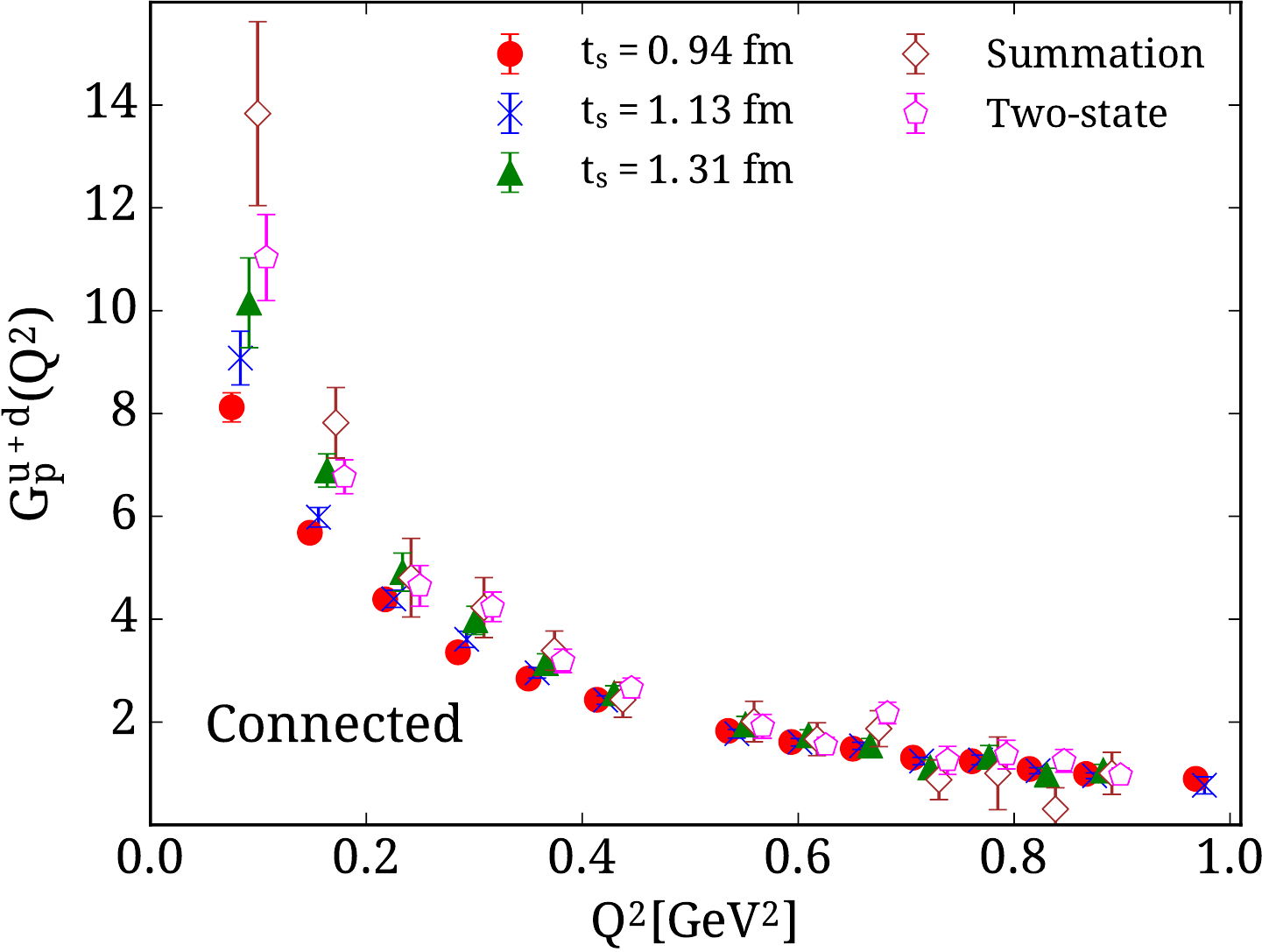}
  \end{minipage}
\hfill
\caption{Results for the connected contribution to $G_A^{u+d}(Q^2)$ (left) and $G_p^{u+d}(Q^2)$ (right). The notation is the same as in Fig.~\ref{Fig:GA_Gp}.}
\label{Fig:GA_Gp_isoscalar_connected}
\end{center}  
\end{figure}

\begin{figure}[!ht]
\begin{center}
  \begin{minipage}[t]{0.47\linewidth}
    \includegraphics[width=\textwidth]{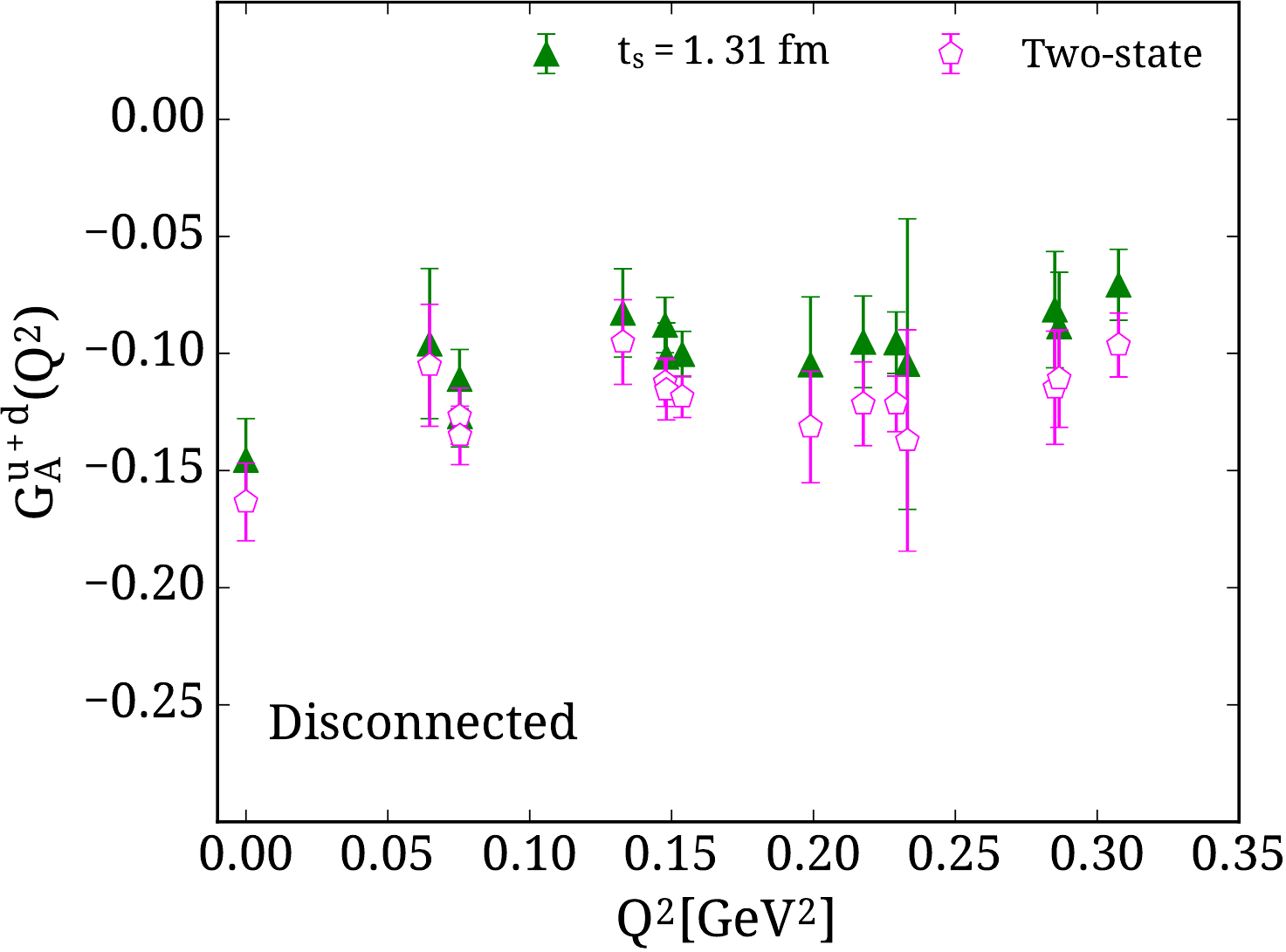}
  \end{minipage}
  \hfill
  \begin{minipage}[t]{0.47\linewidth}
    \includegraphics[width=\textwidth]{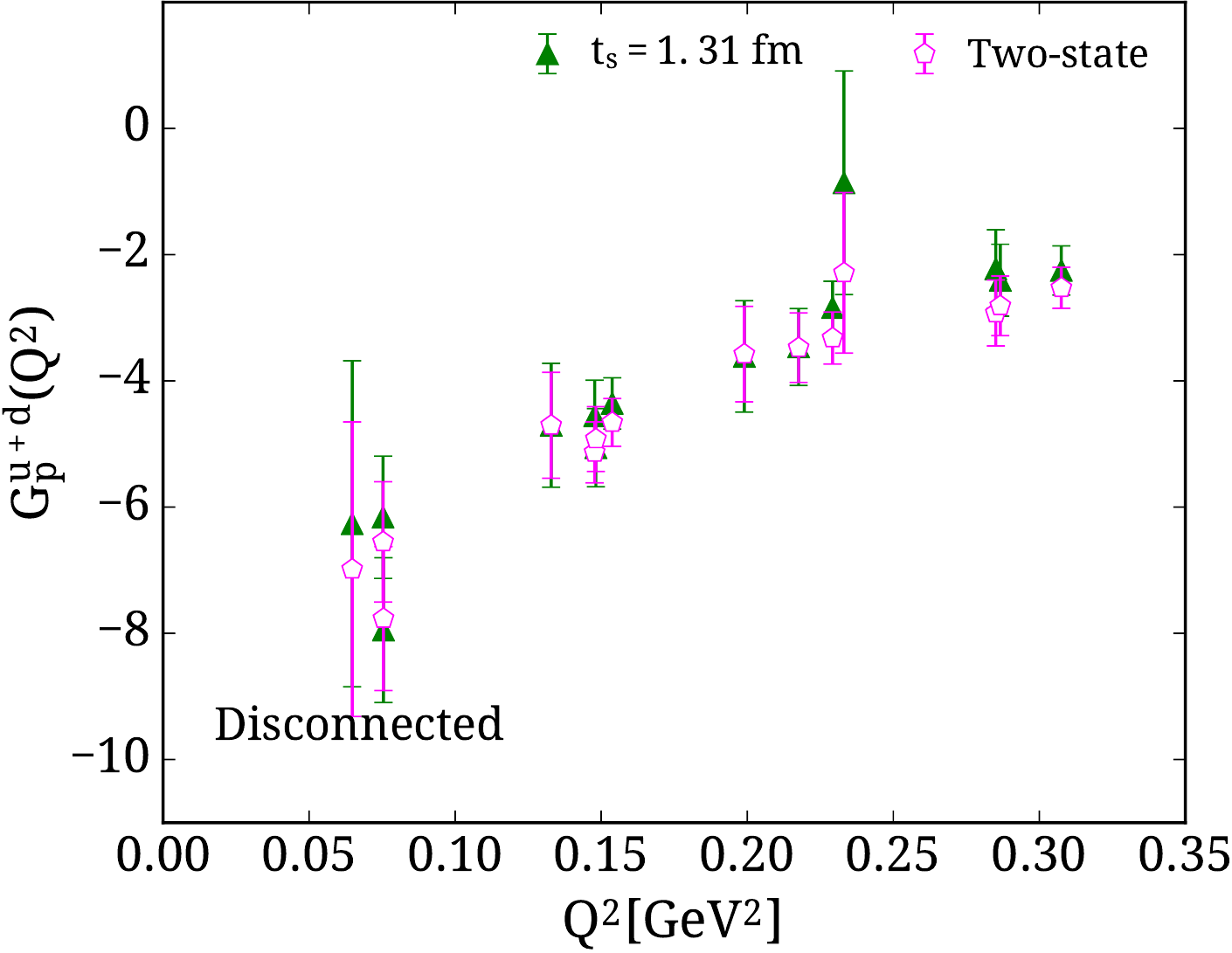}
  \end{minipage}
\hfill
\caption{
  Disconnected contribution to $G_A^{u+d}(Q^2)$ (left) and
  $G_p^{u+d}(Q^2)$ (right). We combine results obtained using sink
  momentum that satisfy $\vec{p}\,'=\vec{0}$ and $\vec{p}^{\prime 2}=(2
  \pi/L)^2$. The notation is the same as in Fig.~\ref{Fig:GA_Gp}.
}
\label{Fig:GA_Gp_isoscalar_disc}
\end{center}  
\end{figure}

\begin{figure}[!ht]
\begin{center}
  \begin{minipage}[t]{0.47\linewidth}
    \includegraphics[width=\textwidth]{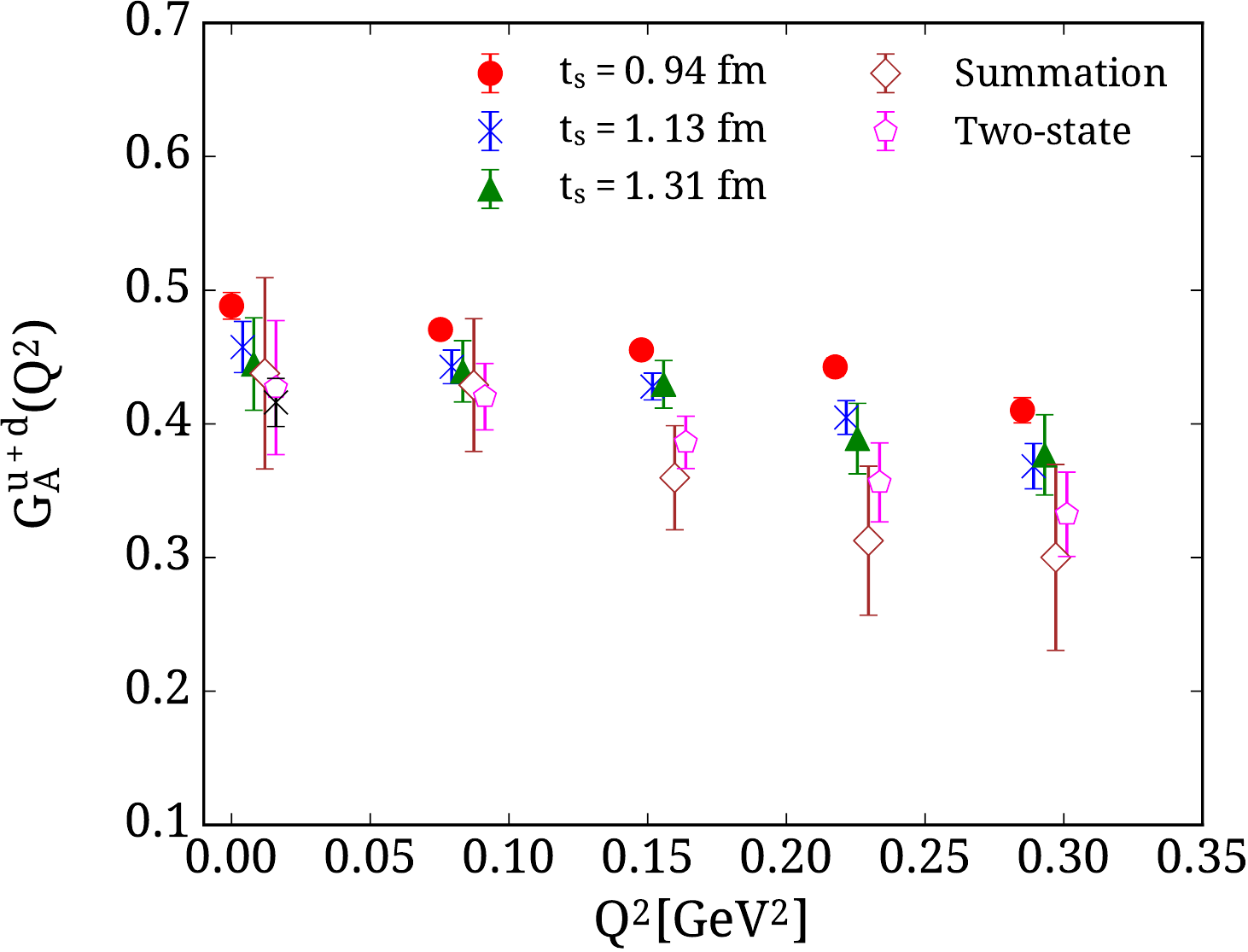}
  \end{minipage}
  \hfill
  \begin{minipage}[t]{0.47\linewidth}
    \includegraphics[width=\textwidth]{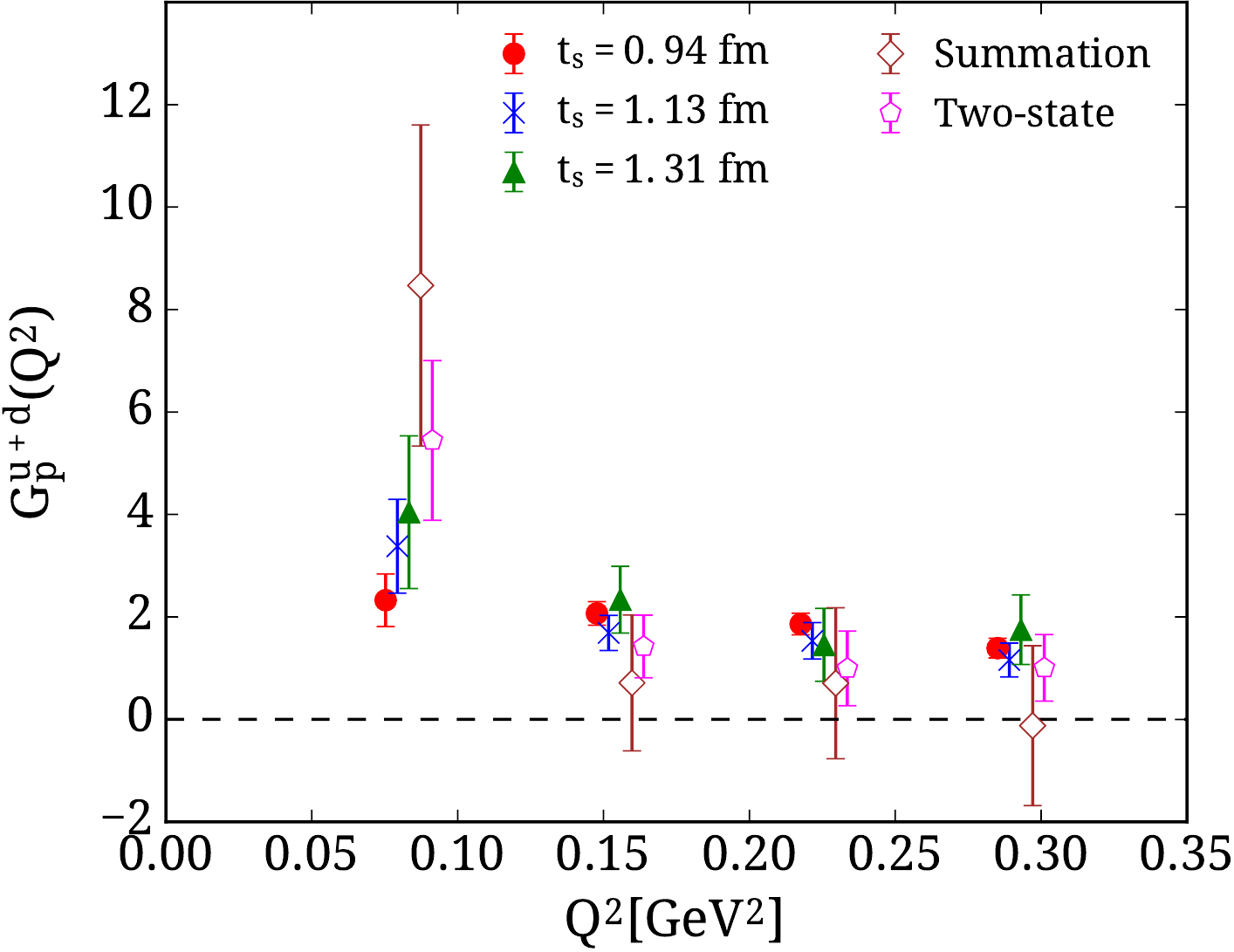}
  \end{minipage}
\hfill
\caption{Total contribution to $G_A^{u+d}(Q^2)$ (left) and $G_p^{u+d}(Q^2)$ (right). The notation is the same as in Fig.~\ref{Fig:GA_Gp}.}
\label{Fig:GA_Gp_isoscalar_total}
\end{center}  
\end{figure}

\begin{figure}[!ht]
\begin{center}
  \begin{minipage}[t]{0.47\linewidth}
    \includegraphics[width=\textwidth]{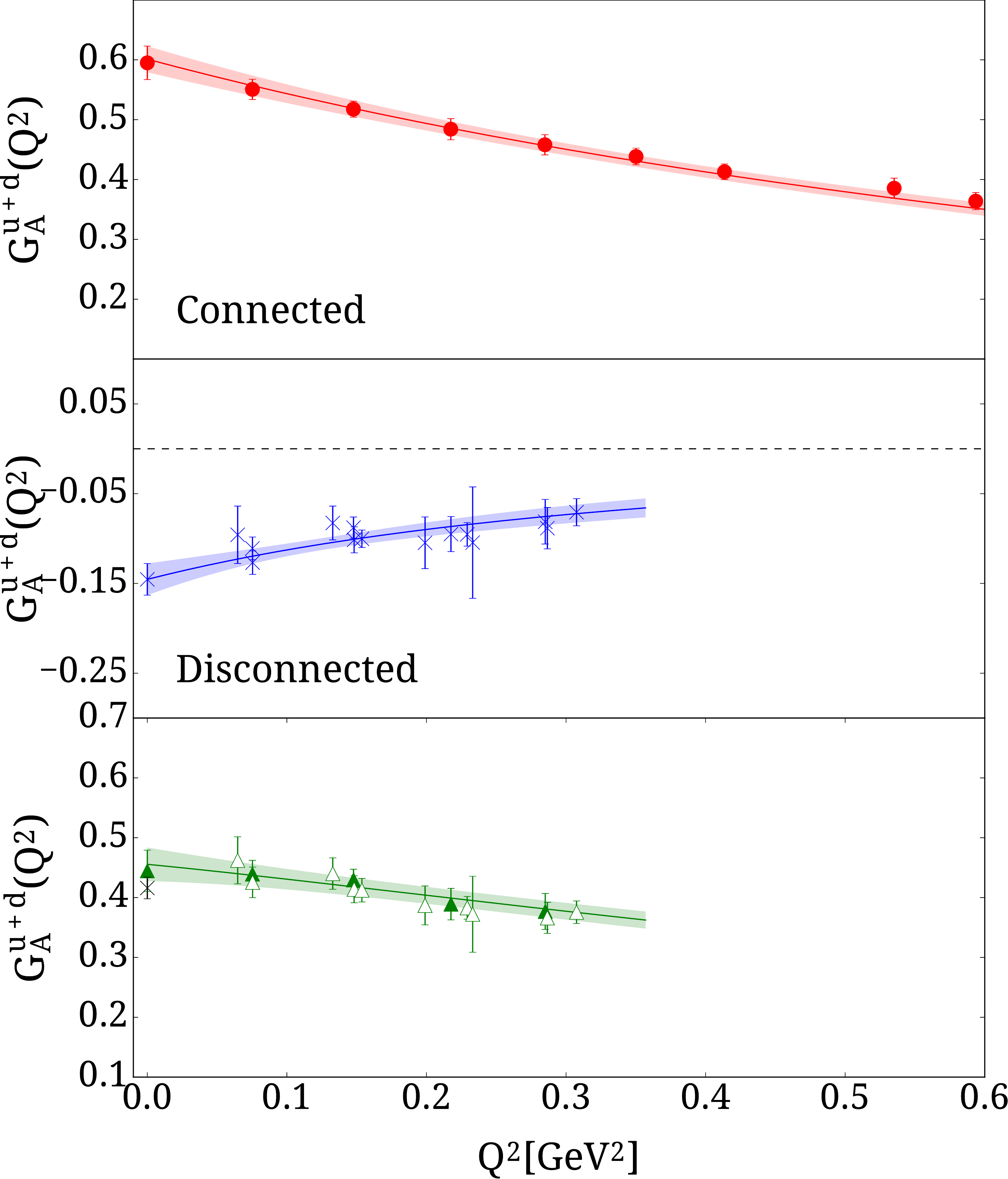}
  \end{minipage}
  \hfill
  \begin{minipage}[t]{0.47\linewidth}
    \includegraphics[width=\textwidth]{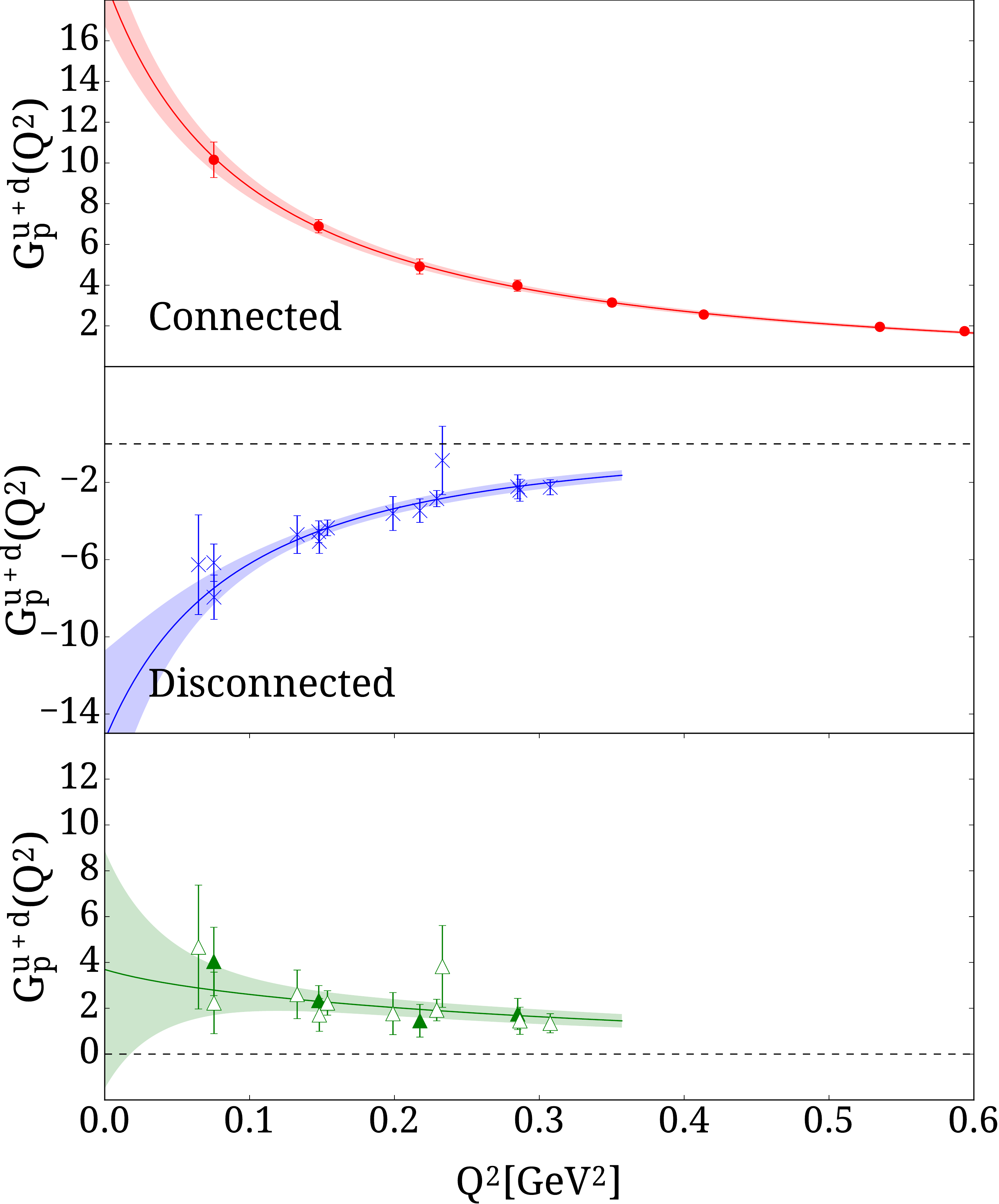}
  \end{minipage}
\hfill
\caption{
  Results for $G_A^{u+d}(Q^2)$ (left) and
  $G_p^{u+d}(Q^2)$ (right). We show their connected contributions
  (squares, upper panels), disconnected contributions (crosses, middle
  panels) and the total (triangles, lower panels).  The solid green
  triangles are obtained by adding connected and the disconnected
  contributions with sink momentum $\vec{p}\,'=\vec{0}$ for which
  lattice results for both are available. The open green triangles
  have been computed by interpolating the connected contributions to
  the additional $Q^2$ values available for the disconnected. The
  solid curves and their associated error bands have been extracted by
  fitting to Eqs.~(\ref{Eq:GA_dipole}) and~(\ref{Eq:Gp_dipole}),
  respectively. The horizontal dashed lines are
  drawn through zero.
}
\label{Fig:GA_Gp_isoscalar_bands}
\end{center}  
\end{figure}
In addition to the dipole form, we fit our results for the axial form factor using the so-called z-expansion~\cite{Hill:2010yb}, given by
\begin{equation}
  G(Q^2) = \sum_{k=0}^{k_{max}} a_k z^k
\end{equation}
where
\begin{equation}
  z = \frac{\sqrt{t_{cut}+Q^2}-\sqrt{t_{cut}} }{\sqrt{t_{cut}+Q^2}+\sqrt{t_{cut}}}
\end{equation}
and $t_{cut}=9m_\pi^2$. In Fig.~\ref{Fig:DipoleVSzExpansion} we
compare  the dipole fit with the z-expansion fit. For the
z-expansion we used $k_{max}=3$, fixing $a_0=g_A$ and imposing
Gaussian priors for the coefficients $a_k$ for $k>1$ with width $w=5
\mathrm{max}(\vert a_0 \vert, \vert a_1 \vert)$. Both fit Ans\"atze
describe the data very well, producing consistent values for the
radius, namely $\langle r_A^2 \rangle=0.266(17)$~fm$^2$ in the case
of the dipole fit and $\langle r_A^2 \rangle=0.265(76)$~fm$^2$ from
the z-expansion. A fit using the z-expansion is more suitable when
precise data are available at a large number of $Q^2$ values. Given 
 the statistical errors and relatively few momenta available
 from our lattice calculation, the z-expansion
 therefore yields  larger errors than a dipole fit. Given the consistency
 between the two fits we thus opt to use the dipole form that yields smaller
 errors
 for all the fits that follow.
\begin{figure}[!ht]
  \begin{center}
    \begin{minipage}[t]{0.47\linewidth}
      \includegraphics[width=\textwidth]{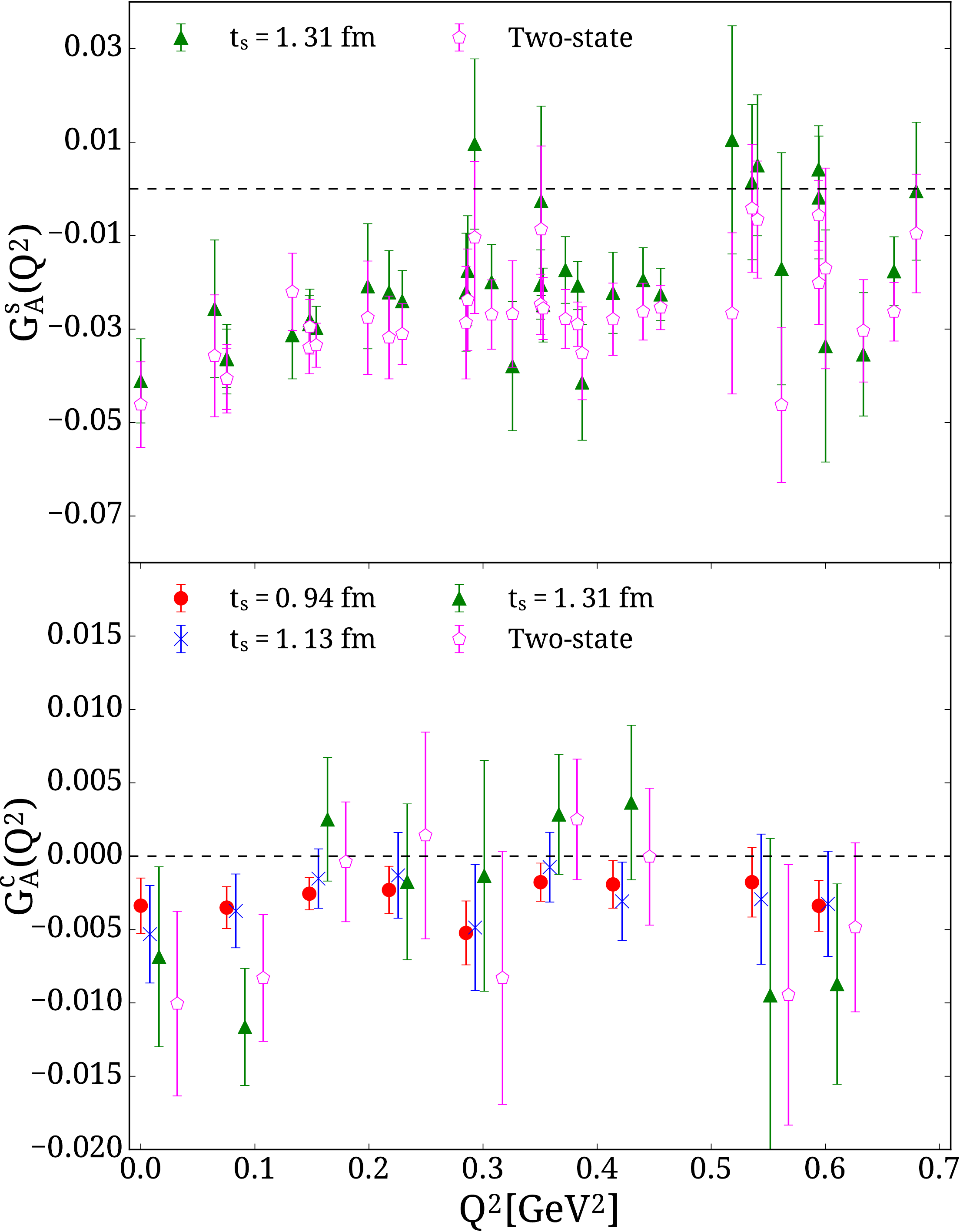}
    \end{minipage}
    \hfill
    \begin{minipage}[t]{0.47\linewidth}
      \includegraphics[width=\textwidth]{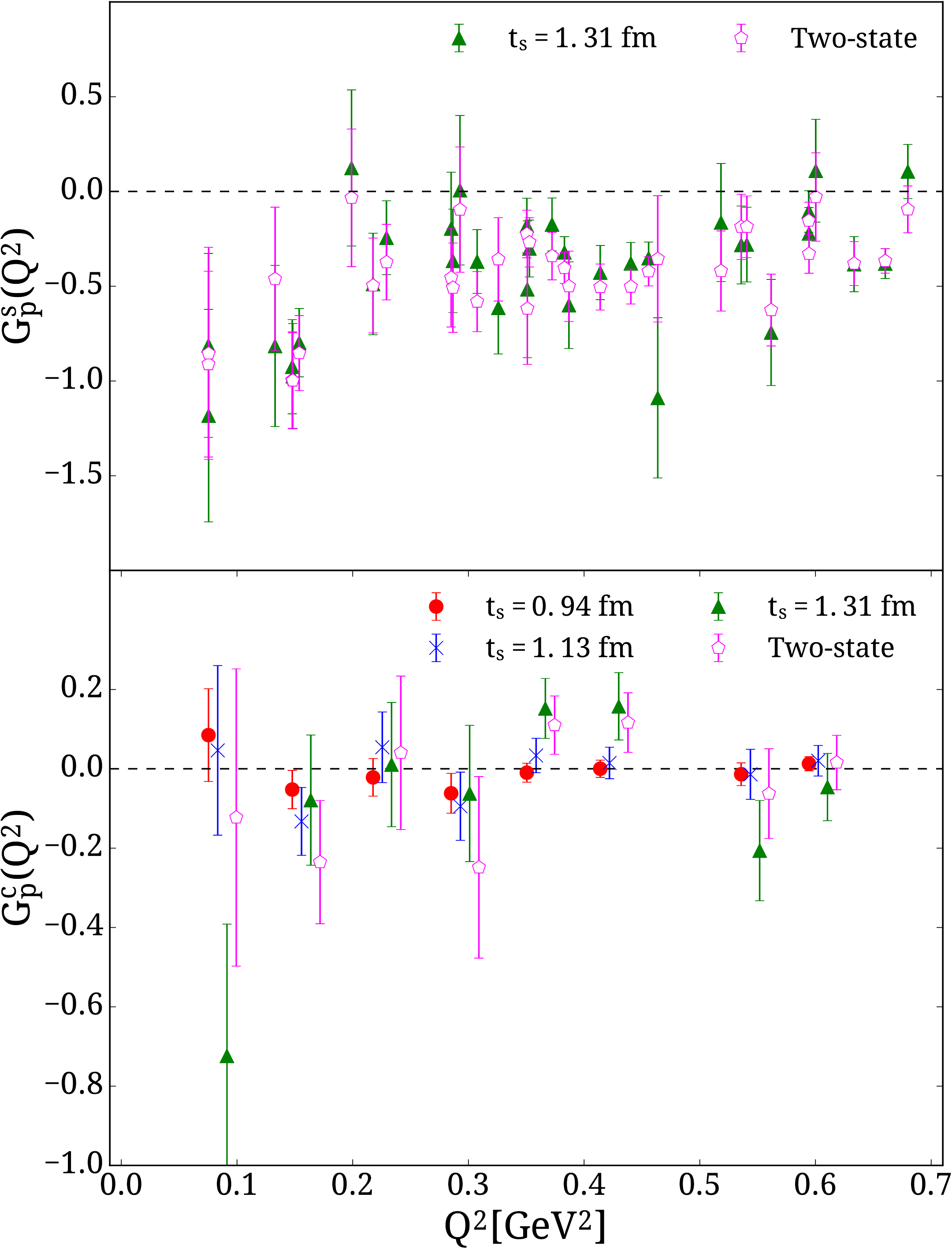}
    \end{minipage}
    \hfill
    \caption{$G_A^s(Q^2)$ and $G_A^c(Q^2)$ (left) and $G_p^s(Q^2)$ and $G_p^c(Q^2)$ (right) versus $Q^2$. The notation is the same as in Fig.~\ref{Fig:GA_Gp}.}
    \label{Fig:GAs_GAc_Gps_Gpc}
  \end{center}  
\end{figure}

PCAC relates the residue of the pion pole  to the pion decay constant $f_\pi$, the nucleon mass $m_N$ and the pion-nucleon coupling constant $g_{\pi NN}$ as follows~\cite{Goldberger:1958vp}
\begin{equation}
\lim_{Q^2 \rightarrow - m_\pi^2} (Q^2 + m_\pi^2) G_p^{u-d}(Q^2)  = 4 m_N f_\pi g_{\pi NN}.
\label{Eq:gpiNN_Gp}
\end{equation}
The relation holds when including the leading correction as obtained
within the chiral perturbative framework used in
Ref.~\cite{Bernard:1995dp}. Using Eq.~(\ref{Eq:Gp_dipole}) we can
relate $g_{\pi NN}$ to the axial form factor as
\begin{equation}
\lim_{Q^2 \rightarrow - m_\pi^2} G_A^{u-d}(Q^2) C = 4 m_N f_\pi g_{\pi NN},
\label{Eq:gpiNN_GA}
\end{equation}
where for this ensemble $f_\pi=89.80$ MeV~\cite{Abdel-Rehim:2015pwa} and $m_N$=0.932(4)~GeV~\cite{Alexandrou:2017xwd}. Using $G_A^{u-d}(-m_\pi^2)$=1.234(35)(20) obtained from our dipole fit and $C=4 m_N^2$, we find  $g_{\pi NN} = 12.81(37)(21)$, which is consistent with the experimental value $g_{\pi NN}=13.12(10)$ measured from pion-nucleon scattering lengths~\cite{Baru:2010xn}. Were we to fit directly  the
lattice data for $G_p^{u-d}(Q^2) $ to the form
\begin{equation}
  G_p^{u-d}(Q^2) = \frac{1}{(1+ Q^2/m_\pi^2)}\frac{G^{u-d}_p(0)}{(1+Q^2/m_p^2)^2}
  \label{Eq:Gp_expanded}
\end{equation}
taking $m_\pi=130$~MeV and omitting the first two $Q^2$ values from
the fit we obtain the solid line in Fig.~\ref{Fig:GA_Gp_bands}, for
which $m_p=1.441(115)(648)$~GeV consistent with the axial mass from
fitting to the axial form factor and 
$G^{u-d}_p(0)=165.62(9.82)(18.46)$ which is smaller than $4 (m_N^2 / m_\pi^2) g_A$. If we then were to use
Eq.~(\ref{Eq:gpiNN_Gp}) we would determine $g_{\pi
  NN}$=8.50(51)(82). This is smaller than the value determined using
pion-pole dominance and our lattice results for $G_A^{u-d}$. Additionally one
can compute also the induced pseudoscalar charge, $g_p^*$, defined as
\begin{equation}
  g_p^* = \frac{m_\mu}{2 m_N} G_p(Q^2 = 0.88 m_\mu^2)
\end{equation}
where $m_\mu$ is the muon mass. We find $g_p^*=7.47(30)(80)$ using our lattice results for $G_A^{u-d}$ and pion-pole dominance. Using the lattice results for $G_p^{u-d}$ the value of $g_p^*$ is lower and calls
for a further study of excited states and volume effects on the lattice determination of $G_p^{u-d}(Q^2)$.

 In order to compute the individual light quark axial form factors one needs, besides the isovector form factors, the isoscalar combination. In Fig.~\ref{Fig:GA_Gp_isoscalar_connected} we illustrate our results for the connected contributions to $G_A^{u+d}(Q^2)$ and $G_p^{u+d}(Q^2)$ using the same analysis as for the isovector. Once more, excited states are clearly more severe for $G_p^{u+d}(Q^2)$ at low $Q^2$ where the pion pole dominates and tends to decrease its value leading to a milder $Q^2$-dependence.

In Fig.~\ref{Fig:GA_Gp_isoscalar_disc} we show the disconnected
contributions to $G_A^{u+d}(Q^2)$, which are clearly non-zero and
negative. 
The form factors for the disconnected contributions are obtained
combining final nucleon states with $\vec{p}^\prime = \vec{0}$, the
same as in the case of the connected contributions, and in addition
all sink momenta which satisfy $\vec{p}^{\prime 2}=(2 \pi/L)^2$. Since
more $Q^2$ values are available, we plot, in
Fig.~\ref{Fig:GA_Gp_isoscalar_disc}, the sink-source separation
$t_s=1.31$~fm and two-state fit methods alone for better clarity.
The disconnected contributions reduce the value of $G_A^{u+d}(Q^2)$
and for zero momentum transfer result in a value compatible with the
experimental one.  As already mentioned, the disconnected
contributions to $G_p^{u+d}(Q^2)$ are particularly large and reduce
its value especially at low values of $Q^2$. 
Adding the connected and disconnected contributions obtained using
$\vec{p}\,'=\vec{0}$ for which common $Q^2$-values are available, yields
the result shown in Fig.~\ref{Fig:GA_Gp_isoscalar_total}. 
We note that, due to the fact that the disconnected part is computed
with much higher statistics as compared to the connected, the error in
the total quantity is computed by adding the individual errors in
quadrature.  In Fig.~\ref{Fig:GA_Gp_isoscalar_bands} we show the
resulting dipole fits to the isoscalar form factor $G_A^{u+d}(Q^2)$
using Eq.~(\ref{Eq:GA_dipole}) for the connected, disconnected and
total value. The parameters extracted are collected in
Table~\ref{Tab:Values_dipole_fit}. The axial mass extracted by fitting
$G_A^{u+d}(Q^2)$ is $m_A^{u+d}=1.736(244)(374)$~GeV. Although the
central value is larger, within the large statistical and systematic
errors it is in agreement with the one extracted for the isovector
case. In Table~\ref{Tab:Values_dipole_fit} we also list the
corresponding axial radii, obtained from the dipole masses via
Eq.(\ref{Eq:AxialRadius}).

For $G_p(Q^2)$ we fit using Eq.~(\ref{Eq:Gp_dipole}) for the connected
and disconnected separately, allowing $C$ and $m_p$ to vary. We obtain
the curves shown in Fig.~\ref{Fig:GA_Gp_isoscalar_bands} and
consistent pole masses, namely $m_p^\textrm{u+d,conn} =
0.324(22)(12)$~GeV for the connected and 
$m_p^\textrm{u+d,disc} = 0.331(81)(36)$~GeV 
for the disconnected. We show the total isoscalar $G_p(Q^2)$ in Fig.~\ref{Fig:GA_Gp_isoscalar_bands}. As can be seen the errors are large, especially in the small $Q^2$ region,  and do not allow us to reliably quote  a value for the pole mass of $G_p(Q^2)$.

In Fig.~\ref{Fig:GAs_GAc_Gps_Gpc} we show the strange and the charm form factors, which only take disconnected contributions. 
For the strange quark contributions, we use sink momenta $\vec{p}\,'=\vec{0}$ and $\vec{p}\,'^2=(2\pi/L)^2$, while for the charm quark where errors are large only the $\vec{p}\, '=\vec{0}$ case yields reasonable results. 
We observe a very good signal for $G_A^s(Q^2)$ up to momentum transfer $Q^2=0.5$~GeV$^2$. As already noted our results from the plateau method with $t_s=1.31$~fm are consistent with the two-state fit.  $G_A^s(Q^2)$ can be well fitted to a dipole form and we obtain 
$m_A^s=0.921(228)(90)$~GeV. 
The results for $G_A^s(Q^2)$ are compatible with the experimental
values measured for $Q^2>0.45$~GeV$^2$, which however carry large
errors~\cite{Pate:2008va}.  $G_A^c(Q^2)$ is more noisy in particular
for the larger time separations. For the smallest source-sink
separation of $t_s=0.94$~fm we obtain a non-zero negative value for
the whole range of $Q^2$. However, for larger values of $t_s$ the
results become noisy forbidding us to reach a conclusion on excited
states contributions. For $G_p^s(Q^2)$ we obtain a non-zero negative
contribution, which is about six times smaller in magnitude compared
to the disconnected $G_p^{u+d}(Q^2)$. In the case of $G_p^c(Q^2)$
results are compatible with zero even for the smallest source-sink
time separation. We do not display the results produced with the
summation method since these are very noisy.
\begin{figure}[!ht]
\begin{center}
  \begin{minipage}[t]{0.47\linewidth}
    \includegraphics[width=\textwidth]{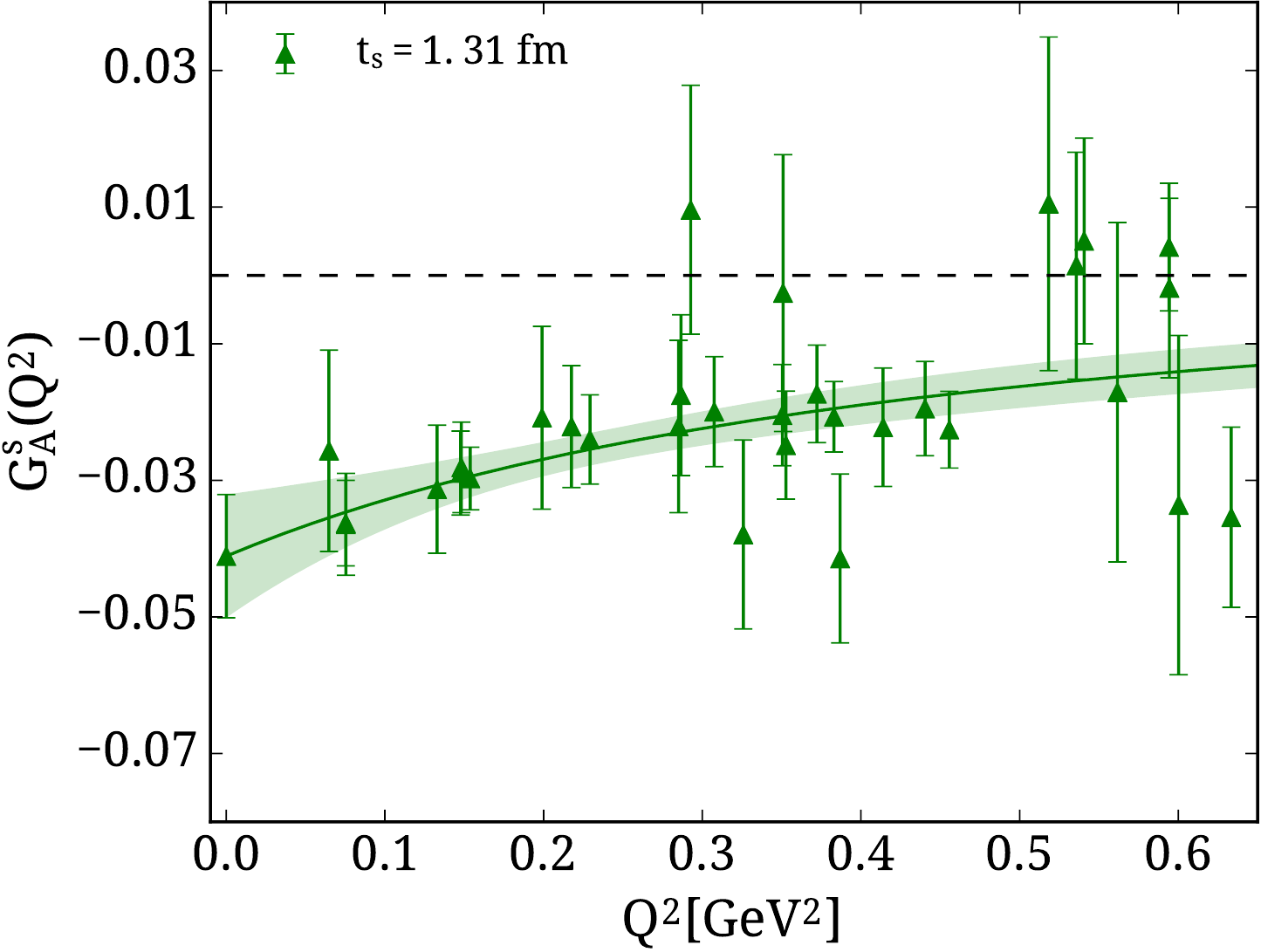}
  \end{minipage}
  \hfill
  \begin{minipage}[t]{0.47\linewidth}
    \includegraphics[width=\textwidth]{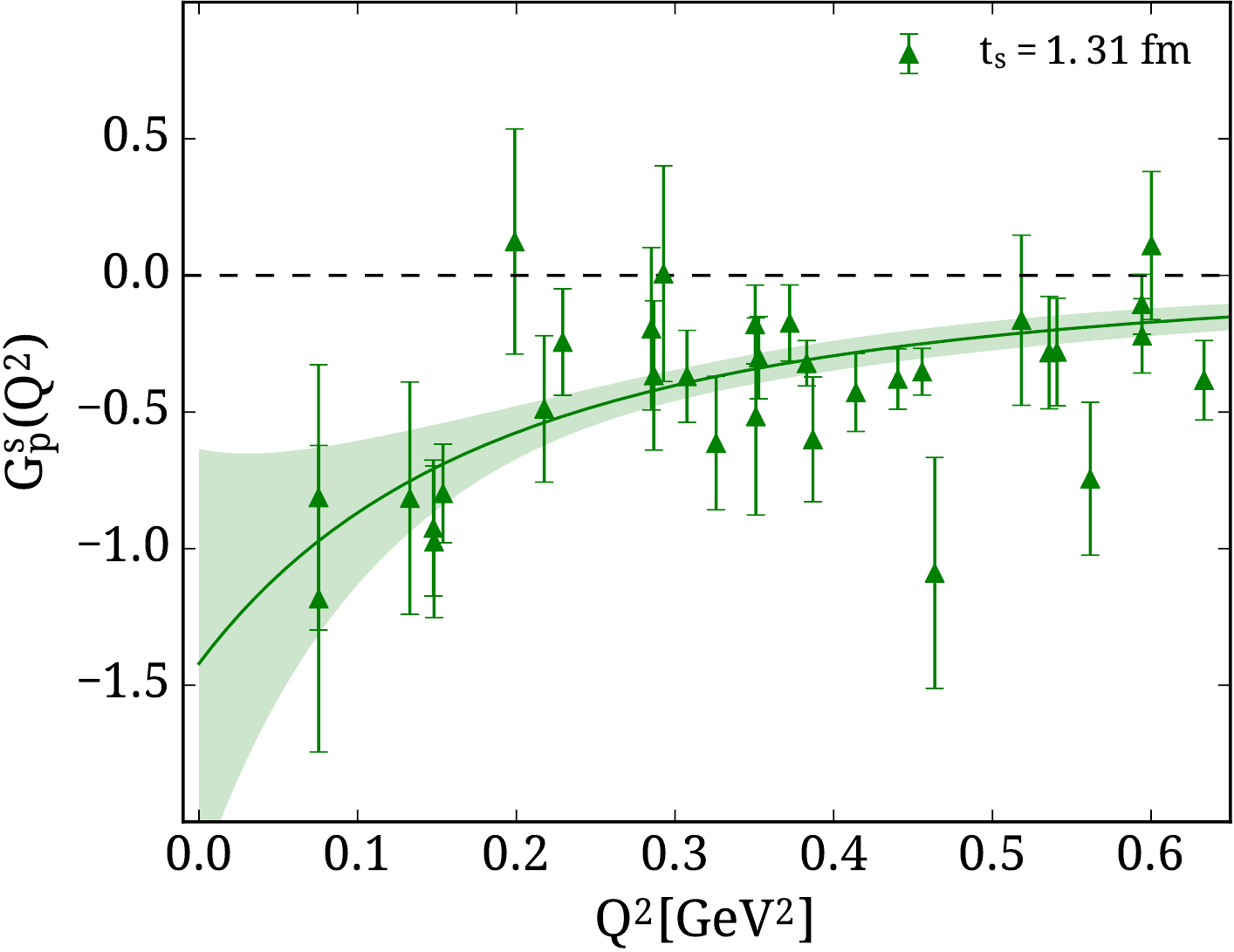}
  \end{minipage}
\hfill
\caption{ Fits to  $G_A^s(Q^2)$  using  Eq.~(\ref{Eq:GA_dipole}) (left) and $G_p^s(Q^2)$ using Eq.~(\ref{Eq:Gp_dipole}) (right).}
\label{Fig:GAsc_Gpsc_bands}
\end{center}  
\end{figure}
In Fig.~\ref{Fig:GAsc_Gpsc_bands} we show fits to $G_A^s(Q^2)$ using
the dipole form of Eq.~(\ref{Eq:GA_dipole}) and $G_p^s(Q^2)$ using
Eq.~(\ref{Eq:Gp_dipole}).  For $G_p^c(Q^2)$ the results are compatible
with zero and no fit is attempted. Axial masses and corresponding
radii extracted from the dipole fits are tabulated in
Table~\ref{Tab:Values_dipole_fit}.

 \begin{table}[!ht]
	\begin{center}
		\begin{tabular}{cccc}
		\hline\hline
		Form factor & $m_A$ [GeV] & $\langle r_A^2\rangle$ $[$fm$^2]$ & $\chi^2$/d.o.f \\
		\hline
                $G_A^{u-d}$       & 1.322(42)(17)  & 0.266(17)(7) & 0.35 \\
                $G_A^{u+d}$  & 1.736(244)(374)& 0.155(43)(96) & 0.64 \\
                $G_A^u$      & 1.439(28)(114) & 0.225(28)(40) & 0.61 \\
                $G_A^d$      & 1.243(49)(133) & 0.301(24)(55) & 0.42 \\
                $G_A^s$     & 0.921(228)(90) & 0.549(272)(93)& 0.30 \\
                \hline\hline
		\end{tabular}
	\end{center}
	\caption{Extracted values for the axial masses and
          corresponding axial radii using dipole fits to
          Eq.~(\ref{Eq:GA_dipole}) with their associated
          $\chi^2/\textrm{d.o.f}$.  The central value and statistical error is
          from fits to results using the plateau method at
          $t_s=1.31$~fm. The first error is statistical while the
          second is the systematic due to excited states, taken as the
          difference between the central value and the value extracted
          from the two-state fit.}
\label{Tab:Values_dipole_fit}
\end{table}

\subsection{Comparison with other studies}

The axial and induced pseudo-scalar form factors have  been studied by several   lattice QCD groups using recent dynamical simulations.
Preliminary lattice QCD results using an ensemble with close to physical pion mass has been presented by the PNDME collaboration~\cite{Jang:2016kch}. They use a mixed action approach of $N_f=2+1+1$ HISQ staggered fermions and clover-improved Wilson valence fermions. This action has ${\cal O}(a)$ lattice artifacts, which are shown to be sizeable for $a=0.09$~fm as compared to  their results at $a=0.06$~fm. Their preliminary results on $G_A^{u-d}$ using an ensemble at pion mass $m_\pi= 130$~MeV and $a=0.06$~fm are in agreement with ours. This shows that lattice artifacts for our ${\cal O}(a)$ improved action computed with $a=0.0938$~fm are small. On the other hand, their results on $G_p^{u-d}(Q^2)$ for the same ensemble are larger at low $Q^2$-values than ours. Given that their spatial box length is $L\sim 5.76$~fm as compared to $L\sim 4.51$~fm of our lattice, these preliminary results may indicate that $G_p^{u-d}(Q^2)$ suffers from sizeable finite volume effects. Additional lattice QCD results on the isovector axial form factors at higher than physical pion mass have been computed recently by two groups: LHPC has obtained results on the isovector axial form factors using $N_f=2+1$ clover-improved Wilson fermions with $m_\pi=317$~MeV~\cite{Green:2017keo}, which includes the isoscalar form factors and using a mixed action for $m_\pi=356$~MeV~\cite{Bratt:2010jn}. CLS has presented preliminary results using an  ensemble of $N_f=2$ clover fermions  at a pion mass of $m_\pi \sim 340$~MeV~\cite{vonHippel:2016nfg}. In what follows we restrict ourselves to showing published results only.

\begin{figure}[!ht]
  \begin{minipage}[t]{0.47\linewidth}
    \includegraphics[width=\textwidth]{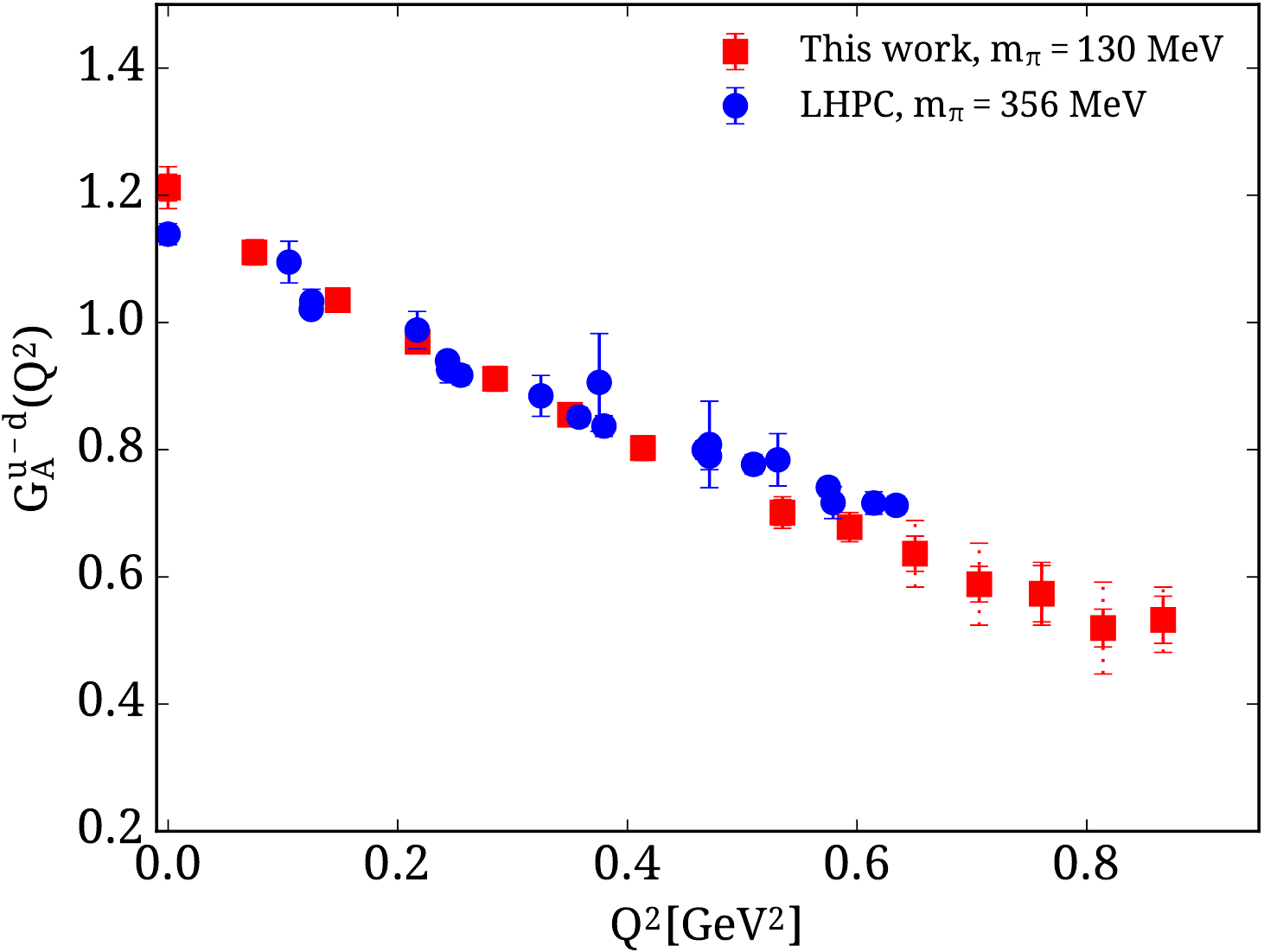}
  \end{minipage}
  \hfill
  \begin{minipage}[t]{0.47\linewidth}
    \includegraphics[width=\textwidth]{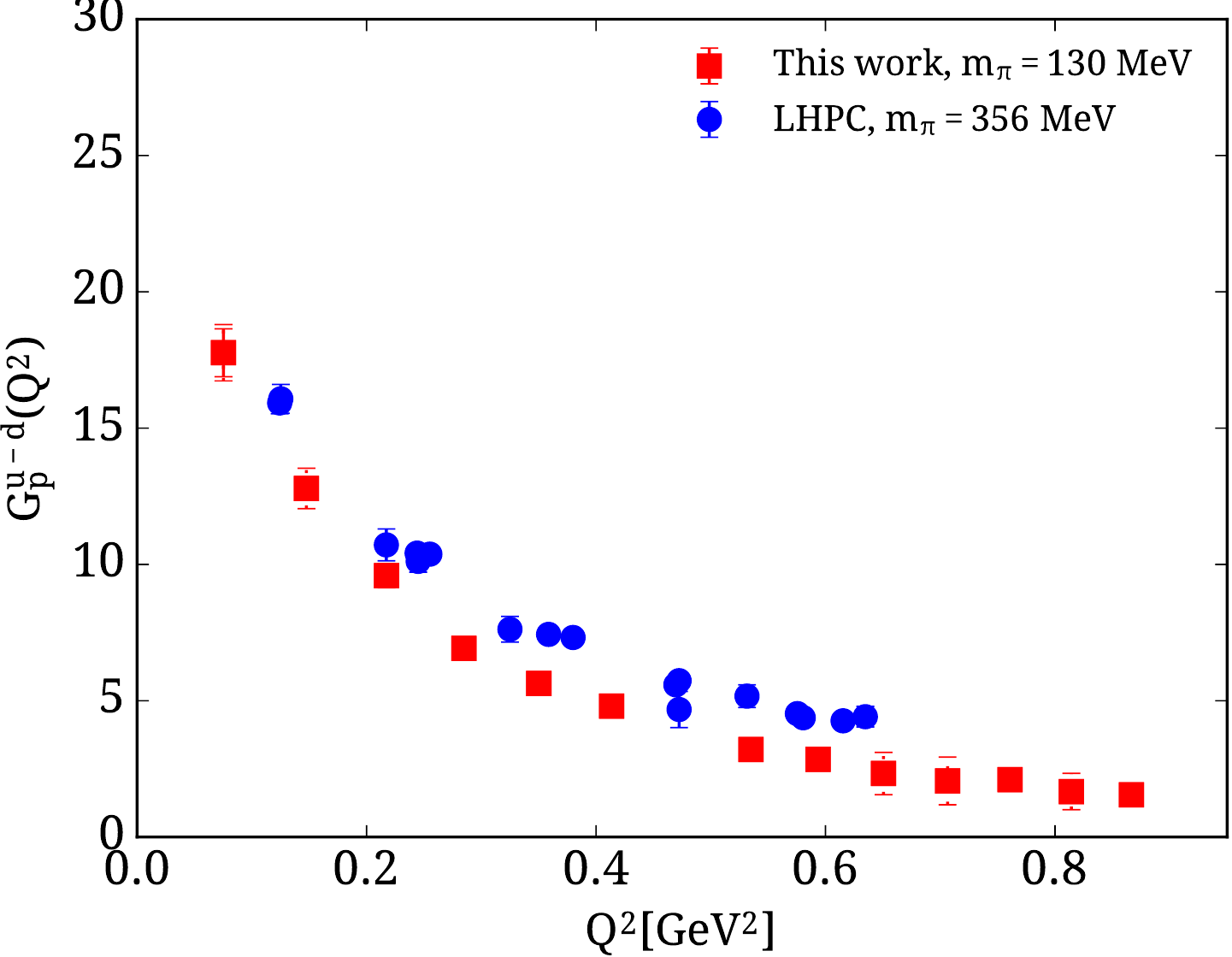}
  \end{minipage}
  \caption{ Comparison for the isovector axial (left) and induced pseudo-scalar (right) form factors with the results from LHPC. The red squares show the results from this work and  the blue circles results from LHPC at $m_{\pi}=356$ MeV~\cite{Bratt:2010jn}. The dotted red error bar shows our systematic error.}
  \label{Fig:GA_Gp_comp}
\end{figure}

In Fig.~\ref{Fig:GA_Gp_comp} we compare our results for the isovector axial form factors to the published LHPC results, which have been produced using $N_f=2+1$ Asqtad staggered sea quarks on a $28^3\times 64$ lattice and Domain-Wall valence fermions for $m_\pi=356$~MeV~\cite{Bratt:2010jn}. Our results for $G_A^{u-d}$ at the physical point  show a steeper $Q^2$-dependence leading to a larger value of $g_A$. For $G_p^{u-d}(Q^2)$, the LHPC results tend to be larger in particular at the smallest $Q^2<0.2$~GeV$^2$. The length of their lattice is $L=3.36$~fm yielding $Lm_\pi\sim 6$ as compared to $Lm_\pi\sim 3.3$~fm for our lattice. This may again point to finite volume effects that need to be investigated.

\section{Conclusions}
\label{sec:conclusions}
Results on the nucleon axial form factors are presented for one ensemble of two degenerate twisted mass clover-improved fermions  tuned to reproduce approximately the  physical value of the pion mass. Using improved techniques we evaluate both connected and  disconnected contributions to both axial and induced pseudo-scalar form factors. Our study includes an   investigation of excited state effects by computing the nucleon three-point functions at several sink-source time separations. Lattice matrix elements are non-perturbatively renormalized by computing both the singlet and non-singlet renormalization functions. 

We find that the isovector axial form factor  $G_A^{u-d}(Q^2)$ is described well by a dipole form with an axial mass $m_A=1.322(42)(17)$~GeV, which is larger than the historical  world average but is in agreement with a recent value produced from the MiniBooNE experiment~\cite{AguilarArevalo:2010zc}. We can relate, via PCAC, the axial form factor to the $\pi N$ coupling constant. We find that 
$g_{\pi NN}=12.81(37)(21)$
consistent with the experimental value of $g_{\pi NN}=13.12(10)$~\cite{Baru:2010xn}. Similarly, we can deduce $G^{u-d}_p(Q^2)$ from our results on $G_A^{u-d}(Q^2)$ assuming pion pole dominance yielding agreement with experiment. However, a direct extraction of  the  isovector induced pseudo-scalar form factor has a weaker $Q^2$-dependence as compared to what is expected from pion pole dominance. Thus, although  one can describe well the data using a pion pole behavior for its $Q^2$-dependence, one extracts a pole mass larger than the ensemble value of $m_\pi=130$~MeV. $G_p(Q^2)$ at low $Q^2$ is shown to have more severe excited states effects, which tend to lower its value. Comparison to preliminary lattice results obtained on a larger volume~\cite{Jang:2016kch} indicate that volume effects may also increase its value at low $Q^2$. We plan to check for such volume effects in a future analysis using a larger lattice. 

An important conclusion of this work is that disconnected contributions to both isoscalar and strange form factors are non-negligible. For the isoscalar $g_A^{u+d}$ these contributions need to be taken into account to bring agreement with the experimental value. For $G_p^{u+d}(Q^2)$ the disconnected contributions are particularly large and of the same order as the connected part but with the opposite sign leading to a weaker $Q^2$-dependence for the isoscalar pseudo-scalar form factor. Both strange from factors $G_A^s(Q^2)$ and $G_p^s(Q^2)$ are found to be negative and non-zero, with  the magnitude of  $G_A^s(Q^2)$ of the same order as that for the light disconnected contributions. Both  charm from factors tend to be negative but given the large errors  they remain compatible with zero.

\section{ACKNOWLEDGMENTS}
We would like to thank the members of the ETMC for a most enjoyable collaboration. We acknowledge funding from the European Union's Horizon 2020 research and innovation programme under the Marie Sklodowska-Curie grant agreement No 642069. This work was partly  supported by a grant from the Swiss National Supercomputing Centre (CSCS) under project IDs \texttt{s540} and \texttt{s625} on the Piz Daint system, by a Gauss allocation on SuperMUC with ID \texttt{44060} and in addition used computational resources  from the John von Neumann-Institute for Computing on the Jureca and
the BlueGene/Q Juqueen systems at the research center in J\"ulich. We also acknowledge PRACE for awarding us access to the Tier-0 computing resources Curie, Fermi and
SuperMUC based in CEA, France, Cineca, Italy and LRZ, Germany, respectively. 
We thank the staff members at all sites for their kind and sustained
support.  K. H. and Ch. K. acknowledge support from the Cyprus Research Promotion Foundation under contract T$\Pi$E/$\Pi\Lambda$HPO/0311(BIE)/09. 

\clearpage

\appendix
\begin{center}
{\bf Appendix A: Table of Results}
\end{center}

\begin{table}[!ht]
\renewcommand*{\arraystretch}{1.4}
	\begin{center}
		\begin{tabular}{l|c|c|c|c|c}
			$Q^2[GeV^2]$ & $G_A^{u-d}$ & $G_A^{u+d}$(Conn.) & $G_A^{u+d}$(Tot.) & $G_A^u$(Tot.) & $G_A^d$(Tot.) \\ 
\hline\hline0.0000 & 1.212(33)(22) & 0.595(29)(1) & 0.445(35)(18) & 0.827(30)(5) & -0.380(15)(23) \\ 
0.0753 & 1.110(20)(16) & 0.551(17)(11) & 0.439(23)(20) & 0.772(18)(2) & -0.339(12)(16) \\ 
0.1477 & 1.035(14)(17) & 0.518(14)(10) & 0.430(18)(44) & 0.728(13)(12) & -0.308(10)(29) \\ 
0.2174 & 0.970(18)(10) & 0.484(18)(13) & 0.389(27)(33) & 0.676(19)(8) & -0.294(13)(23) \\ 
0.2849 & 0.911(20)(5) & 0.458(17)(1) & 0.377(31)(45) & 0.641(20)(29) & -0.270(17)(29) \\ 
0.3502 & 0.855(18)(9) & 0.438(14)(1) & - & - & - \\ 
0.4135 & 0.802(20)(9) & 0.413(14)(3) & - & - & - \\ 
0.5351 & 0.701(25)(21) & 0.385(17)(23) & - & - & - \\ 
0.5936 & 0.678(23)(18) & 0.364(15)(1) & - & - & - \\ 
0.6506 & 0.636(28)(53) & 0.321(18)(17) & - & - & - \\ 
0.7064 & 0.588(28)(65) & 0.289(24)(35) & - & - & - \\ 
0.7609 & 0.573(50)(45) & 0.307(42)(29) & - & - & - \\ 
0.8143 & 0.520(30)(73) & 0.265(23)(7) & - & - & - \\ 
0.8666 & 0.533(37)(52) & 0.301(24)(22) & - & - & - \\ 
0.9683 & 0.096(1.665)(161) & 0.056(470)(169) & - & - & - \\ 

			\hline
		\end{tabular}
	\end{center}
	\caption{Our values for the axial form factor for various values of $Q^2$. The first column gives $Q^2$ in $\mathrm{GeV^2}$, the second is the isovector axial form factor, the third is the connected contribution to the isoscalar axial form factor, the forth column gives the total value of the isoscalar, while the fifth and the sixth columns are the total contributions from the up and down quarks separately.}
		\label{Table:GA_light_conn_tot}
\end{table}

\begin{table}[!ht]
\renewcommand*{\arraystretch}{1.4}
	\begin{center}
		\begin{tabular}{l|c|c|c|c|c|c}
			$Q^2[GeV^2]$ & $G_p^{u-d}$ & $G_p^{u+d}$(Conn.) & $G_p^{u+d}$(Tot.) & $G_p^u$(Tot.) & $G_p^d$(Tot.) \\ 
\hline\hline0.0753 & 17.766(879)(1.032) & 10.155(874)(878) & 4.045(1.491)(1.401) & 10.817(990)(909) & -7.016(725)(66) \\ 
0.1477 & 12.788(417)(739) & 6.893(324)(125) & 2.338(651)(915) & 7.519(415)(148) & -5.410(340)(729) \\ 
0.2174 & 9.588(391)(441) & 4.917(371)(269) & 1.455(713)(460) & 5.394(452)(181) & -4.218(360)(204) \\ 
0.2849 & 6.923(326)(194) & 3.977(276)(264) & 1.750(679)(745) & 4.314(384)(297) & -2.673(360)(626) \\ 
0.3502 & 5.636(220)(56) & 3.146(183)(43) & - & - & - \\ 
0.4135 & 4.801(191)(310) & 2.557(148)(107) & - & - & - \\ 
0.5351 & 3.214(161)(384) & 1.954(157)(38) & - & - & - \\ 
0.5936 & 2.850(147)(107) & 1.741(108)(188) & - & - & - \\ 
0.6506 & 2.330(164)(777) & 1.544(136)(632) & - & - & - \\ 
0.7064 & 2.060(156)(874) & 1.116(150)(140) & - & - & - \\ 
0.7609 & 2.107(248)(168) & 1.324(220)(44) & - & - & - \\ 
0.8143 & 1.669(115)(665) & 0.974(127)(273) & - & - & - \\ 
0.8666 & 1.552(135)(202) & 1.058(122)(91) & - & - & - \\ 
0.9683 & 0.278(6.813)(159) & -0.013(2.984)(609) & - & - & - \\ 

			\hline
		\end{tabular}
	\end{center}
	\caption{Our values for the induced pseudo-scalar form factor for various values of $Q^2$. The notation is the same as in Table~\ref{Table:GA_light_conn_tot}.}
	\label{Table:Gp_light_conn_tot}
\end{table}

\begin{table}[!ht]
\renewcommand*{\arraystretch}{1.4}
	\begin{center}
		\begin{tabular}{l|c|c}
			$Q^2[GeV^2]$ & $G_A^{u+d}$(Disc.) & $G_p^{u+d}$(Disc.) \\ 
\hline\hline0.0000 & -0.150(20)(19) & - \\ 
0.0647 & -0.096(33)(10) & -6.264(2.584)(720) \\ 
0.0753 & -0.111(13)(16) & -6.160(969)(392) \\ 
0.0754 & -0.127(14)(9) & -7.947(1.147)(184) \\ 
0.1329 & -0.083(19)(13) & -4.702(981)(1) \\ 
0.1477 & -0.088(12)(25) & -4.555(565)(579) \\ 
0.1482 & -0.101(15)(15) & -5.058(621)(138) \\ 
0.1538 & -0.100(10)(19) & -4.356(407)(303) \\ 
0.1990 & -0.105(29)(27) & -3.611(882)(37) \\ 
0.2176 & -0.095(20)(27) & -3.462(608)(12) \\ 
0.2292 & -0.095(14)(27) & -2.838(418)(483) \\ 
0.2331 & -0.105(63)(33) & -0.862(1.771)(1.427) \\ 
0.2851 & -0.081(25)(34) & -2.227(621)(698) \\ 
0.2866 & -0.088(24)(23) & -2.405(570)(406) \\ 
0.3075 & -0.071(16)(26) & -2.251(389)(275) \\ 

			\hline
		\end{tabular}
	\end{center}
	\caption{Our values for the disconnected contributions to $G_A^{u+d}$ and $G_p^{u+d}$ as a function of $Q^2$.}
	\label{Table:GA_Gp_disc}
\end{table}

\begin{table}[!ht]
\renewcommand*{\arraystretch}{1.4}
	\begin{center}
		\begin{tabular}{l|c|c}
			$Q^2[GeV^2]$ & $G_A^{s}$ & $G_p^{s}$ \\ 
\hline\hline0.0000 & -0.0427(100)(93) & - \\ 
0.0647 & -0.0257(148)(101) & -0.812(486)(99) \\ 
0.0753 & -0.0363(63)(44) & -1.183(561)(329) \\ 
0.0754 & -0.0364(75)(42) & -0.815(426)(354) \\ 
0.1329 & -0.0313(94)(93) & -0.925(249)(72) \\ 
0.1477 & -0.0289(62)(51) & -0.975(278)(24) \\ 
0.1482 & -0.0281(67)(15) & -0.798(181)(55) \\ 
0.1538 & -0.0297(46)(38) & 0.124(412)(158) \\ 
0.1990 & -0.0208(134)(68) & -0.488(268)(8) \\ 
0.2176 & -0.0221(90)(97) & -0.244(196)(130) \\ 
0.2292 & -0.0240(66)(71) & -0.196(297)(260) \\ 
0.2851 & -0.0221(127)(66) & -0.366(274)(142) \\ 
0.2866 & -0.0175(118)(63) & 0.007(395)(103) \\ 
0.2927 & 0.0096(183)(201) & -0.370(169)(212) \\ 
0.3075 & -0.0199(81)(70) & -0.614(245)(256) \\ 
0.3259 & -0.0379(139)(112) & -0.180(145)(45) \\ 
0.3505 & -0.0205(75)(43) & -0.516(361)(102) \\ 
0.3509 & -0.0026(203)(61) & -0.301(151)(33) \\ 
0.3528 & -0.0249(80)(8) & -0.174(140)(167) \\ 
0.3723 & -0.0173(72)(106) & -0.321(84)(83) \\ 
0.3830 & -0.0207(52)(83) & -0.600(229)(100) \\ 
0.3869 & -0.0414(124)(63) & -0.428(143)(77) \\ 
0.4140 & -0.0222(87)(57) & -0.379(111)(123) \\ 
0.4404 & -0.0195(69)(68) & -0.352(86)(70) \\ 
0.4557 & -0.0226(57)(29) & -1.089(423)(734) \\ 
0.5183 & 0.0105(245)(372) & -0.164(312)(256) \\ 
0.5358 & 0.0014(167)(57) & -0.283(206)(96) \\ 
0.5407 & 0.0050(151)(117) & -0.280(198)(94) \\ 
0.5617 & -0.0171(249)(292) & -0.744(281)(119) \\ 
0.5942 & 0.0042(94)(99) & -0.106(110)(49) \\ 
0.5944 & -0.0018(132)(184) & -0.221(137)(108) \\ 
0.6002 & -0.0336(249)(167) & 0.109(272)(139) \\ 
0.6334 & -0.0354(132)(51) & -0.383(146)(3) \\ 
0.6603 & -0.0176(74)(87) & -0.381(79)(15) \\ 
0.6798 & -0.0005(148)(91) & 0.106(143)(201) \\ 

			\hline
		\end{tabular}
	\end{center}
	\caption{Our values for the $G_A^{s}$ and $G_p^{s}$ as a function of $Q^2$.}
	\label{Table:GAp_s}
\end{table}

\begin{table}[!ht]
\renewcommand*{\arraystretch}{1.4}
	\begin{center}
		\begin{tabular}{l|c|c}
			$Q^2[GeV^2]$ & $G_A^{c}$ & $G_p^{c}$ \\ 
\hline\hline0.0000 & -0.00338(189)(668) & - \\ 
0.0753 & -0.00351(144)(481) & 0.085(117)(208) \\ 
0.1477 & -0.00256(110)(218) & -0.052(49)(184) \\ 
0.2176 & -0.00230(161)(372) & -0.022(48)(62) \\ 
0.2851 & -0.00523(218)(308) & -0.062(51)(187) \\ 
0.3505 & -0.00178(131)(429) & -0.010(24)(120) \\ 
0.4140 & -0.00193(162)(189) & 0.000(22)(117) \\ 
0.5358 & -0.00178(238)(768) & -0.014(29)(49) \\ 
0.5944 & -0.00338(174)(147) & 0.013(18)(4) \\ 

			\hline
		\end{tabular}
	\end{center}
	\caption{Our values for the $G_A^{c}$ and $G_p^{c}$ as a function of $Q^2$.}
	\label{Table:GAp_c}
\end{table}

\clearpage
\bibliographystyle{unsrtnat}
\bibliography{refs}

\end{document}